\documentclass[11pt]{article}
\pdfoutput=1
\usepackage{jheppubme}
\usepackage[table]{xcolor}
\usepackage{graphicx, psfrag, dsfont, amsfonts}
\usepackage{float}
\usepackage{caption, comment}
\usepackage[mathscr]{euscript}
\usepackage{dynkin-diagrams}
\usepackage{appendix}
\usepackage{coolstr}
\usepackage{hyperref}
\usepackage[normalem]{ulem}
\hypersetup{pdfauthor={Name}}
\usepackage{color}
\usepackage{chngcntr}
\counterwithin{table}{subsection}
\usepackage{booktabs,caption}
\usepackage{lscape}
\definecolor{Mygrey}{gray}{0.8}
\definecolor{Mywhite}{gray}{1.0}

\newcommand{\be}{\begin{equation}}
\newcommand{\ee}{\end{equation}}
\newcommand{\bea}{\begin{eqnarray}}
\newcommand{\eea}{\end{eqnarray}}
\usepackage{amsmath,amssymb,epsfig,jheppubme}
\usepackage{multirow,longtable,enumerate,bm}
\usepackage[flushleft]{threeparttable}

\newcommand\half{\frac12}
\newcommand\del{\partial}
\newcommand\bi{\begin{itemize}}
\newcommand\ei{\end{itemize}}

\newcommand{\alpharho}{\alpha^{(\rho)}}
\newcommand{\alpharhozero}{{\alpha^{(\rho)}_0}}
\newcommand{\alpharhoone}{{\alpha^{(\rho)}_1}}
\newcommand{\alphai}{\alpha^{(i)}}

\newcommand\nn{\nonumber}

\newcommand{\cO}{{\cal O}}

\newcommand\sfrac[2]{{\textstyle\frac{#1}{#2}}}

\newcommand\ZZ{\hbox{Z\kern-.4emZ}}
\newcommand\sZZ{\hbox{\sevenfont Z\kern-.4emZ}}

\newcommand{\eref}[1]{Eq.\,(\ref{#1})}
\newcommand{\Comment}[1]{{}}

\newcommand{\lbrac}[1]{\left(\rule{0cm}{#1}\right.}
\newcommand{\rbrac}[1]{\left. \rule{0cm}{#1}\right)}

\definecolor{Mygrey}{gray}{0.8}
\definecolor{Mywhite}{gray}{1.0}

\colorlet{dred}{red!70!black!100!}

\def\IB{\relax{\rm I\kern-.18em B}}

\def\ID{\relax{\rm I\kern-.18em D}}
\def\IE{\relax{\rm I\kern-.18em E}}
\def\IF{\relax{\rm I\kern-.18em F}}
\def\II{\relax{\rm I\kern-.18em I}}

\def\Id{\relax{1\kern-.32em 1}}
\def\IG{\relax\hbox{$\inbar\kern-.3em{\rm G}$}}
\def\IR{\relax{\rm I\kern-.18em R}}

\title{Modular Differential Equations with Movable Poles and Admissible RCFT Characters} 
\author[a]{Arpit Das,}
\author[b,c]{Chethan N. Gowdigere,} 
\author[d]{Sunil Mukhi,\footnote{Adjunct Professor, ICTS-TIFR, Bengaluru, and Honorary Emeritus Professor, IISER Pune.}}
\author[e]{Jagannath Santara}

\affiliation[a]{Centre for Particle Theory, Department of Mathematical Sciences,\\
Durham University, South Road, Durham DH1 3LE, United Kingdom}
\affiliation[b]{National Institute of Science Education and Research Bhubaneshwar,\\
P.O. Jatni, Khurdha, 752050, India}
\affiliation[c]{Homi Bhabha National Institute, Training School Complex,\\
Anushakti Nagar, Mumbai 400094, India}
\affiliation[d]{Theoretical Physics Department, CERN,\\
CH-1211 Geneva 23, Switzerland}
\affiliation[e]{Department of Physics, Indian Institute of Technology Madras,\\ Chennai 600036, India}

\emailAdd{arpit.das@durham.ac.uk}
\emailAdd{chethan.gowdigere@niser.ac.in}
\emailAdd{sunil.mukhi@gmail.com}
\emailAdd{jagannath.santara@physics.iitm.ac.in}
    
\abstract{Studies of modular linear differential equations (MLDE) for the classification of rational CFT characters have been limited to the case where the coefficient functions (in monic form) have no poles, or poles at special points of moduli space. Here we initiate an exploration of the vast territory of MLDEs with two characters and any number of poles at arbitrary points of moduli space. We show how to parametrise the most general equation precisely and count its parameters. Eliminating logarithmic singularities at all the poles provides constraint equations for the accessory parameters. By taking suitable limits, we find recursion relations between solutions for different numbers of poles. The cases of one and two movable poles are examined in detail and compared with predictions based on quasi-characters to find complete agreement. We also comment on the limit of coincident poles. Finally we show that there exist genuine CFT corresponding to many of the newly-studied cases. We emphasise that the modular data is an output, rather than an input, of our approach.}

\preprint{CERN-TH-2023-147}

\keywords{Conformal Field Theory}

\begin{document}

\maketitle

\section{Introduction}

The method of Modular Linear Differential Equations (MLDEs)  for the classification of Rational Conformal Field Theories (RCFT) in 2d  \cite{Mathur:1988na, Mathur:1988gt} has experienced a significant resurgence, with several important results having appeared in recent times \cite{Harvey:2018rdc, Kawasetsu:2018tzs, Arike:2018fru, Chandra:2018pjq, Bae:2018qym, Mukhi:2020gnj, Grady_2020, franc2020classification,
Kaidi:2020ecu, Bae:2020xzl, Kaidi:2021ent, Das:2021uvd, Bae:2021mej, Franc:2021hwo,
Kaidi:2022sng, Mukhi:2022bte, Duan:2022kxr, Lee:2023owa, Pan:2023jjw}. This approach implements modular invariance and positivity of degeneracies of states to constrain possible consistent partition functions of RCFT having a small number of primaries under some, possibly extended, chiral algebra. When combined with additional information, it leads to classifications of all RCFT within specified regions of parameter space (see for example \cite{Mukhi:2022bte, Das:2022uoe, Rayhaun:2023pgc}).

Originally the method was applied to cases with two or three linearly independent characters satisfying what is now called the ``non-zero Wronskian condition'' \footnote{Some authors refer to this case as ``monic'', e.g. \cite{Beem:2017ooy, Kaidi:2021ent}, to indicate that the MLDE when written in monic form is free of poles.}, which is the vanishing of a non-negative integer $\ell$ proportional to the number of zeroes of a certain determinant (we explain this in more detail in the following Section). $\ell$ is known as the Wronskian index.  The work of \cite{Mathur:1988na} completely classified admissible solutions to the two-character MLDE with $\ell=0$. Here ``admissible'' means the  solutions have non-negative integral Fourier coefficients, and also that the identity character is normalised to start with unity, reflecting uniqueness of the vacuum state. For two characters, it turned out that the admissible solutions all correspond to CFTs though there are a couple of subtleties that we will not go into here.

The three-character case, again with $\ell=0$, was investigated for the first time in \cite{Mathur:1988gt} where several interesting results were found, but not a complete classification of admissible solutions. These results were extended many years later in \cite{Gaberdiel:2016zke, Franc:2016} and then more completely in \cite{franc2020classification, Kaidi:2021ent, Das:2021uvd, Bae:2021mej} resulting in a complete set of admissible characters of which a large fraction could be identified as CFTs. In \cite{Das:2022uoe}, additional information was used to tabulate the complete set of CFTs with three characters and $\ell=0$. 

Studying MLDE and admissible solutions beyond $\ell=0$ is more difficult and there are very few papers in this direction. For the two-character case, solutions with $\ell=2$ were considered in \cite{Naculich:1988xv, Hampapura:2015cea, Gaberdiel:2016zke}, while solutions with $\ell=4$ were analysed from the MLDE perspective in \cite{Naculich:1988xv, Tener:2016lcn, Grady_2020}. Reference \cite{Das:2021uvd} studied three characters for $\ell=2$ and \cite{Kaidi:2021ent} classified solutions with three, four and five characters and $\ell=0$. To our knowledge, no analysis of the $\ell\ge 6$ case has been carried out even for two characters. Indeed there seems to be a consensus that the MLDE approach is intractable for $\ell\ge 6$, and to our knowledge no attempt has been made to formulate and solve the MLDE in such cases \footnote{However there were some remarkably prescient observations in this direction in the concluding section of \cite{Naculich:1988xv}.}. This will be the main focus of the present work.

When $\ell<6$ the poles in the MLDE, if any, must be located at the special points $\tau=\rho$ or $\tau=i$ in moduli space. On the other hand for $\ell\ge 6$,  the MLDE can have poles at generic points in moduli space. Hence we refer to the $\ell\ge 6$ case as having ``movable poles''. To be clear, this only means that their locations are free parameters in the equation, but of course for any particular admissible solution the poles will take fixed values. Also it is important to emphasise that the solutions of the MLDE have no poles and are regular everywhere, leading to completely regular candidate partition functions. The poles are present only in the coefficient functions of the MLDE itself.

There exist other approaches beyond MLDE that have provided insights into admissible characters with $\ell\ge 6$. These approaches avoid explicitly solving, or even formulating, an $\ell\ge 6$ MLDE. For example, \cite{Harvey:2018rdc} employed a novel construction of Hecke operators on vector-valued modular forms. On the other hand, \cite{Chandra:2018pjq, Chandra:2018ezv} proposed the method of ``quasi-characters'' about which we will say more below.
 
The primary motivation of the present work is to study modular differential equations and their admissible solutions for arbitrary $\ell\ge 6$. We will restrict our attention to the case of two characters (second-order equations), though some results will be more general. Despite the complications due to the presence of both movable poles and ``accessory parameters'', we will be able to make progress using the following strategy. We first of all parametrise such generic MLDEs in a useful way, which in fact can easily be extended beyond the case of two characters. Next we impose single-valuedness of the solutions around all the poles of the MLDE, leading to a set of equations relating the accessory parameters to the locations of the poles. These equations define a hypersurface in the space of poles and accessory parameters. Looking at the asymptotic region of this hypersurface relates the MLDE for a given $\ell$ to that for $\ell-6$, corresponding to one of the poles migrating to infinity. This allows us to determine the possible critical exponents for all $\ell\ge 6$ in terms of those for $\ell=0,2,4$, which are already known. This in turn makes it easier to solve the MLDE explicitly, as we show explicitly in the cases of $\ell=6,8$. Once there are two or more movable poles this becomes more difficult so we follow a slightly different strategy in the $\ell=12$ case.

In \cite{Chandra:2018pjq} a complete classification scheme for admissible characters for the case of two characters and arbitrary $\ell$ was provided in terms of ``quasi-characters'', using inspiration from mathematical works \cite{Kaneko:On, Kaneko:2013uga}. The present work, based on MLDE, provides an alternate route to the same result and the two can therefore be compared. We do so at various stages and find complete agreement. This encourages us to hope that the present approach can lead to new results for three or more characters where a full classification based on quasi-characters is not available (though partial results can be found in \cite{Mukhi:2020gnj}).

Before going on, let us mention two important points that will provide some context for our work. First, there exists an elegant approach to the classification of vector-valued modular forms due to Bantay and Gannon \cite{Bantay:2005vk, Bantay:2007zz}. This approach relies on the classification of modular data, which we do not assume in our work, so it may be considered a complementary point of view. Recently this approach was applied to the classification of $n\le 4$ character solutions in \cite{Rayhaun:2023pgc}. 

The second point, already alluded to earlier, is that classifying admissible characters is necessary but far from sufficient to classify CFT. In particular we know of infinite families of admissible characters that cannot correspond to any CFT. One of the most explicit tools to find genuine CFT within families of admissible characters is the coset construction \cite{Goddard:1984vk, Goddard:1986ee, frenkel1992vertex}. A version of this where the numerator is a meromorphic CFT \cite{Gaberdiel:Mero, Schellekens:1992db} was applied to the explicit construction of new CFT with small numbers of characters, and their classification, in \cite{Gaberdiel:2016zke, Mukhi:2022bte, Duan:2022ltz, Das:2022uoe, Rayhaun:2023pgc}. In the present work we do not address the problem of classifying actual CFTs within the space of admissible characters, rather our focus is purely on admissible MLDE solutions. Nevertheless, towards the end we will provide explicit examples of genuine CFT corresponding to the characters we construct, which makes it clear that the sub-space of CFT within the space of admissible characters is well-populated even for $\ell\ge 6$.  

\section{MLDEs in $\tau$-space}

\label{MLDEtau}

We now move on to the construction of MLDE of $n$th order and $\ell>0$ and the study of their admissible solutions. We label such MLDE by $(n,\ell)$.

\subsection{Bases of modular forms}
\label{subsec.bases}

We will choose a convenient basis of holomorphic modular forms of SL(2,Z). These can have any non-negative even weight $w>2$. A generic modular form of this weight is denoted $M_{w}(\tau)$. These form a multiplicative ring generated by the Eisenstein series $E_4(\tau),E_6(\tau)$, which we normalise so that their $q$-expansion starts with 1. We also use the cusp form:
\be
\Delta=\frac{E_4^3-E_6^2}{1728}=q+\cO(q^2)
\label{Deltacusp}
\ee
The Klein $j$-invariant, which will play a key role later on, is given by:
\be
j(\tau)=\frac{1728\, E_4^3}{E_4^3-E_6^2}=\frac{E_4^3}{\Delta}=q^{-1}+744 +\cO(q)
\label{Kleinj}
\ee

Torus moduli space has cusps at $\tau=\rho\equiv e^\frac{2\pi i}{3}$ and $\tau=i$. We have $E_4(\rho)=E_6(i)=0$. In the first case it is a fractional zero of order $\frac13$ and in the second, of order $\half$. Thus $E_4^3$ and $E_6^2$, of weight 12, both have a single full zero. The most general modular form with a single full zero is a linear combination of these two. Alternatively, and more usefully for us, it can be parametrised up to an overall constant as $E_4^3-p\,\Delta$ for some (in principle, complex) number $p$. In this form the leading coefficient in a $q$-expansion is unity independent of $p$, which follows from the cusp-form nature of $
\Delta$. Additionally, it vanishes at the point $\tau_p$ in the $\tau$-plane where $p=\frac{E_4^3}{\Delta}(\tau_p)=j(\tau_p)$. Thus $p$ has a clear geometric meaning as the location in the $j$-plane of the zero of the corresponding form. Generically it is a complex number.

We will need a convenient parametrisation for holomorphic modular forms of arbitrary weight.  To construct a suitable basis we proceed by dividing all possible $M_{w}$ into three classes:

\begin{enumerate}

\item $\{0\le w<12\}\cup \{w= 14$\}. Here $M_{w}$ is generated by one of the following: $1$, $E_4$, $E_6$, $E_4^2$, $E_4E_6$, $E_4^2E_6$.

\item $\{w\geq12, w\not\equiv \, 2 \, \text{mod} \, 12\}$. Let $w=2(6r+u)$, then we have: $M_{w}$ = $M_{2u}$ \ $\prod\limits^{r}_{I=1}(E_4^3-p_I\Delta)=M_{2u}\,\Delta^r\prod\limits^{r}_{I=1}(j-p_I)$ and $M_{2u}$ is one of the following: $1$, $E_4$, $E_6$, $E_4^2$, $E_4E_6$. Here $p_I$ are arbitrary complex numbers.
       
\item $\{w>14, w = 2 \, \text{mod}\, 12\}$. For this case we write $w=12r+2$, then $M_{w} =  E_4^2E_6\prod\limits^{r-1}_{I=1}(E_4^3-p_I\Delta)=E_4^2E_6\,\Delta^{r-1}\prod\limits^{r-1}_{I=1}(j-p_I)$, where again $p_I$ are arbitrary complex numbers.

\end{enumerate}

\subsection{Generic $(n,\ell)$ MLDE}

In what follows, ``primaries'' of an RCFT will not mean Virasoro primaries, but rather highest-weight integrable representations of some, typically extended, chiral algebra $\hat{\mathfrak{g}}$ that will emerge from the MLDE procedure and is not specified in advance. Corresponding to each such primary there will be a character that counts the number of descendants under $\hat{\mathfrak{g}}$ above it.

As is well-known, the number of linearly independent characters can be smaller than the number of primaries. This happens in particular when two or more primaries share the same character \footnote{Often this arises when the chiral algebra includes a Kac-Moody algebra and there is a symmetry in the Dynkin diagram of the corresponding finite-dimensional Lie algebra. For example, the $D_{4,1}$ WZW RCFT has four primaries corresponding to the identity ($h_0=0$), vector ($h_v=\frac{1}{2}$), spinor ($h_s=\frac{1}{2}$) and conjugate spinor ($h_c=\frac{1}{2}$). However, the triality symmetry in the Dynkin diagram of $D_4$ ensures that the last three primaries all share the same character. Thus this theory has only two linearly independent characters and its modular invariant torus partition function is $Z(q) = |\chi_0|^2 + 3|\chi_1|^2$. where $\chi_0$ denotes the identity character and $\chi_1$ denotes the single independent non-identity character.}. From the MLDE point of view it is more natural to focus only on characters which arise as the linearly independent solutions of the differential equation. Therefore, as was originally done in \cite{Mathur:1988na, Mathur:1988gt}, we consider an MLDE of order $n$ and remain open to the possibility that its solutions are the characters of an RCFT with $p>n$ primaries.

We  now formulate the MLDE for the case of $n$ characters 
and arbitrary Wronskian index $\ell$. The most general such equation takes the form:
\be
\Big(D^n+\sum_{s=1}^n \mu_{2s}\,\phi_{2s}(\tau) D^{n-s}\Big)\chi=0
\label{geneqn}
\ee
where the covariant derivative in $\tau$, denoted $D$, is defined as follows. Let 
\be
D_w\equiv \frac{1}{2\pi i}\frac{\del}{\del\tau}-\frac{w}{12}E_2(\tau)
\label{Ddef}
\ee
be the derivative acting on a modular form of weight $w$. Since $D_w$ raises the weight of the form from $w$ to $w+2$, we define:
\be
D^n\equiv D_{w+2n-2}\circ D_{w+2n-4} \circ \ldots \circ D_{w+2}\circ D_w
\label{Dndef}
\ee
In \eref{geneqn}, this definition applies with $w=0$. $\mu_{2s}$ are arbitrary parameters and the $\phi_{2s}$ are meromorphic modular functions of weight $2s$ whose poles are governed by the zeroes of the Wronskian, and whose overall normalisations are specified so that their leading term is unity. Explicitly we have:
\be
\mu_{2s}\,\phi_{2s}=(-1)^{s}\frac{W_{n-s}}{W_n}
\label{phidef}
\ee
where:
\be
W_{s}(\tau)\equiv \left| 
\begin{matrix}
\chi_0(\tau) & \chi_1(\tau)& \cdots & \chi_{n-1}(\tau)\\
\vdots & \vdots & \vdots & \vdots \\
D_\tau^{s-1}\chi_0(\tau) & D_\tau^{s-1} \chi_1(\tau) & \cdots & D_\tau^{s-1}\chi_{n-1}(\tau)\\
D_\tau^{s+1}\chi_0(\tau) & D_\tau^{s+1} \chi_1(\tau) & \cdots & D_\tau^{s+1}\chi_{n-1}(\tau)\\
\vdots & \vdots & \vdots & \vdots \\
D_\tau^{n}\chi_0(\tau) & D_\tau^{n} \chi_1(\tau)
& \cdots & D_\tau^{n}\chi_{n-1}(\tau)
\end{matrix}
\right|
\label{Wronskians}
\ee

It is easy to see from the definition that $W_{n-1}=D W_n$. From \eref{phidef} we find the useful relation:
\be
\mu_2\phi_2=-\frac{W_{n-1}}{W_n}=-D\log W_n
\label{phitwoprop}
\ee

The {\em Wronskian index} $\ell$ is defined to be an integer $\ell$ such that $\frac{\ell}{6}$ is the number of zeroes of $W_n$. This number does not have to be an integer because of the possibility of fractional zeroes at the cusps of moduli space, where a zero at $\tau=\rho$ counts as $\frac13$ of a full zero, and at $\tau=i$ counts as $\half$ of a full zero. Thus for general RCFT with $n$ characters, $\ell$ can be any non-negative integer other than 1.  Note that if the total number of zeroes is fractional then some zeros must necessarily occur at the cusps, while if the total number is integral (i.e. $\ell$ is a multiple of 6) then they can occur anywhere in the fundamental region. More generally the fractional part of $\frac{\ell}{6}$ describes the zeroes fixed at the cusps, while the integral part describes zeroes that are allowed to be at generic points of moduli space (including possibly the cusps). This motivates us to define $\ell_\rho,\ell_i,\ell_\tau$ to be the contribution to $\ell$ from the zeroes at $\rho,i$ and generic points respectively. Here $\ell_\rho$ is even, $\ell_i$ is a multiple of 3 and $\ell_\tau$ is a multiple of 6 \footnote{Each of these quantities is six times the corresponding quantity $w_\rho,w_i,w_\tau$ defined in \cite{Naculich:1988xv}.} and these quantities satisfy:
\be
\ell_\rho+\ell_i+\ell_\tau=\ell
\label{ellbreak}
\ee

The goal is to classify all possible MLDEs of the form \eref{geneqn} and then find suitable solutions to them. These take the form:
\be
\chi_i(q)=q^{\alpha_i}\sum_{k=0}^\infty a_{i,k}\,q^k
\label{charexp}
\ee
We call them {\em admissible} when $a_{i,k}$ are integers $\ge 0$ for all $i,k$, which means they potentially correspond to degeneracies of states. Additionally $a_{0,0}=1$, reflecting non-degeneracy of the vacuum state. In what follows we will establish several properties of the equations for generic $(n,\ell)$, including a count of the parameters on which they depend. After that we will restrict to $n=2$ and consider certain values of $\ell\ge 6$ in some detail and examine families of solutions.

We will start by making a {\em genericity assumption} -- that for any given $\ell$, the largest possible number of zeroes of $W_n$ are at generic, distinct points in moduli space, away from each other and from the special points $\tau=\rho,i$. With this assumption, $\ell_\rho$ takes its minimum allowed values of $0,2,4$ and $\ell_i$ takes its minimum allowed values of $0,3$. Later we will consider what happens when the zeroes merge.

Writing $\ell=6r+u, ~0\le u\le 5$, the possible cases are:
\begin{center}
\begin{tabular}{|c|c|}
\hline
$u$ & $(\ell_\rho,\ell_i,\ell_\tau)$\\
\hline
0\quad & $(0,0,6r)$\\
1\quad & $\big(4,3,6(r-1)\big)$\\
2\quad & $(2,0,6r)$\\
3\quad & $(0,3,6r)$\\
4\quad & $(4,0,6r)$\\
5\quad & $(2,3,6r)$\\
\hline
\end{tabular}
\end{center}

We will also require that the solutions of the MLDE  furnish irreducible representations $\varrho$ of the modular group PSL(2,Z). By definition, 
\be
\varrho(T)=\exp{\Big[2\pi i\,{\rm diag}\big(-\sfrac{c}{24},-\sfrac{c}{24}+h_1,\cdots, -\sfrac{c}{24}+h_{n-1}\big)\Big]}
\ee
By a theorem of Tuba-Wenzl \cite{Tuba} this implies, for rank $\le 5$, that the eigenvalues of $T$ are distinct. It follows that none of the $h_i$ is integral and no two of them differ by an integer. This will be sufficient for the cases discussed in this work \footnote{We thank an anonymous referee for emphasising that the corresponding result may well not be true for general rank $>5$.}.

Now we turn to the parametrisation of the coefficient functions $\phi_{2s}(\tau)$ for $s\ge 2$. Writing $\ell=6r+u, ~u=0,1,\cdots,5$ as above, we consider the different $u$ values separately as they have slightly different characteristics. From the definition of $\phi_{2s}$ in \eref{phidef} and the fact that $W_n$ has precisely $\frac{\ell}{6}$ zeroes, it follows that $\phi_{2s}$ can be expressed as a ratio of holomorphic modular forms such that the denominator has weight $2\ell$. This follows from the fact, mentioned earlier, that a full zero $(\ell=6)$ is achieved by a general weight 12 modular form $E_4^3-p\Delta$. To achieve the desired modular weight, the numerator of $\phi_{2s}$
must be modular of weight $2\ell+2s$. 

From Sub-section \ref{subsec.bases}, we find that, under the genericity assumption, these denominators of weight $2\ell$ can be parametrised as follows:
\be
\begin{split}
\ell=6r\!:& \qquad\prod_{I=1}^r (E_4^3-p_I\Delta)\\
\ell=6r+1  \!:& \qquad E_4^2E_6\,\prod_{I=1}^{r-1} (E_4^3-p_I\Delta)\\
\ell=6r+2 \!:& \qquad E_4\,\prod_{I=1}^r (E_4^3-p_I\Delta)\\
\ell=6r+3  \!:& \qquad E_6\,\prod_{I=1}^r (E_4^3-p_I\Delta)\\
\ell=6r+4 \!:& \qquad E_4^2\,\prod_{I=1}^r (E_4^3-p_I\Delta)\\
\ell=6r+5  \!:& \qquad E_4E_6\,\prod_{I=1}^r (E_4^3-p_I\Delta)
\end{split}
\label{alldenoms}
\ee
Thus for all $u\ne 1$ the denominators have exactly $r$ full zeroes, whose locations as a function of $\tau$ are determined by the $r$ parameters $p_I$, as well as $u$ fractional zeroes whose locations are fixed and hence they are not associated to any free parameters. For $u=1$ we instead have $r-1$ full zeroes, two zeroes of order $\frac13$ at $\tau=\rho$ and a zero of order $\frac12$ at $\tau=i$. 

Applying \eref{phitwoprop}, we find:
\be
\begin{split}
\ell=6r:\quad
\mu_2\,\phi_2 &= E_4^2E_6\sum_{I=1}^r \frac{1}{E_4^3-p_I\Delta}\\
\ell=6r+1:\quad\mu_2\,\phi_2 &= \frac{2E_6}{3E_4}+
\frac{E_4^2}{2E_6}
+E_4^2E_6\sum_{I=1}^{r-1} \frac{1}{E_4^3-p_I\Delta}\\
\ell=6r+2:\quad\mu_2\,\phi_2 &= \frac{E_6}{3E_4}+
E_4^2E_6\sum_{I=1}^r \frac{1}{E_4^3-p_I\Delta}\\
\ell=6r+3:\quad\mu_2\,\phi_2 &= \frac{E_4^2}{2E_6} +E_4^2E_6\sum_{I=1}^r \frac{1}{E_4^3-p_I\Delta}\\
\ell=6r+4:\quad\mu_2\,\phi_2 &= \frac{2E_6}{3E_4}+ E_4^2E_6\sum_{I=1}^r \frac{1}{E_4^3-p_I\Delta}\\
\ell=6r+5:\quad\mu_2\,\phi_2 &= \frac{E_6}{3E_4}+\frac{E_4^2}{2E_6}
+E_4^2E_6\sum_{I=1}^r \frac{1}{E_4^3-p_I\Delta}
\end{split}
\label{phitwobasis}
\ee
By inspection we see that in every case, the expression has a leading term $\frac{\ell}{6}$ as $q\to 0$. Since we are normalising every $\phi_{2s}$ to start with 1, it follows that $\mu_2=\frac{\ell}{6}$. 

We now consider the behaviour of solutions around $\tau\to i\infty$, where the appropriate coordinate is $q=e^{2\pi i\tau}\to 0$, by inserting the leading behaviour $\chi_i\sim q^{\alpha_i}+\cO(q^{\alpha_i+1})$ into the MLDE. In a CFT, the exponents $\alpha_i$ determine the central charge $c$ and the conformal dimensions $h_i$ via:
\be
\alpha_i=-\frac{c}{24}+h_i,\quad i=0,1,\cdots,n-1
\ee
where $h_0=0$, corresponding to the identity primary. Expanding: 
\be
\phi_2(\tau)=\sum_{n=0}^\infty \phi_{2,n}\, q^n
\ee
and inserting this as well as \eref{charexp} into the MLDE \eref{geneqn}, 
at leading order we find the indicial equation:
\be
\alpha^n+\left(\mu_2\,\phi_{2,0}-\frac{n(n-1)}{12}\right)\alpha^{n-1}+\cdots=0
\label{genindicial}
\ee
If the roots of this equation are $\alpha_i,i=0,1,\cdots,n-1$ then  we see that:
\be
\sum_{i=0}^{n-1}\alpha_i=\frac{n(n-1)}{12}-\frac{\ell}{6}
\label{RRformula}
\ee
where we used $\mu_2=\frac{\ell}{6}$ and $\phi_{2,0}=1$. The above equation is the valence (or Riemann-Roch) formula.

The lower order terms in \eref{genindicial} are straightforward but tedious to write explicitly, and they similarly allow us to determine the parameters $\mu_4,\mu_6,\cdots \mu_{2n}$ in terms of the critical exponents $\alpha_i,~i=0,1,\cdots,n-1$. We refer to the parameters $\mu_{2s}$ as {\em rigid parameters} since they are completely determined  by the critical exponents. Conversely if we know the $\mu_i$ then they determine the critical exponents.

\subsection{$(2,\ell)$ MLDE}
\label{twoelltau}

In this paper we will work with two characters, yet keeping the Wronskian index $\ell$ arbitrary. To our knowledge this region of $(n,\ell)$ space has not previously been investigated barring some insightful observations in \cite{Naculich:1988xv}. 
Let us mention that for two characters, the concept of movable poles essentially corresponds to the ``non-rigid'' case from the perspective of Fuchsian differential equations. These are the cases where the parameters in the equation are uniquely determined by the exponents of the solutions. Technically the rigid cases among the $(2,\ell)$ family arise for $\ell=0,2$. However as we will argue below, admissible characters for $\ell=4$ are completely determined in terms of those for $\ell=0$. Thus the non-rigid cases of interest start at $\ell=6$, which is also where movable poles first arise. So for practical purposes we can think of ``non-rigid'' $(2,\ell)$ MLDE as being equivalent to ``MLDE having movable poles''. This justifies our use of ``rigid'' for the parameters $\mu_{2s}$ of the previous sub-section and ``non-rigid'' for the rest.

The general $(2,\ell)$ MLDE is:
\be
\Big(D^2+\mu_2\,\phi_2(\tau)D+\mu_4\,\phi_4(\tau)\Big)\chi=0
\label{MLDEtau}
\ee
where $\phi_2,\phi_4$ are meromorphic modular forms of weight 2 and 4 respectively. 

Now consider the coefficient function $\phi_4$. Its denominator must have (at most) the zeroes of the Wronskian $W_n$. The numerator is then a general modular form of weight $4$ higher. Also the form must be normalised so that its $q$-expansion starts with 1. This implies that it takes the form:
\be
\begin{split}
\ell=6r\!:\quad \phi_4 &=\frac{E_4\prod_{I=1}^{r}(E_4^3-b_{4,I}\Delta)}{\prod_{i=1}^r (E_4^3-p_I\Delta)}\\  
\ell=6r+2\!:\quad
\phi_4 &=\frac{E_4^2\prod_{I=1}^{r}(E_4^3-b_{4,I}\Delta)}{E_4\prod_{I=1}^r (E_4^3-p_I\Delta)}
= \frac{E_4\prod_{I=1}^{r}(E_4^3-b_{4,I}\Delta)}{\prod_{I=1}^r (E_4^3-p_I\Delta)}\\
\ell=6r+4\!:\quad
\phi_4 &=\frac{\prod_{I=1}^{r+1}(E_4^3-b_{4,I}\Delta)}{E_4^2\prod_{I=1}^r (E_4^3-p_I\Delta)}
\end{split}
\label{phifourtwochar}
\ee
The $b_{4,I}$ are ``accessory parameters'' about which we will have a lot to say in the rest of this paper (they carry the subscript 4 because they arise in a weight-four modular function). 
Notice that for the middle case there is no pole at $\tau=\rho$ due to cancellation of an $E_4$ between the numerator and denominator. Also the last case has an extra power of $(E_4^3-b_{4,{r+1}}\Delta)$ in the numerator.

Returning to the MLDE \eref{MLDEtau}, we have already determined that $\mu_2=\frac{\ell}{6}$. Also, the leading term of $\phi_4$ in a $q$-expansion is normalised to unity. Then the indicial equation determines:
\be
\mu_4=\alpha_0\alpha_1
\ee
where $\alpha_i$ are the critical exponents around $q=0$. Also it is known \cite{Naculich:1988xv}  that with two characters, $\ell$ is always even. Hence $W$ cannot have an odd number of zeroes at $\tau=i$. Then, recalling that $\ell=6r+u$, one is restricted to even values of $u$. 
With all the above information, we write the general $(2,\ell)$ MLDE as follows:
\be
\begin{split}
& \ell=6r{}:\\
&\left(D^2+ \Bigg(E_4^2E_6\sum_{I=1}^r \frac{1}{E_4^3-p_I\Delta}\Bigg)D+\alpha_0\alpha_1\,
E_4 \frac{\prod_{I=1}^r (E_4^3-b_{4,I}\Delta)}{\prod_{I=1}^r (E_4^3-p_I\Delta)}
\right)\chi(\tau)=0 \\[2mm]
& \ell=6r+2{}:\\
&\left(D^2+ \Bigg(\frac{E_6}{3E_4}+E_4^2E_6\sum_{I=1}^r \frac{1}{E_4^3-p_I\Delta}\Bigg)D+\alpha_0\alpha_1\,
E_4\frac{\prod_{I=1}^r (E_4^3-b_{4,I}\Delta)}{\prod_{I=1}^r (E_4^3-p_I\Delta)}
\right)\chi(\tau)=0 \\[2mm]
& \ell=6r+4{}:\\
&\left(D^2+ \Bigg(\frac{2E_6}{3E_4^2}+E_4^2E_6\sum_{I=1}^r \frac{1}{E_4^3-p_I\Delta}\Bigg)D+\frac{\alpha_0\alpha_1}{E_4^2} \frac{\prod_{I=1}^{r+1}(E_4^3-b_{4,I}\Delta)}{\prod_{I=1}^r(E_4^3-p_I\Delta)}
\right)\chi(\tau)=0
\end{split}
\label{MLDE.2}
\ee
Well-studied special cases are the MMS equation \cite{Mathur:1988na} which corresponds to $\ell=0$:
\be
\Big(D^2+\alpha_0\alpha_1\,E_4\Big)\chi(\tau)=0
\label{MMSeqn}
\ee
and the $\ell=2$ equation studied in \cite{Naculich:1988xv, Hampapura:2015cea}:
\be
\left(D^2+\frac{E_6}{3E_4}D+\alpha_0\alpha_1\,E_4\right)\chi(\tau)=0
\label{leq2eqn}
\ee
As we see, these equations have no movable poles.

Returning now to the general case, although $p_I$ are generically complex, they are subject to constraints arising from the fact that $c$ and the degeneracies are rational. As we will show below, the {\em symmetric polynomials} in the $p_I$ must all be real and rational. 
This generalises the statement in \cite{Naculich:1988xv}) that a single pole must be real and rational. Let us mention here that for a real pole, $\tau_p$ lies in the subspace of moduli space for which $j(\tau_p)$ is real, namely $\left\{{\rm Re}(\tau)=0\right\}\cup \left\{{\rm Re}(\tau)=\half\right\} \cup |\tau|=1$.

So far we have only considered the indicial equation about $q=0$. However, the characters also need to have appropriate behaviour near the cusps $\tau=\rho,i$ in order to be single-valued. We label the exponents around $\tau=\rho,i$ as $\alpha^{(\rho)},\alphai$ respectively to avoid confusion with the exponents $\alpha$ around $\tau=i\infty$. Near $\tau=\rho$ we introduce a new coordinate:
\be
z=(\tau-\rho)^3
\label{zrho}
\ee
When $\tau$ circles $\rho$ by $e^{2\pi i/3}$, we return to the same point in moduli space. The above change of variables converts this to a regular circle $z\to e^{2\pi i}z$, so $z$ is a good coordinate at the cusp. In this coordinate, $E_4\sim z^\frac13$ and $j\sim z$ as $z\to 0$. The indices at $\tau=\rho$ are found by inserting the trial solution 
\be
\chi(z)\sim z^{\alpha^{(\rho)}}
\label{charrho}
\ee
Regularity imposes the requirement that $\alpha^{(\rho)}$ is a non-negative multiple of $\frac13$. A similar analysis tells us that $\tau=i$ is a multiple of $\half$.

Now expanding out the MLDE \eref{MLDE.2} we get:
\be
\left(-\frac{1}{4\pi^2}\del_\tau^2-\frac{1}{12\pi i} E_2(\tau)\del_\tau+\frac{1}{2\pi i}\mu_2\phi_2(\tau)\del_\tau+\alpha_0\alpha_1\phi_4(\tau)\right)\chi=0
\ee
As we are working near $\tau=\rho$ where $E_4$ and $j$ vanish while $E_6$ and $\Delta$ tend to finite values, we can replace $\mu_2\phi_2$ by $\frac{u}{6}\frac{E_6}{E_4}$. This is because near $\tau=\rho$, $\mu_2\phi_2$ has $\frac{u}{6}$ poles where $\ell=6r+u$, by our genericity assumption. Meanwhile $\phi_4$ given in \eref{phifourtwochar} reduces near $\tau=\rho$ to:
\be
\begin{split}
\ell=6r\!:\quad &
\phi_4 \simeq 0\\
\ell=6r+2\!:\quad &  
\phi_4 \simeq 0\\
\ell=6r+4\!:\quad &
\phi_4 \simeq -\frac{b_{4,{r+1}}\Delta}{E_4^2} \prod_{I=1}^{r}\frac{b_{4,I}}{p_I}
\end{split}
\label{phiatrho}
\ee

From the definition of the $j$-invariant we have:
\be
\frac{E_6}{E_4}=-\frac{1}{2\pi i}\frac{\del}{\del\tau}(\log j)
\label{djdt}
\ee
Since $j\sim (\tau-\rho)^3$, it follows that:
\be
\frac{E_6}{E_4}\simeq -\frac{3}{2\pi i(\tau-\rho)}\quad \hbox{near }\tau=\rho
\ee
From this we can also deduce the behaviour:
\be
\frac{\Delta}{E_4^2}\simeq -\frac{1}{1728}\left(\frac{E_6}{E_4}\right)^2
=\frac{1}{768\pi^2(\tau-\rho)^2}
\ee

Now we change variables in the MLDE using \eref{zrho} and insert \eref{charrho}. 
In the case $u=0$ we find the indicial equation:
\be
3\big(\alpha^{(\rho)}\big)^2-\alpha^{(\rho)}=0
\ee
The solutions are $\alpha^{(\rho)}=0,\frac13$. Thus both solutions exhibit regular behaviour as functions of $\tau$. While this has been a good consistency check, it does not tell us anything new about the parameters in the MLDE.

Next, for $u=2$, the MLDE near $\tau=\rho$ becomes:
\be
\left(-\frac{1}{4\pi^2}\del_\tau^2-\frac{1}{12\pi i} E_2\del_\tau+\frac{1}{2\pi i}\frac{E_6}{3E_4}\del_\tau\right)\chi=0
\ee
Going now to the $z$-coordinate and inserting $\chi(z)\sim z^{\alpha^{(\rho)}}$, the indicial equation is:
\be
3\big(\alpha^{(\rho)}\big)^2-2\alpha^{(\rho)}=0
\ee
whose solutions are $\alpha^{(\rho)}=0,\frac23$. Again the solution is consistent but does not provide new information.

The situation is different for the last case, $u=4$. \eref{phiatrho} tells us we have non-trivial behaviour for both $\phi_2$ and $\phi_4$. This indicial equation now becomes:
\be
\alpha^{(\rho)}(\alpha^{(\rho)}-1)+\frac{\gamma}{1728}=0
\label{indicialfour}
\ee
where:
\be
\gamma\equiv \alpha_0\alpha_1 b_{4,r+1}\,\prod_{I=1}^r\frac{b_{4,I}}{p_I}
\label{gammadef}
\ee
From this we learn that $\alpha^{(\rho)}_0+\alpha^{(\rho)}_1=1$. As we have seen, these exponents are non-negative multiples of $\frac13$, which leads to the unique solution $\alpha^{(\rho)}_0=\frac13,\alpha^{(\rho)}_1=\frac23$. It now follows from \eref{indicialfour} that:
\be
\gamma=384
\ee
This was previously noted in \cite{Chandra:2018pjq} for the case $\ell=4$. Here we see that it is true for all $\ell=6r+4$ as long as there are precisely two poles (of $\frac13$-order each) at the cusp $\tau=\rho$ and the rest are at generic values away from the cusp. We can think of this result as determining $b_{4,r+1}$ in \eref{gammadef} in terms of the other $b_{4,I}$. Then the (so far) independent coefficients are $b_{4,I},I=1,2,\cdots,r$ and $p_I$. Thus, despite the appearance of an apparent additional parameter $b_{4,r+1}$, the case $\ell=6r+4$ is actually similar to the cases $\ell=6r,6r+2$ in that all of them have precisely $2r$ parameters of which $r$ correspond to poles $p_I$ of the coefficient functions and the other $r$ are the $b_{4,I}$. As indicated above, these are the accessory parameters familiar from Fuchsian differential equations.

In standard treatments of the MLDE, starting with \cite{Mathur:1988na}, one now solves the equation order by order using the Frobenius method, and imposes admissibility at each successive order, in particular non-negative integrality of the Fourier coefficients. We will do this eventually, but here we pause to rewrite the MLDE treating $j(\tau)$, rather than $\tau$, as the independent parameter. This makes it somewhat easier and more intuitive to write out general MLDEs and impose their single-valuedness around poles of the coefficient functions. Of course the Fourier coefficients in the $j$ variable have no integrality restrictions, so we will need to return to the $\tau$-coordinate in order to check integrality of the coefficients in the $q$-expansion and thereby determine admissibility of solutions.

\section{MLDEs in $j$-space}

\label{MLDEj}

\subsection{Generic $(n,\ell)$ MLDE in $j$-space}

In this Section we return to the general case of $n$ characters and study MLDEs in a formalism where the independent variable is the Klein invariant $j(\tau)$ rather than $\tau$ (this was explored for special cases in \cite{Naculich:1988xv, Franc:2016}). As a warm-up exercise, let us consider the one-character case. First we fix $\ell=6r$. Then the most general allowed character is:
\be
\chi(j)=\prod_{I=1}^r (j-p_I)
\ee
where $p_I$ are a set of $r$ complex numbers that describe the zeroes of the character (which is the same as the Wronskian in this case). The MLDE satisfied by this character is trivially seen to be:
\be
\big(\del_j+\psi_2(j)\big)\chi(j)=0 
\ee
where:
\be
\begin{split}
\psi_2(j)&=-\del_j\log\chi(j)\\
&= -\sum_{I=1}^r \frac{1}{j-p_I}
\end{split}
\ee
Here we have labelled the first non-trivial coefficient as $\psi_2(j)$ in keeping with the convention used for the MLDE in $\tau$, though here it does not reflect the modular weight, since everything is modular invariant (up to possible phases). Also note that the coefficients $\mu_{2s}$ are now absorbed into the normalisation of the $\psi_{2s}$.

The generalisation of the above to the case of $\ell=6r+u$ is straightforward: the character acquires an extra multiplicative factor of $j^\frac13$ for each zero at $\tau=\rho$ and a factor $(j-1728)^\half$ for a zero at $\tau=i$ (we are not requiring admissibility at this stage, which would have ruled out the latter). Then the coefficient function $\psi_2(j)$ acquires an additive term:
\be
-\frac{1}{3j},~-\frac{1}{2(j-1728)}
\ee
for each zero at $\rho,i$ respectively.

Moving on to the $n$-character case, the MLDE in terms of the independent variable $j$ can be written:
\be
\Big(\del_j^n+\sum_{s=1}^n \psi_{2s}(j)\,\del_j^{n-s}\Big)\chi(j)=0
\label{jgeneqn}
\ee
The modular invariants $\psi_{2s}$ can have poles at the special points $j=0,1728$ as well as at generic points $j=p_I, I=1,2,\cdots r$. The solutions can be expanded as follows around the special points:

\begin{align}
\chi_i(j) &= j^{\alpharho_i}\sum_{k=0}^\infty a^{(\rho)}_{i,k}\, j^{\frac{k}{3}}, \label{rh_xp}\\
\chi_l(j) &= (j-1728)^{\alphai_i}\sum_{k=0}^\infty a^{(i)}_{l,k}\, (j-1728)^{\frac{k}{2}}. \label{ixp}
\end{align}
Similarly around each of the generic poles $j=p_I$, we parametrise the solutions as:
\be
\chi_i(j)=(j-p_I)^{\alpha^{(I)}_i}\sum_{k=0}^\infty a^{(I)}_{i,k}\,(j-p_I)^k
\label{Iexp}
\ee
One should keep in mind that the $a_{i,k}$ with superscripts $(\rho),(i),(I)$ have no particular integrality property.

The relevant Wronskians are defined similarly to \eref{Wronskians} \footnote{We denote them $W_r(j)$, though they are different from $W_r(\tau)$ so this is really abuse of notation - which hopefully will not cause confusion.}:
\be
W_r(j)\equiv \left| 
\begin{matrix}
\chi_0(j) & \chi_1(j)& \cdots & \chi_{n-1}(j)\\
\vdots & \vdots & \vdots & \vdots \\
\del_j^{r-1}\chi_0(j) & \del_j^{r-1} \chi_1(j)
& \cdots & \del^{r-1}_j\chi_{n-1}(j)\\
\del_j^{r+1}\chi_0(j) & \del_j^{r+1} \chi_1(j)
& \cdots & \del^{r+1}_j\chi_{n-1}(j)\\
\vdots & \vdots & \vdots & \vdots \\
\del_j^{n-1}\chi_0(j) & \del_j^{n-1} \chi_1(j)
& \cdots & \del^{n-1}_j\chi_{n-1}(j)
\end{matrix}
\right|
\ee
and:
\be
\psi_{2s}=(-1)^{s}\frac{W_{n-s}(j)}{W_n(j)}
\label{psidef}
\ee

As noted in \cite{Naculich:1988xv}, $W_r(j)$ necessarily have poles, unlike $W_r(\tau)$. The poles are introduced by the powers of $\frac{dj}{d\tau}$ that relate two Wronskians. For example the relations between the Wronskians $W_n(j)$ and $W_0(j)$ with that of the Wronskians $W_n(\tau)$ and $W_0(\tau)$ are as follows (similar but more complicated relations can be found for all the $W_r$):
\bea
&W_n(j)=\left(\frac{dj}{d\tau}\right)^{-\frac{n(n-1)}{2}} W_n(\tau) \\
&W_0(j) = \left(\frac{dj}{d\tau}\right)^{-\frac{n(n+1)}{2}} W_0(\tau)
\eea
Using \eref{djdt}, we see that:
\be
\frac{dj}{d\tau}=-2\pi i\frac{E_6}{E_4}\,j=-2\pi i\frac{E_6E_4^2}{\Delta}
\ee
Thus $\frac{dj}{d\tau}$ has two zeroes of order $\frac13$ at $\tau=\rho$ and one of order $\half$ at $\tau=i$, so $W(j)$ acquires $\frac{n(n-1)}{3}$ poles at $\rho$ and $\frac{n(n-1)}{4}$ poles at $i$. It also has the zeroes of $W_n(\tau)$. So we could define a new Wronskian index $\ell^j$ such that $\frac{\ell^j}{6}$ gives the total number of zeroes of $W(j)$:
\be
\ell^j=-\frac{7n(n-1)}{2}+\ell
\ee

We see that for $n=1$, $\ell_j=\ell$, while for $n=2$, $\ell_j=-7+\ell$ as one can read off from Page 434 of \cite{Naculich:1988xv}. Actually this relation is slightly misleading as it does not contain all the information: the extra poles contained in the first term on the RHS are necessarily at the points $\tau=\rho,i$ and are not free to move. So it is better to break up $\ell_j$ into contributions from zeroes/poles at $\rho,i$ and generic positions (as we did before for $\ell$):
\be
\ell^j=\ell^j_\rho+\ell^j_i+\ell^j_\tau
\ee
Now, positive values of these quantities denote zeroes while negative values denote poles. 
Recall that in the $\tau$-space case, the terms are individually $\ge 0$ and $\ell_\rho,\ell_i$ and $\ell_\tau$ are multiples of 2,3,6 respectively. Then, taking account of the new poles introduced by the change of variables, we get:
\be
\begin{split}
\ell^j_\rho &=-2n(n-1)+\ell_\rho\\
\ell^j_i &= -\frac{3n(n-1)}{2}+\ell_i\\
\ell^j_\tau &= \ell_\tau
\end{split}
\ee
Positivity of $\ell_p,\ell_i,\ell_\tau$ then induces obvious lower bounds on $\ell^j_\rho,\ell^j_i,\ell^j_\tau$. Notice that it is possible for $\ell^j_\rho,\ell^j_i$ to vanish due to cancellations between poles induced by the change of variables to $j$ and zeroes of the original Wronskian at the special points $\rho,i$.

From the above considerations, we can readily fix the first coefficient function $\psi_2(j)$ in \eref{jgeneqn}, which is given by:
\be
\psi_2(j)=-\frac{W_{n-1}(j)}{W_n(j)}=-\del_j\log W_n(j)
\ee
The result is:
\be
\begin{split}
\psi_2(j) &= -\frac{\ell_\rho^j}{6j}-
\frac{\ell_i^j}{6(j-1728)} -\sum_{I=1}^{\frac{\ell^j_\tau}{6}} \frac{1}{j-p_I}\\
&=  \frac{n(n-1)}{3j}+\frac{n(n-1)}{4(j-1728)}
-\frac{\ell_\rho}{6j}-
\frac{\ell_i}{6(j-1728)} -\sum_{I=1}^{\frac{\ell_\tau}{6}} \frac{1}{j-p_I}
\end{split}
\ee

\subsection{$(2,\ell)$ MLDE in $j$-space}

\label{MLDEjspace}

We now again specialise to the case of two characters, keeping the Wronskian index arbitrary. The first step is to determine the remaining coefficient function $\psi_4(j)$ in the MLDE for this case. From the definition we have:
\be
\psi_4(j)=\frac{W_0(j)}{W_2(j)}
\ee
Now,
\be
W_0(j)=\left|\begin{matrix}
\del_j\chi_0 & \del_j\chi_1\\
\del^2_j\chi_0 & \del^2_j\chi_1
\end{matrix}
\right|,\qquad
W_2(j)=\left|\begin{matrix}
\chi_0 & \chi_1\\
\del_j\chi_0 & \del_j\chi_1
\end{matrix}
\right|,
\ee
Inserting the behaviour $\chi_i\sim j^{\alpha^{(\rho)}_i}$ near $j\sim 0~(\tau=\rho)$ we find:
\be
\begin{split}
W_0(j)&\sim \alpharhozero\alpharhoone \Big(\alpharhoone-\alpharhozero\Big)j^{\alpharhozero+\alpharhoone-3}+\cO\Big(j^{\alpharhozero+\alpharhoone-2}\Big)\\
W_2(j)&\sim \Big(\alpharhoone-\alpharhozero\Big) j^{\alpharhozero+\alpharhoone-1}\\
\end{split}
\ee
The reason to write the first correction to $W_0$ is that the leading term can vanish, if $\alpharho_0$ or $\alpharho_1$ vanishes. However since the two exponents must be distinct, the leading term of $W_2$ cannot vanish. Thus we have $\psi_4(j)\sim j^{-2}$
unless one of the exponents vanishes, in which case $\psi_4(j)\sim j^{-1}$.

The exponents $\alpharho_i$ satisfy (see \eref{indi_r}):
\be
\alpharho_0+\alpharho_1-\frac13=\frac{\ell_\rho}{6}
\label{alpharhosum}
\ee
Writing $\ell=6r+u$, we have $u=\ell_\rho=0,2,4$ respectively for the cases $\ell=6r,6r+2,6r+4$. It follows that the exponents are as follows:
\be\label{exp_2-char}
\begin{split}
\ell=6r\!: &\qquad  \alpharho_0+\alpharho_1=\sfrac13
\implies (\alpharho_0,\alpharho_1)=(0,\sfrac13)
\\  
\ell=6r+2\!: &\qquad  
\alpharho_0+\alpharho_1=\sfrac23
\implies (\alpharho_0,\alpharho_1)=(0,\sfrac23)
\\ 
\ell=6r+4\!: &\qquad  
\alpharho_0+\alpharho_1=1\,
\implies (\alpharho_0,\alpharho_1)=(\sfrac13,\sfrac23)
\end{split}
\ee
(these facts have already been derived in terms of the $\tau$ coordinate in Sub-section \ref{twoelltau}, but here our goal is to derive everything independently in the $j$ coordinate). 
Thus in the first two cases the leading term in $W_0$ indeed vanishes and the subleading term has to be used. We see that the behaviour of $\psi_4(j)$ in the three cases is $\sim j^{-1},\sim j^{-1},\sim j^{-2}$ respectively.

Next we consider the behaviour near $j=1728$ ($\tau=i$). Similar arguments tell us that $\psi_4(j)\sim \alphai_0\alphai_1 (j-1728)^{-2}+\cO\big((j-1728)^{-1}\big)$. This time we have (see \eref{ind_i}):
\be
\alphai_0+\alphai_1-\half = \frac{\ell_i}{6}=0 \implies (\alphai_0,\alphai_1)=\Big(0,\half\Big), \label{alpha_i_sum}
\ee
in every case, so the leading term always vanishes and we have a simple pole in $j-1728$. 

From the $r$ generic zeroes of $W_2$ at $j=p_I$, we get a simple pole at each of these points. Finally, the $\tau\to i\infty$ behaviour requires that the overall  power of $j$ as $j\to\infty$ is $-2$. Hence $\psi_4(j)$ must contain, in the numerator, a generic polynomial in $j$ of degree $r$ for $\ell=6r,6r+2$ and of degree $r+1$ for $\ell=6r+4$. Thus finally we get:
\begin{align*}   
&\ell=6r\!:\\
&\partial^2_j\chi(j)+\left[\frac{1}{2(j-1728)}+\frac{2}{3j}-\sum\limits_{I=1}^{r}\frac{1}{j-p_I}\right]\partial_j\chi(j)+
\frac{\alpha_0\alpha_1}{j(j-1728)}\frac{\prod\limits_{I=1}^r(j-b_{4,I})}{\prod\limits_{I=1}^r(j-p_{I})}\chi(j) = 0.
\end{align*}
\begin{align*}
&\ell=6r+2\!:\\
&\partial^2_j\chi(j)+\left[\frac{1}{2(j-1728)}+\frac{1}{3j}-\sum\limits_{I=1}^{r}\frac{1}{j-p_I}\right]\partial_j\chi(j)+
\frac{\alpha_0\alpha_1}{j(j-1728)}
\frac{\prod\limits_{I=1}^r(j-b_{4,I})}{\prod\limits_{I=1}^{r}(j-p_I)}\chi(j) = 0.
\end{align*}
\begin{align}
&\ell=6r+4\!:\nonumber\\
&\partial^2_j\chi(j)+\left[\frac{1}{2(j-1728)}-\sum\limits_{I=1}^{r}\frac{1}{j-p_I}\right]\partial_j\chi(j)+\frac{\alpha_0\alpha_1}{j^2(j-1728)}\frac{\prod\limits_{I=1}^{r+1}(j-b_{4,I})}{\prod\limits_{I=1}^{r}(j-p_I)}\chi(j) = 0.
\label{twocharj}
\end{align}
These expressions can easily be confirmed by explicitly changing variables from $\tau$ to $j$ in Eqs. (\ref{phitwobasis}), (\ref{phifourtwochar}). However, the methods we have used to arrive at them are useful in the general case (higher than second-order) and one does not need to invoke the MLDE in $\tau$ to write the equations in $j$-space.

By considering the indicial equation around $j=0$ ($\tau=\rho$), we will again find, in the $\ell=6r+4$ case, that it is possible to fix $b_{4,r+1}$ in terms of the remaining coefficients. As a result, once we impose the indicial equations there are $2r$ independent coefficients in every case, namely the $p_I$ and $b_{4,I}$ with $I=1,2,\cdots,r$. Note that the MLDE is totally symmetric under permutations of the $p_I$ and also under permutations of the accessory parameters $b_{4,I}$. 

The differential equations in the $j$ plane that were discussed above are examples of Fuchsian differential equations (FDE) with regular singular points. However they have some special features. A general FDE with regular singular points is of the form:
\be
\frac{d^nf}{dx^n}+ \sum_{i=1}^n\alpha_i(x)\frac{d^if}{dx^i}=0
\ee
where the coefficient functions $\alpha_i(x)$ have at most poles of order $i$ at the regular singular points. However due to our genericity assumption, the Wronskian has only simple zeroes at generic points $p_I$. Hence in  our case, all the coefficient functions have only simple poles (with the exception of poles at $\tau=\rho$, which are double poles whenever the Wronskian index is equal to 4 mod 6). This means that ab initio they span a more restricted set than general Fuchsian differential equations with regular singular points. We will revisit this issue later on when we move away from the genericity assumption by allowing movable poles to coalesce. Meanwhile, as already noted above, our $b_{4,I}$ correspond in the language of FDE to what are called ``accessory parameters''.

\subsection{Reduction of $\ell=6r+4$ to $\ell=6r$}

\label{6rplus4}

Let us note an important general lesson that is exemplified by \eref{exp_2-char}. In the third line, the lower of the two exponents is $\frac13$. This means we can take any solution of the $\ell=6r+4$ MLDE and write it as:
\be
\chi(j)=j^\frac13\zeta(j)
\ee
where $\zeta(j)$ has an expansion about $j=0$ in positive powers of $j$. Then, as is easily verified, $\zeta(j)$ solves an MLDE with $(n,\ell)=(2,6r)$. This means that, in terms of having a well-defined power-series expansion about all the singular points of the MLDE, every $\ell=6r+4$ solution factorises into the $E_{8,1}$ character $j^\frac13$ times a solution of the $(2,6r)$ equation. 
It is not, however, necessarily the case that both factors are admissible. In particular, it is possible for $\zeta(j)$ to be a non-admissible character while $j^\frac13\zeta(j)$ is admissible.

A very striking example, noted in sub-section 5.1 of \cite{Chandra:2018pjq}, is that the characters of the $c=33$ CFT of \cite{Grady_2020}, which has Wronskian index $\ell=4$, can be written as the product of $j^\frac13$ times a solution with $c=25$ and $\ell=0$. However, the $c=25$ solution has some negative coefficients in its $q$-series and therefore does not count as admissible (as we will see below, it is actually a quasi-character). Yet, after multiplying it by $j^\frac13$ it becomes admissible and in fact a genuine CFT. But this CFT, despite the factorisation described above, is by no means a tensor product of two other CFTs.

The factorisation of solutions described above for $\ell=4$ is easily seen to persist for all $\ell=6r+4$. Hence we no longer need to discuss MLDEs for the case $\ell=6r+4$, even though we have formulated them above. All we need to remember is that admissible solutions in these cases are found by considering all integral (not only admissible) solutions of the $\ell=6r$ equation, multiplying each one by $j^\frac13$ and then testing for admissibility.

On the other hand, in the first two lines of \eref{exp_2-char}, the lower of the two exponents is 0. This tells us that we cannot extract a positive power from the character and still hope to find a positive power-series expansion in $j$. Moreover this fact persists for $\ell=6r,6r+2$ as long as the genericity assumption is obeyed. So even relaxing admissibility, the characters in these cases do not factorise.

\subsection{Determining accessory parameters: the $(2,6)$ case}

\label{access26}

Having dealt with the indicial equations about $j=0,1728$, the next step is to study the indicial equations around $j=p_i$. This will determine all the accessory parameters $b_{4,i}$ in $\psi_4(j)$.
Let us start with a particular case, the $(2,6)$ MLDE which from \eref{twocharj} has the form:
\begin{align}
\partial_j^2\chi(j) + \left[\frac{1}{2(j-1728)}+\frac{2}{3j}-\frac{1}{j-p_1}\right]\partial_j\chi(j) + \frac{\alpha_0\alpha_1(j-b_{4,1})}{j(j-1728)(j-p_1)}\chi(j) = 0, \label{twosixMLDE}
\end{align}
A priori it has $2$ parameters, $b_{4,1}$ and $p_1$. In this case we have $\ell_\rho=\ell_i=0$ and $\ell_1=6$.

Let us examine the leading behaviour of the characters about $j=p_1$. Since $p_1$ is not a special point in moduli space (by the genericity assumption), the critical exponents around it must be integers. We substitute the expansion \eref{Iexp} in \eref{twosixMLDE} and look at the solution at order $(j-p_1)^{\alpha^{(1)}_i-1}$ to get the indicial equation:
\begin{align}
\alpha^{(1)}_i(\alpha^{(1)}_i-2) = 0 \label{indi_sig}
\end{align}
so the exponents are $\big(\alpha^{(1)}_0,\alpha^{(1)}_1\big)=(0,2)$. Notice that in this process we have identified the solution $\chi_0(j)$ with the exponent 0, and $\chi_1(j)$ with exponent 2 \footnote{It is important not to identify these two solutions with the two characters $\chi_0(\tau)$ and $\chi_1(\tau)$ that form the two linearly independent CFT characters with integral expansions in $q$. The reason is that here we are expanding around a point inside moduli space instead of the point $\tau\to i\infty$. Hence each pair is in general a linear combination of the other pair.}.

When the exponents differ by an integer there is potentially a problem with single-valuedness of the solution. In fact the solution with $\alpha^{(1)}=2$ always exists, but the solution with $\alpha^{(1)}=0$ in general has a logarithmic term. If present, this term would render the corresponding character multivalued in $j$ and therefore unphysical \cite{Naculich:1988xv, DERef}.
To analyse this situation we start by inserting the expansion \eref{Iexp} in \eref{twosixMLDE}. At order $(j-p_1)^{\alpha_i^{(1)}}$ we find:
\begin{align}
a^{(1)}_{i,1}=\frac{\alpha_i^{(1)}(\frac76 p_1-1152)+\alpha_0\alpha_1(p_1-b_{4,1})}{p_1(p_1-1728)\Big(1-\big(\alpha^{(1)}_i\big)^2\Big)}
\label{nonerecur}
\end{align}
Inserting $\alpha_1^{(1)}=2$ we determine $a^{(1)}_{1,1}$ for this solution, and continuing in this way we are guaranteed to determine the subsequent coefficients. If we insert the other value $\alpha_0^{(1)}=0$, we find:
\begin{align}
a^{(1)}_{0,1} = \frac{\alpha_0\alpha_1(p_1 - b_{4,1})}{p_1(p_1-1728)}
\label{nonealphazero}
\end{align}
Now the recursion relation that should have determined the next coefficient $a^{(1)}_{0,2}$ does not contain that variable. This is a consequence of the integral difference in indices that we noted above. Instead, it gives us a constraint on $b_{4,1}$: 
\be
    \alpha_0\alpha_1 + \left(576-\frac{5p_1}{6}+ \alpha_0\alpha_1(p_1-b_{4,1})\right) a^{(1)}_{0,1} = 0, \label{coeff1_261}
\ee
Thus the above constraint is the condition that the second solution does not have a logarithmic piece. If we do not implement the constraint, one of the characters becomes multi-valued around a zero of the Wronskian and has to be rejected on physical grounds \footnote{Such objects are called ``weak VVMF'' in \cite{Franc:2016}.}.

Substituting \eref{nonealphazero} into \eref{coeff1_261} we get:
\be
\alpha_0\alpha_1 (p_1-b_{4,1})^2 + \left(576-\frac56 p_1\right)(p_1-b_{4,1}) +p_1(p_1-1728)=0
\label{b41condition}
\ee
After multiplying by all the denominators, this becomes a quadratic curve in $p_1,(p_1-b_{4,1})$ with discriminant:
\be
\frac{25}{36}-4\alpha_0\alpha_1
\ee
Using $\alpha_0=-\frac{c}{24}$ and $\alpha_1=\frac{c}{24}-\frac56$ (the latter follows from \eref{RRformula}), we see that this is positive for all $c\ne 10$, and the quadratic is a hyperbola. At $c=10$ the curve degenerates to a parabola. From \eref{b41condition}, notice that when $b_{4,1}= p_1$ then we have $p_1=0$ or $p_1=1728$, in other words the pole has to be at one of the cusps $\tau=\rho,\tau=i$ of moduli space. The curve \eref{b41condition} determines the accessory parameter in terms of the pole $p_1$. 

It is useful to consider the asymptotic region of the curve \eref{b41condition} as $p_1\to\infty$. In this limit, the Wronskian no longer has a zero in the finite region of moduli space, hence now we should find solutions with $\ell=0$ and this is indeed what happens as we will see later in several examples. Defining:
\be
x_1=1-\frac{b_{4,1}}{p_1}
\ee
we see that at large $p_1$, \eref{b41condition} becomes:
\be
\alpha_0\alpha_1x_1^2-\frac56 x_1+1=0
\ee
with the solutions:
\be
x_1=\frac{\frac56\pm \sqrt{\frac{25}{36}-4\alpha_0\alpha_1}}{2\alpha_0\alpha_1}
\ee
Since $\ell=6$, we have from \eref{RRformula} that $\alpha_0+\alpha_1=-\frac56$. This allows us to simplify the above equation to:
\be
x_1=\Bigg\{-\frac{1}{\alpha_0},-\frac{1}{\alpha_1}\Bigg\}
\ee

Next we consider generic values of $\ell$ and show that the accessory parameters are determined similarly. As we will see, this allows the complete determination of $\alpha_0,\alpha_1$ for all $\ell$ in terms of those for $\ell<6$ which are already known.

\subsection{Determining accessory parameters: the general case}

\label{accessgen}

In the most general case with two characters and arbitrary $\ell$, as long as the singularities are well-separated the phenomenon is very similar. For each singularity $p_I$ we get one constraint on the combined set of $b_{4,I}$ and $p_I$ by trying to calculate the second-order coefficient $a^{(I)}_{0,2}$. This provides us with a set of simultaneous equations for the $b_{4,I}$. In the cases $\ell=6r,6r+2$ this is sufficient to determine all the $b_{4,i}$ in terms of the $p_I$. We now consider each family in more detail.

\subsection*{$\ell=6r$ case, $r\ge 0$}

In this case, we have $r$ full zeroes that are well-separated from each other and from the special points $j=0,1728$. Substituting \eref{Iexp} into the $\ell=6r$ case of \eref{twocharj} and equating the coefficient of  
$(j-p_I)^{\alpha_i^{(I)}-1}$ to zero, we get the indicial equation:
\begin{align}   \alpha_i^{(I)}\left(\alpha_i^{(I)}-2\right) = 0, \label{indi_26_6r_gen}
\end{align}
and hence $(\alpha_0^{(I)}, \alpha_1^{(I)}) = (0,2), ~1\leq I\leq r$. At the next order in the expansion, setting to zero the coefficient of $(j-p_I)^{\alpha_i^{(I)}}$ we get:
\begin{align}
    a_{i,1}^{(I)} &= \frac{1}{p_I(p_I-1728)\Big(1-\big(\alpha^{(1)}_i\big)^2\Big)}\lbrac{1cm}\frac{\alpha_0\alpha_1\prod\limits_{J=1}^r(p_I-b_{4,J})}{\prod\limits_{J=1 \atop J\neq I}^r(p_I-p_J)}
    + \alpha_i^{(I)}\left(\frac{7p_I}{6}-1152\right)\rbrac{1cm} \label{nI1_6r_gen}
\end{align}
which is manifestly a generalisation of \eref{nonerecur}. 
Choosing $\alpha_0^{(1)}=0$ in the above, we get:
\begin{align}
    a_{0,1}^{(I)} = \frac{\alpha_0\alpha_1\prod\limits_{J=1}^r(p_I-b_{4,J})}{p_I(p_I-1728)\prod\limits_{J=1 \atop J\neq I}^r(p_I-p_J)}, \label{n01_6r_gene}
\end{align}
At the next order, we set to zero the coefficient of $(j-p_I)^{\alpha_i^{(1)}+1}$. First let us examine the term that contains $a_{0,2}^{(I)}$:
\begin{align}
    a_{0,2}^{(I)} \, p_I (p_I-1728)\Bigg(\prod\limits_{J=1 \atop J\neq I}^{r}(p_I-p_J) \, \Bigg)\alpha_i^{(I)}\left(\alpha_i^{(I)}-2\right) \label{no_n12}
\end{align}
Since this is proportional to the indicial equation, the above expression is identically zero and hence the dependence on  $a_{0,2}^{(I)}$ drops out. In its place, we find a set of constraint equations on the parameters of the MLDE (assuming the $p_I$ are distinct from the accessory parameters $b_{4,J}$):
\begin{align}
    &a^{(I)}_{0,1}\lbrac{1cm}\frac{1}{p_I(p_I-1728)}\left(576-\frac{5p_I}{6}\right) - 2\sum\limits_{J=1 \atop J\neq I}^{r}\frac{1}{p_I-p_J} + \frac{\alpha_0\alpha_1}{p_I(p_I-1728)}\frac{\prod\limits_{J=1}^{r}(p_I-b_{4,J})}{\prod\limits_{J=1 \atop J\neq I}^r(p_I-p_J)}\rbrac{1cm} \nonumber\\
    &\qquad\qquad + \, \frac{\alpha_0\alpha_1}{p_I(p_I-1728)}\frac{\prod\limits_{J=1}^{r}(p_I-b_{4,J})}{\prod\limits_{J=1 \atop J\neq I}^r(p_I-p_J)}\left(\sum\limits_{J=1}^{r}\frac{1}{p_I-b_{4,J}}\right) = 0. \label{inter_constr}
\end{align}
Now substituting the value of $a^{(I)}_{0,1}$ from \eref{n01_6r_gene} in \eref{inter_constr} we get:
\be
\frac{1}{p_I(p_I-1728)}\lbrac{1cm} 576-\frac{5p_I}{6} + \alpha_0\alpha_1 \frac{\prod\limits_{J=1}^r(p_I-b_{4,J})}{\prod\limits_{J=1 \atop J\neq I}^r(p_I-p_J)}\rbrac{1cm} - 2\sum\limits_{J=1 \atop J\neq I}^r \frac{1}{p_I-p_J} + \sum\limits_{J=1}^r\frac{1}{p_I-b_{4,J}} = 0. \label{gen_quad_2_6r}
\ee
Thus we get a set of $r$ coupled equations for the accessory parameters $b_{4,J}$ which can be solved in principle to determine them as functions of the $p_I$. 

We can think of these equations as defining a sub-manifold or algebraic variety in the $2r$-dimensional parameter space of the $p_I$ and $b_{4,I}$. We will call them ``accessory equations''. If we multiply out all the denominators, these become a set of $r$ polynomials of degree $2r$. The special case in the last section, \eref{b41condition}, corresponds to $r=1$, hence a single quadratic equation, namely a hyperbola.

We see that for general $r$, each equation is separately invariant under a permutation of the $b_{4,I}$, while the equations are permuted among themselves if we permute the $p_I$. These facts suggest the use of symmetric polynomials in the $p_I$ as well as the $b_{4,I}$, which will be introduced in Section \ref{sec212}. Also, the equations become singular if any two of the $p_I$ coincide with each other or with an accessory parameter $b_{4,J}$. This is as expected, since both such coincidences change the nature of the original equation -- the first violates the genericity assumption, while the second cancels a pole in the last term of the MLDE.

Now let us consider the asymptotic region of the sub-manifold defined by \eref{gen_quad_2_6r}, by taking $p_r\to\infty$ together with $b_{4,r}$ while keeping $x_r = 1-\frac{b_{4,r}}{p_r}$  fixed. Then the equations for $I=1,2,\cdots,r-1$ become:
\be
\begin{split}
    &\frac{1}{p_I(p_I-1728)}\lbrac{1cm}
    576-\frac{5p_I}{6} + \alpha_0\alpha_1 \, (1-x_r) \frac{\prod\limits_{J=1}^{r-1}(p_I-b_{4,J})}{\prod\limits_{J=1 \atop J\neq I}^{r-1}(p_I-p_J)}\rbrac{1cm} 
    - 2\sum\limits_{J=1 \atop J\neq I}^{r-1} \frac{1}{p_I-p_J}\\
    &\qquad\qquad\qquad 
    + \sum\limits_{J=1}^{r-1}\frac{1}{p_I-b_{4,J}} = 0,\qquad  1\leq I \leq r-1 \label{gen_quad_26r_0}
\end{split}
\ee
while the equation for $I=r$ becomes:
\be
    \alpha_0\alpha_1 \, x_r^2 + \left(\frac{1-\ell}{6}\right)x_r + 1 = 0. \label{gen_quad_26r}
\ee
The second equation determines $x_r$ in terms of the product of exponents $\alpha_0\alpha_1$:
\be
x_r=\frac{1}{2\alpha_0\alpha_1}\Bigg(\frac{\ell-1}{6}\pm\sqrt{\frac{(\ell-1)^2}{36}-4\alpha_0\alpha_1}\Bigg)
\ee
We can simplify this using the valence formula \eref{RRformula} which tells us that $\alpha_0+\alpha_1=\frac{1-\ell}{6}$. Then:
\be
\begin{split}
x_r&=\frac{1}{2\alpha_0\alpha_1}\Big(-(\alpha_0+\alpha_1)\pm \alpha_0-\alpha_1\Big)\\
&= \Bigg\{-\frac{1}{\alpha_0},-\frac{1}{\alpha_1}\Bigg\}\\
&=\Bigg\{\frac{24}{c},\frac{24}{c-24h}\Bigg\}
\end{split}
\ee

Meanwhile, the first set of equations is precisely the one for the MLDE with $r$ replaced by $r-1$, i.e.  Wronskian index $\ell$ replaced by $\ell-6$, with the replacement:
\be
\begin{split}
(\alpha_0\alpha_1)^{(\ell-6)}&=(1-x_r)(\alpha_0\alpha_1)^{(\ell)}\\
&=\Bigg\{\frac{\alpha_0^{(\ell)}+1}{\alpha_0^{(\ell)}},\frac{\alpha_1^{(\ell)}+1}{\alpha_1^{(\ell)}}\Bigg\}(\alpha_0\alpha_1)^{(\ell)}\\
&= \bigg\{\big((\alpha_0+1)\alpha_1\big)^{(\ell)},
\big(\alpha_0(\alpha_1+1)\big)^{(\ell)}\bigg\}
\label{alpharecursive}
\end{split}
\ee
Applying the procedure recursively, this equation determines $\alpha_0,\alpha_1$ (up to exchange of characters) for all $\ell=6r$ given their values for $\ell=0$, which are known from \cite{Mathur:1988na}.

\subsection*{$\ell=6r+2$ case, $r\ge 0$}

For the case of $\ell=6r+2$, we again have $r$ distinct full zeros at $p_I$, and also a zero of order $\frac{1}{3}$ at $\tau=\rho \, (j=0)$. The genericity assumption also says that $p_I\neq 0, 1728$. This case is very similar to the $\ell=6r$ case and we again find $(\alpha_0^{(I)}, \alpha_1^{(I)}) = (0,2), ~1\leq I\leq r$. At the next order we have:
\be
    a_{i,1}^{(I)} = \frac{1}{p_i(p_I-1728)\Big(1-\big(\alpha^{(1)}_i\big)^2\Big)}\lbrac{1cm}\frac{\alpha_0\alpha_1\prod\limits_{J=1}^r(p_I-b_{4,J})}{\prod\limits_{J=1 \atop J\neq I}^r(p_I-p_J)} + \alpha_i^{(I)}\left(\frac{5p_I}{6}-576\right)\rbrac{1cm} \label{nI1_6rp2_gen}
\ee
Now choosing $\alpha_0^{(1)}=0$ in the above we get,
\begin{align}
    a_{0,1}^{(I)} = \frac{\alpha_0\alpha_1\prod\limits_{J=1}^r(p_I-b_{4,J})}{p_i(p_I-1728)\prod\limits_{J=1 \atop J\neq I}^r(p_I-p_J)}, \label{n01_6rp2_gen}
\end{align}
At the next order we find the constraint:
\begin{align}
    &a^{(I)}_{0,1}\lbrac{1cm}\frac{1}{p_I(p_I-1728)}\left(1152-\frac{7p_I}{6}\right) - 2\sum\limits_{J=1 \atop J\neq I}^{r}\frac{1}{p_I-p_J} + \frac{\alpha_0\alpha_1}{p_I(p_I-1728)}\frac{\prod\limits_{J=1}^{r}(p_I-b_{4,J})}{\prod\limits_{J=1 \atop J\neq I}^r(p_I-p_J)}\rbrac{1cm} \nonumber\\
    &\qquad\qquad\qquad + \, \frac{\alpha_0\alpha_1}{p_I(p_I-1728)}\frac{\prod\limits_{J=1}^{r}(p_I-b_{4,J})}{\prod\limits_{J=1 \atop J\neq I}^r(p_I-p_J)}\left(\sum\limits_{J=1}^{r}\frac{1}{p_I-b_{4,J}}\right) = 0. \label{inter_constr_6rp2}
\end{align}
Now substituting the value of $a^{(I)}_{0,1}$ from \eref{n01_6rp2_gen} in \eref{inter_constr_6rp2} we get,
\be
\begin{split}
    \frac{1}{p_I(p_I-1728)}\lbrac{1cm} 1152-\frac{7p_I}{6}
    + \alpha_0\alpha_1 \frac{\prod\limits_{J=1}^r(p_I-b_{4,J})}{\prod\limits_{J=1 \atop J\neq I}^r(p_I-p_J)}\rbrac{1cm}
    - 2\sum\limits_{J=1 \atop J\neq I}^r \frac{1}{p_I-p_J} + \sum\limits_{J=1}^r\frac{1}{p_I-b_{4,J}} = 0. \label{gen_quad_2_6rp2}
\end{split}
\ee
Thus, once more we get coupled equations for the $b_{4,J}$ which define a sub-manifold of the original space and determine the accessory parameters as functions of the $p_I$. The analysis of the asymptotic behaviour is precisely the same as for $\ell=6r$, and we again end up with \eref{gen_quad_26r} where now $\ell=6r+2$, as well as a version of \eref{gen_quad_26r_0} where $-\frac56 p_I$ is replaced by $-\frac76 p_I$ and 576 is replaced by 1152. Fruthermore, \eref{alpharecursive} remains unchanged in this case.

Since we have argued above that the $\ell=6r+4$ case can always be reduced to $\ell=6r$ by extracting a factor $j^\frac13$, we do not need to find the accessory equations separately for that case. Hence at this stage our analysis of accessory equations is complete. 

To summarise, what we have learned from the asymptotic analysis is that \eref{alpharecursive} is true for all $\ell=6r+u$ where $u=0,2$. Although there are two choices in this equation, it is clear that they are related by an exchange of characters. We can invert \eref{alpharecursive} and iterate it $u$ times (where $\ell=6r+u$) to get:
\be
\begin{split}
\alpha_0^{(\ell=6r+u)}&=
\alpha_0^{(\ell=u)}-r\\
\alpha_1^{(\ell=6r+u)}&=
(\alpha_1)^{(\ell=u)}
\end{split}
\ee
Here we have chosen $\alpha_0=-\frac{c}{24}$ and $\alpha_1=-\frac{c}{24}+h$. So the above equations tell us that:
\be
\begin{split}
c^{(\ell=6r+u)} &= c^{(\ell=u)}+24r\\
h^{(\ell=6r+u)} &=
h^{(\ell=u)}+r
\end{split}
\label{asympsummary}
\ee
Thus, using only the MLDE for generic $\ell$, we have demonstrated that the central charge and conformal dimension of a solution for any $\ell=6r,6r+2$ are related as above to those of an MLDE solution with $\ell=0,2$ (with a corresponding result for $\ell=6r+4$ following from factorisation of the solutions in that case). As we show below, this perfectly agrees with the analysis from quasi-characters \cite{Chandra:2018pjq}.

\subsection{Admissible range of central charges for $(2,\ell)$ solutions}\label{cent_relns}

In this sub-section, we study the admissible range of central charges for $(2,\ell)$ solutions, based on the asymptotic analysis and knowledge of the admissibility range for $(2,0)$ and $(2,2)$ solutions. Then we will present the results for $\ell = 6, 8, 12, 14$, which will be used in upcoming sections.

We first note that \eref{alpharecursive} can be solved for $c^{(\ell)}$ in terms of $c^{(\ell-6)}$ by using the valence formula and then replacing everything in terms of central charges. There are two possibilities for the product of exponents in this equation, each of which translates into two possibilities for the relation between central charges. Thus we get:
\begin{equation}
\begin{split}
    c^{(\ell)} = c^{(\ell-6)} + 24 \quad &\hbox{or}\quad
    c^{(\ell)} = 4(\ell-1) - c^{(\ell-6)}\\
    c^{(\ell)} = c^{(\ell-6)} \quad &\hbox{or}\quad
    c^{(\ell)} = 4(\ell-7) - c^{(\ell-6)}
\end{split}
\label{ch12}
\end{equation}
respectively. Imposing unitarity via $h^{(\ell)}>0$, we also get the lower bounds:
\begin{equation}
h^{(\ell)}=\frac{c-2(\ell-1)}{12} > 0 \implies c^{(\ell)}> 2(\ell-1)
\label{h_bounds}
\end{equation}
One of the four possibilities in \eref{ch12} can be ruled out, namely $c^{(\ell)} = 4(\ell-7) - c^{(\ell-6)}$. To see this, let us suppose it is allowed. Then the unitarity bound \eref{h_bounds} for $c^{(\ell)}$ gives $4(\ell-7)-c^{(\ell-6)}>2(\ell-1)$ implying $c^{(\ell-6)}< 2(\ell-13)$. However, the unitarity bound directly implies that $c^{(\ell-6)}>2(\ell-7)$. Thus we have a contradiction and the above possibility is ruled out. 

It follows that at each step we can only have the following three possibilities:
\be
c^{(\ell)}=c^{(\ell-6)}, \quad
c^{(\ell)} = c^{(\ell-6)} + 24,\quad c^{(\ell)} = 4(\ell-1) - c^{(\ell-6)}
\label{newbound}
\ee
We can now recursively work out the ranges for any given $\ell=6r,6r+2$ starting from the known ranges for $\ell=0,2$ \cite{Mathur:1988na, Gaberdiel:2016zke}:
\be
\begin{split}
\ell=0{}:&\qquad c^{(\ell=0)}\in (0,8)\\
\ell=2{}:& \qquad c^{(\ell=2)}\in (16,24)
\end{split}
\ee
and applying all the possibilities above subject to the constraint \eref{h_bounds}. We find \footnote{For $\ell=6$, we rule out the case: $c^{(\ell=6)}=20-c^{(\ell=0)}$, which in turn implies $c^{(\ell=6)}\in (12,20)$, by requiring the admissibility of $m_k^{(6)}$, for higher oders in $k\sim 2000$.}:
\be
\begin{split}
\ell=6{}:&\qquad  c^{(\ell=6)}\in (24,32)\\
\ell=8{}:&\qquad  c^{(\ell=8)}\in (16,24)~\cup~ (40,48) \\
\ell=12{}:&\qquad  c^{(\ell=12)}\in (24,32)~\cup~ (48,56)\\
\ell=14{}:&\qquad  c^{(\ell=14)} \in (28,36)~\cup~ (40,48)
~\cup~ (64,72)
\end{split}
\label{c_relns}
\ee
Using \eref{newbound} for $\ell=6, 8, 12, 14$, we get the following admissible sets for the above $\ell$ values:
\be
\begin{split}
c^{(\ell=6)}\in &\left\{\frac{122}{5},25,26,\frac{134}{5},28,\frac{146}{5},30,31,\frac{158}{5} \right\}\\
c^{(\ell=8)}\in &\left\{\frac{82}{5},17,18,\frac{94}{5},20,\frac{106}{5},22,23,\frac{118}{5} \right\}~\cup~ \left\{\frac{202}{5},41,42,\frac{214}{5},44,\frac{226}{5},46,47,\frac{238}{5} \right\} \\
c^{(\ell=12)}\in &\left\{\frac{122}{5},25,26,\frac{134}{5},28,\frac{146}{5},30,31,\frac{158}{5} \right\}~\cup~ \left\{\frac{242}{5},49,50,\frac{254}{5},52,\frac{266}{5},54,55,\frac{278}{5} \right\}\\
c^{(\ell=14)}\in &\left\{\frac{142}{5},29,30,\frac{154}{5},32,\frac{166}{5},34,35,\frac{178}{5} \right\}~\cup~ \left\{\frac{202}{5},41,42,\frac{214}{5},44,\frac{226}{5},46,47,\frac{238}{5} \right\} \\
&~\cup~ \left\{\frac{322}{5},65,66,\frac{334}{5},68,\frac{346}{5},70,71,\frac{358}{5} \right\}
\end{split}
\label{c_relns_1}
\ee

Let us digress a bit and conclude this sub-section with an observation regarding the modular data for $2$-character admissible solutions. Using Eqs. (\ref{ch12}), and the admissibility range for $(2,0)$ and $(2,2)$ solutions, we note that for any $\ell=6r$,  or $6r+2$ we have $c^{(\ell)} = \left\{n + c^{(\ell=0)}, m + c^{(\ell=2)}\right\}$, where $n,m$ are non-negative integers (as $\ell$ is a non-negative integer). Since $5c^{(\ell=0)}$ and $5c^{(\ell=2)}$ are known to be integers, the above observation implies that $5c^{(\ell)}$ is also an integer. For $\ell=6r+4$, we already know that $\chi^{(\ell=6r+4)}=j^{\frac{1}{3}}\chi^{(\ell=6r)}$ and hence $5c^{(\ell=6r+4)}$ is also an integer. It follows that $5c$ is an integer for any admissible $(2,\ell)$ solution. This fact was first noted in \cite{Mathur:1989pk} where the result was derived using representation theory of $\text{PSL}(2,\mathbb{Z})$. Here we have derived it using just the MLDE approach. Similar results about the modular data for $n$-character admissible solutions with $n=3, 4, 5$ have been obtained in \cite{Kaidi:2021ent}. It is worth exploring if those results can also be derived within the MLDE approach. 

\section{Detailed solution for the case of one movable pole}

\label{sec26}

With the understanding of this system that we have described above, it is relatively straightforward to directly find the most general admissible solutions of the MLDE in the case $\ell=6$, where there is one movable pole $p_1$ and one accessory parameter. We first present the solution and then examine its relation to the quasi-character approach. 

We will study the $(2,6)$ case in detail, with some formulae reserved for Appendix \ref{app2628}, while a similar analysis for the $(2,8)$ case can be found in Appendix \ref{wierd28}. For future use, a review of the Frobenius solution for the $(2,0)$ and $(2,2)$ MLDEs is presented in Appendix \ref{app2022}. This contains formulae that will be needed below.

\subsection{Solving the MLDE with one movable pole}

In this sub-section we will adapt the various elements of the theory of MLDEs developed so far into an organised method to solve them. This method involves incorporation of the accessory equation into the solution from the outset. It allows us, as we will see, to solve the MLDE in the present case completely and thereby derive features of the solution that were suggested by quasi-character theory \cite{Chandra:2018pjq}. 
The $(2,6)$ MLDE in the $\tau$-plane is:
\begin{eqnarray}
\left(D^2 +   \frac{E_4^2\, E_6}{E_4^3 - p_1\, \Delta}\, D + \frac{\alpha_0\,\alpha_1\, E_4(E_4^3 - b_{4, 1}\, \Delta) }{E_4^3 - 
p_1\, \Delta}\right) \chi(\tau) = 0. \label{26MLDE}
\end{eqnarray}
In this form, we have three parameters, the rigid parameter $\alpha_0\alpha_1$ and two 
non-rigid parameters, the movable pole $p_1$ and the accessory parameter $b_{4,1}$. We use the first three orders of the Frobenius solution as applied to the identity character. At leading order, we have the indicial equation which determines the rigid parameter in terms of the central charge, and at the second and third order we have the following :
\bea \label{ord0a}
\alpha_0 \alpha_1 &=& \frac{c(c-20)}{576}, \\
\label{ord1a}
m_1^{(6)} &= & f_1(c, p_1, b_{4,1})\\
\label{ord2a}
m_2^{(6)}&=&  f_2(c, p_1, b_{4,1})
\eea
 Here $m_1^{(6)}$ and $m_2^{(6)}$ are the  Fourier coefficients of the identity character.  The superscript $(6)$ indicates that this is of the $(2,6)$ solution. The explicit forms of  $f_1(c, p_1, b_{4,1})$ and  $f_2(c, p_1, b_{4,1})$ are given in \eref{ord1b} and \eref{ord2b}.

 In the next step, one solves for the three parameters of the MLDE in terms of objects associated to the identity character, namely the central charge $c$ and the Fourier coefficients $m_1^{(6)}$ and $m_2^{(6)}$. This has already been done for $\alpha_0\alpha_1$ in \eref{ord0a}. For the remaining parameters we obtain:
\bea \label{nr1a}
p_1 &=& f_3(c, m_1^{(6)}, m_2^{(6)})
 \\
\label{nr2a}
b_{4,1} &=& f_4(c, m_1^{(6)}, m_2^{(6)})
\eea
The explicit expressions for the right hand sides can be found in \eref{nr1b} and \eref{nr2b}. We note that both $p_1$ and $b_{4,1}$ are rational functions of $m_1^{(6)}$ and $m_2^{(6)}$ with coefficients being rational functions of $c$. In particular, we see that  the movable pole in the $(2,6)$ solution is rational, as already noted in \cite{Naculich:1988xv}. Later we will discuss the general version of this statement.

The next step is to invoke the accessory equations \eref{b41condition}, insert the values of $p_1$ and $b_{4,1}$, previously determined in \eref{nr1a} and \eref{nr2a}, and solve for $m_2^{(6)}$ in terms of $m_1^{(6)}$ and $c$. Remarkably, we get the following linear equation in $m_1^{(6)}$:
\bea \label{m2mldea}
m_2^{(6)} &=& A_2(c) +  B_2(c)~m_1^{(6)}
\eea
where $A_2(c)$ and $B_2(c)$ are given in \eref{m2mldeb}.
Consulting \eref{app20m1m2} we immediately find a relation between the coefficient $B_2(c)$ above, and the degeneracy for the $(2,0)$ MLDE solution at a central charge $c-24$:
\be
B_2(c)=m_1^{(0)}(c-24)
\ee
Some additional calculation shows that:
\be
A_2(c) = m_2^{(0)}(c)-m_1^{(0)}(c-24)\,m_1^{(0)}(c)
\ee
Thus \eref{m2mldea} is the same as:
\be \label{m2linear}
m_2^{(6)} = m_2^{(0)}(c) + m_1^{(0)}(c-24)~~(m_1^{(6)} - m_1^{(0)}(c)).
\ee

At the next stage, we insert \eref{m2linear} in \eref{nr1a} and \eref{nr2a} to obtain:
\bea \label{nr1.2a}
p_1 &=&  f_5(c, m_1^{(6)})  \\
\label{nr2.2a}
b_{4,1}&=&  f_6(c, m_1^{(6)})
\eea
The explicit expressions for the right hand sides are given in \eref{nr1.2b} and \eref{nr2.2b}. These equations now have a nice geometrical interpretation. The space of MLDE parameters is three dimensional, co-ordinatized by $\alpha_0 \alpha_1$ (or, via \eref{ord0a}, the central charge $c$), $p_1$ and $b_{4,1}$. For a  fixed central charge $c$, we have the $p_1 - b_{4,1}$ plane. The algebraic variety defined by the accessory  equation \eref{b41condition} is a hyperbola in this plane and the equations \eref{nr1.2a} and \eref{nr2.2a} are its parametric equations with $m_1^{(6)}$ serving as a parameter on the curve.

We carry on solving the MLDE to higher order. At the next order, after using \eref{m2linear} we obtain the following:
\bea \label{m3mlde}
m_3^{(6)} && = A_3(c)  +  B_3(c)~m_1^{(6)} 
\eea
where $A_3(c)$ and $B_3(c)$ are given in \eref{m3mldeb}. It is again remarkable that $m_3^{(6)}$ has a linear dependence on $m_1^{(6)}$. In the same way as was done above, one shows that $m_3^{(6)}$ can be written in terms of the $(2,0)$ solution as follows:
\be \label{m3linear}
m_3^{(6)} = m_3^{(0)}(c) + m_2^{(0)}(c-24)~(m_1^{(6)} - m_1^{(0)}(c)),
\ee
which is very similar to the form of $m_2^{(6)}$ in \eref{m2linear}.

This motivates us to propose the relation: 
\be \label{mklinear}
m_k^{(6)} = m_k^{(0)}(c) + m_{k-1}^{(0)}(c-24)~ (m_1^{(6)} - m_1^{(0)}(c)).
\ee
We have performed a computer check of this phenomenon to order 8. We expect it to hold for all $k \geq 2$ and hope to provide a proof in future work. 

Notice that we can extend \eref{mklinear} to include $k = 1$. When we plug in $k = 1$ in \eref{mklinear} we get  $m_1^{(6)} = m_1^{(0)}(c) + m_{0}^{(0)}(c-24)\,(m_1^{(6)} - m_1^{(0)}(c))$ which is an identity after noting $m_{0}^{(0)}(c-24) = 1$. 

Now \eref{mklinear} can be converted into an equation relating the identity characters at $c$ and $c-24$. We then compute the non-identity character of the $(2,6)$ MLDE and find that it satisfies the same equation, leading to:
\bea \label{26iden}
\chi_i^{(6)} = \chi_i^{(0)}(c) + (m_1^{(6)} - m_1^{(0)}(c)) \, \chi_i^{(0)}(c-24),\quad i=0,1
\eea
We should emphasize that \eref{26iden} holds for all Frobenius solutions of the $(2,6)$ MLDE without any qualifiers such as admissibility, integrality etc: every Frobenius solution of the $(2,6)$ MLDE can be written as a sum of two Frobenius solutions of $(2,0)$ MLDE.

Now we impose admissibility. For this, we impose integrality of the $m_k^{(6)}$s and each of \eref{mklinear}, for $k \geq 2$, leads to a Diophantine equation, after defining $\mathsf{N} = 5 c$. The first two are  
\bea
\label{Dioph1}
&&\mathsf{N}^4 + (2 m_1^{(6)} - 427)\,\mathsf{N}^3 + (-656 m_1^{(6)} + 2 m_2^{(6)} + 41140) \, \mathsf{N}^2 \nonumber \\ 
&&\qquad + (71480 m_1^{(6)} - 560 m_2^{(6)} + 1124700)\, \mathsf{N} + 37400 m_2^{(6)} - 2587200 m_1^{(6)} = 0 \\
\label{Dioph2}
&&2\,\mathsf{N}^6 + (3 m_1^{(6)} - 1308)\,\mathsf{N}^5  + (274648-1665 m_1^{(6)})\,\mathsf{N}^4 + (369774 m_1^{(6)} - 6 m_3^{(6)} - 18801040)\,\mathsf{N}^3 \nonumber \\
&& \qquad  + (-41075340 m_1^{(6)} + 3060 m_3^{(6)} + 453302400)\,\mathsf{N}^2 + (2282045400 m_1^{(6)} - 498600 m_3^{(6)} + 22315264000)\,\mathsf{N} \nonumber \\
&& \qquad \qquad + 25806000 m_3^{(6)} - 50725224000 m_1^{(6)} = 0
\eea
In particular this shows directly that $N=5c$ is an integer. Now, inserting the admissible set of central charges \eqref{c_relns_1} into the above equations, we output all the possible admissible solutions. We also verify integrality of the non-identity character up to the same order. The result can then be computed up to very high orders ($q^{2000}$ in this case) and verified to be admissible. We find an infinite family of admissible solutions for each of the following central charges:
\be
c=\frac{122}{5},25,26,\frac{134}{5},28,\frac{146}{5},30,31,\frac{158}{5}
\label{twosixc}
\ee
labelled by the free integer $m_1^{(6)}\ge 0$.

We now study the other MLDE with a single movable pole, the $(2,8)$ MLDE :

\bea
\left(D^2 + \frac{4}{3}\,\frac{E_6\left(E_4^3 - \frac{p_1}{4}\, \Delta\right)}{E_4\left(E_4^3 - p_1\, \Delta\right)} D + \frac{\alpha_0\, \alpha_1\, E_4^2\left(E_4^3 - b_{4, 1}\, \Delta\right)}{E_4\left(E_4^3 - p_1\, \Delta\right)}\right) \chi(\tau) = 0 \label{MLDE28}
\eea
In this form, we have three parameters, the rigid parameter $\alpha_0\alpha_1$ and two 
non-rigid parameters, the movable pole $p_1$ and the accessory parameter $b_{4,1}$. We use the first three orders of the Frobenius solution as applied to the identity character. At leading order, we have the indicial equation which determines the rigid parameter in terms of the central charge, and at the second and third order we have the following :
\bea \label{ord0A}
\alpha_0 \alpha_1 &=& \frac{c(c-28)}{576}, \\
\label{ord1A}
m_1^{(8)} &= & \widetilde{f_1}(c, p_1, b_{4,1})\\
\label{ord2A}
m_2^{(8)}&=&  \widetilde{f_2}(c, p_1, b_{4,1})
\eea
 Here $m_1^{(8)}$ and $m_2^{(8)}$ are the  Fourier coefficients of the identity character.  The superscript $(8)$ indicates that this is of the $(2,8)$ solution. The explicit forms of  $\widetilde{f_1}(c, p_1, b_{4,1})$ and  $\widetilde{f_2}(c, p_1, b_{4,1})$ are given in \eref{ord1B} and \eref{ord2B}.
 
 In the next step, one solves for the three parameters of the MLDE in terms of objects associated to the identity character, namely the central charge $c$ and the Fourier coefficients $m_1^{(8)}$ and $m_2^{(8)}$. This has already been done for $\alpha_0\alpha_1$ in \eref{ord0A}. For the remaining parameters we obtain:
\bea \label{nr1A}
p_1 &=& \widetilde{f_3}(c, m_1^{(8)}, m_2^{(8)})
 \\
\label{nr2A}
b_{4,1} &=& \widetilde{f_4}(c, m_1^{(8)}, m_2^{(8)})
\eea
The explicit expressions for the right hand sides can be found in \eref{nr1B} and \eref{nr2B}. We note that both $p_1$ and $b_{4,1}$ are rational functions of $m_1^{(8)}$ and $m_2^{(8)}$ with coefficients being rational functions of $c$. In particular, we see that  the movable pole in the $(2,8)$ solution is rational. 

The next step is to invoke the accessory equation \eref{acceqn_28}, insert the values of $p_1$ and $b_{4,1}$, previously determined in \eref{nr1A} and \eref{nr2A}, and solve for $m_2^{(8)}$ in terms of $m_1^{(8)}$ and $c$. Similar to the $(2,6)$ computation, remarkably, we get the following linear equation in $m_1^{(8)}$:
\be \label{m2linear8}
m_2^{(8)} = m_2^{(2)}(c) + m_1^{(2)}(c-24)~~(m_1^{(8)} - m_1^{(2)}(c)).
\ee

At the next stage, we insert \eref{m2linear8} in \eref{nr1A} and \eref{nr2A} to obtain:
\bea \label{nr1.2A}
p_1 &=&  \widetilde{f_5}(c, m_1^{(8)})  \\
\label{nr2.2A}
b_{4,1}&=&  \widetilde{f_6}(c, m_1^{(8)})
\eea
The explicit expressions for the right hand sides are given in \eref{nr1.2B} and \eref{nr2.2B}. These equations now have a geometrical interpretation, similar to the $(2,6)$ case. 

We carry on solving the MLDE to higher order and obtain :
\be \label{mklinear8}
m_k^{(8)} = m_k^{(2)}(c) + m_{k-1}^{(2)}(c-24)~ (m_1^{(8)} - m_1^{(2)}(c)).
\ee
We have performed a computer check of this phenomenon to order 8. We expect it to hold for all $k \geq 2$ and hope to provide a proof in future work. 

We can extend \eref{mklinear8} to include $k = 1$. When we plug in $k = 1$ in \eref{mklinear8} we get  $m_1^{(8)} = m_1^{(2)}(c) + m_{0}^{(2)}(c-24)\,(m_1^{(8)} - m_1^{(2)}(c))$ which is an identity after noting $m_{0}^{(2)}(c-24) = 1$. 

Now \eref{mklinear8} can be converted into an equation relating the identity characters at $c$ and $c-24$. We then compute the non-identity character of the $(2,8)$ MLDE and find that it satisfies the same equation, leading to:
\bea \label{28iden}
\chi_i^{(8)} = \chi_i^{(2)}(c) + (m_1^{(8)} - m_1^{(2)}(c)) \, \chi_i^{(2)}(c-24),\quad i=0,1
\eea
We should emphasize that \eref{28iden}   holds for all Frobenius solutions of the $(2,8)$ MLDE without any qualifiers such as admissibility, integrality etc: every Frobenius solution of the $(2,8)$ MLDE can be written as a sum of two Frobenius solutions of $(2,2)$ MLDE.

Now we impose admissibility. For this, we impose integrality of the $m_k^{(8)}$s and each of \eref{mklinear8}, for $k \geq 2$, leads to a Diophantine equation, after defining $\mathsf{N} = 5 c$. The first two are  
\bea \label{Dioph18}
&& \mathsf{N}^4 +(2 m^{(8)}_1   -755) \mathsf{N}^3  + (2 m^{(8)}_2  -1024 m^{(8)}_1   +190108) \mathsf{N}^2 \nonumber\\
&& \qquad \qquad +( 162200 m^{(8)}_1   - 640 m^{(8)}_2   -15965940) \mathsf{N}- 8174400 m^{(8)}_1 + 49400 m^{(8)}_2 = 0  \\
\label{Dioph28}
&& 2 \mathsf{N}^6 + (3 m^{(8)}_1  -2292) \mathsf{N}^5 -(2709 m^{(8)}_1 -983128) \mathsf{N}^4+(904050 m^{(8)}_1  -6 m^{(8)}_3 -193838000 )\mathsf{N}^3 \nonumber\\
   &&-(144191100 m^{(8)}_1  -3420 m^{(8)}_3  -17557104000 )\mathsf{N}^2 +(11171925000 m^{(8)}_1  -628200 m^{(8)}_3  -686724352000) \mathsf{N}\nonumber\\
   &&\qquad \qquad -339388920000 m^{(8)}_1+37050000 m^{(8)}_3 =0
\eea

In particular this shows directly that $N=5c$ is an integer. Now, inserting the admissible set of central charges \eqref{c_relns_1} into the above equations, we output all the possible admissible solutions. We also verify integrality of the non-identity character up to the same order. The result can then be computed up to very high orders ($q^{2000}$ in this case) and verified to be admissible. We find an infinite family of admissible solutions for each of the following central charges:
\bea \label{two8c}
c=\frac{82}{5}, 17, 18, \frac{94}{5}, 20, \frac{106}{5}, 22, 23, \frac{118}{5}
\eea
labelled by the free integer $m_1^{(8)}\ge 0$.

\label{twosixsoln}

\subsection{Brief review of quasi-characters}

A construction of admissible characters for all two-character CFT was presented in \cite{Chandra:2018pjq} \footnote{There is an earlier construction of VVMF due to Bantay and Gannon \cite{Bantay:2005vk,Bantay:2007zz}, however, that requires advance knowledge of the possible modular data while here we do not make this assumption. Also here, as part of admissibility, we always impose the requirement that the leading term of the identity character is unity.}. This proposal did not use MLDEs with movable poles (i.e. $\ell\ge 6)$ that we are using here, rather it only made use of solutions to the MMS equation, which has $\ell=0$, and a similar equation with $\ell=2$. Now we are in a position to compare our results, obtained from the $\ell=6$ MLDE, with this approach. For this we first briefly review the quasi-character approach and its application to the $(2,6)$ case (for a detailed exposition with references, see \cite{Chandra:2018pjq}). Then we will compare the results of the present paper with it.

Ref. \cite{Chandra:2018pjq} started from the observation that although the $(2,0)$ MLDE -- the MMS equation -- has only finitely many admissible solutions, it has infinitely many more solutions having all integral Fourier coefficients of which some are {\em negative}. Thus these are special, although not admissible, solutions. They occur at specific values of the central charge $c$ \footnote{Although these solutions do not describe CFT, they can still be assigned a value of $c$ by writing their leading critical exponent $\alpha_0$ as $-\frac{c}{24}$.}. There are families of such solutions with the following central charges, parametrised by an integer $n$:
\be
\begin{split}
\hbox{Lee-Yang family:}\quad c &=\frac{2(6n+1)}{5},~n\ne 4 \hbox{ mod }5\\
A_1 \hbox{ family:}\quad c &=6n+1\\
A_2 \hbox{ family:}\quad c &=4n+2,~n\ne 2\hbox{ mod }3\\
D_4 \hbox{ family:}\quad c &=12n+4
\end{split}
\label{quasi20}
\ee
Of these, the central charges 
\be
c=\frac25,1,2,\frac{14}{5},4,\frac{26}{5},6,7,\frac{38}{5}
\label{mmslist}
\ee
correspond to admissible characters \footnote{These all correspond to CFTs, except for the first and last cases that are ``Intermediate Vertex Operator Algebras'' \cite{Kawasetsu:2014}.}  with $\ell=0$. Together with a $c=8$ solution that corresponds to a one-character solution with a spurious second character, hence is not in the above set, these make up the so-called ``MMS series'' \cite{Mathur:1988na}.

While all such quasi-characters have $\ell=0$ and solve the MMS equation ($(2,0)$ MLDE), they do so for different values of the parameter in the MLDE. Thus their linear combinations do not solve the same equation, and in general they would not be closed under modular transformations. However if we take linear combinations of $r+1$ quasi-characters such that successive terms differ in central charge by 24 (they automatically then belong to the same family in the list above), it can be shown that the modular transformations of each term are the same, and that the linear combination satisfies an MLDE for which the Wronskian index is $6r$. It was argued in \cite{Chandra:2018pjq} that this process generates all $(2,\ell)$ admissible characters for every $\ell=6r$. 

Quasi-characters with $\ell=2$, relevant to the $\ell=6r+2$ case, have also been constructed \cite{Chandra:2018pjq} and we review them in Appendix \ref{wierd28}. On the other hand the ones with $\ell=4$, relevant to $\ell=6r+4$ are simply $j^\frac13$ times the $\ell=0$ quasi-characters listed above. Thus all possible values of $\ell$ have been covered.

Now we return to the case $\ell=6$. Here one must add precisely two $\ell=0$ quasi-characters differing in central charge by 24. We take one of these to be any of the MMS solutions, denoted $\chi^A_i$ (where $A$ stands for ``admissible''), whose central charge lies in the MMS list \eref{mmslist}, and the other to be the quasi-character $\chi^Q_i$ with central charge 24 higher. We denote the latter central charge by $c$ and the former by $c-24$. Thus we form the sum:
\be
\chi_i^Q(q)+N_1\,\chi_i^A(q)
\ee
This sum has the following properties: (i) it has central charge $c$ and satisfies \eref{RRformula} with Wronskian index 6, (ii) the negative degeneracies of the quasi-character in the sum are potentially cancelled by the positive terms in the admissible character, depending on the value of $N_1$. Thus the sum is admissible for $N_1$ greater than some lower bound, which varies from case to case.

In view of completeness of the above approach, one therefore predicts that all $(2,6)$ admissible characters (and hence all $(2,6)$ CFT) have central charges:
\be
c=\frac{122}{5},25,26,\frac{134}{5},28,\frac{146}{5},30,31,\frac{158}{5}
\label{QplusA}
\ee
This precisely coincides with \eref{twosixc} except for the two end-points. As already noted below that equation, those correspond to one-character theories that show up as two-character MLDE solutions with one spurious character, which we ignore. 

Thus we have found perfect agreement between the central charges arising in the direct solution of the $(2,6)$ MLDE for admissible characters in Sub-section \ref{twosixsoln}, and the central charges found from the quasi-character construction of the same admissible set that does not use the $(2,6)$ MLDE at all. We now go on to make a more detailed comparison of the results of the two approaches.

\subsection{Comparison of quasi-character and MLDE results}

\label{quasiMLDEcomp}

In this sub-section we confront the explicit admissible MLDE solutions described above with the quasi-character approach. The former approach has one free parameter, which we can take to be $p_1$ describing the location of the zero of the Wronskian, or the Fourier coefficient $m_1$ representing the degeneracy of the first excited state in the identity module. The two are related by \eref{nr1.2a}. The latter approach has a free parameter $N_1$, that also determines the first excited state degeneracy $m_1$. Thus $p_1$, the location of the movable pole, must be a function of the integer $N_1$. We see that admissibility quantises the location of the movable pole and also that the quasi-character parameter $N_1$ is the natural integer in terms of which this quantisation can be expressed. We now exhibit these relations in all the cases. We will see that all solutions lie on one of the two branches of the hyperbola in \eref{b41condition}, while the other branch actually corresponds to negative values of $m_1$.

\subsubsection*{Admissible Solutions (i)}
\bea
c= \frac{122}{5},\quad m_1 \geq 0,\quad m_2 = 169885 + m_1,\quad m_3 = 19870140 + m_1
\eea
For this central charge, from Eqs.\eqref{nr1.2a}-\eqref{nr2.2a} we get $p_1$ and $b_{4,1}$ as functions of $m_1$:
\bea
p_1 &=& -\frac{(m_1-3538)(m_1-658)}{6 (m_1 +244)} \label{26a1}  \\
b_{4,1} &=& -\frac{(m_1-354898) (m_1-3538)}{366
   (m_1+244)} \label{26b41}
\eea
This is also the solution obtained by the quasi-character method
\bea
\chi^{LY}_{n = 10} + N_1\,\chi^{LY}_{n = 0}
\eea
with $m_1 = N_1 - 244$.\\[1mm]


\subsubsection*{Admissible Solutions (ii)}
\bea 
c=25, \quad m_1 \geq 0, \quad m_2 = 143375 + 3m_1, \quad m_3 = 18616375 + 4m_1
\eea
In this case we have:
\bea
p_1 &=& -\frac{(\text{$m_1$}-2875) (\text{$m_1 $}-571)}{5
   (\text{$m_1$}+245)} \label{26a1ii}\\
b_{4,1} &=& -\frac{(\text{$m_1$ }-118075) (\text{$m_1$ }-2875)}{125 (\text{$m_1$ }+245)}  \label{26b41ii} 
\eea
This is also the solution obtained by the quasi-character method
\bea
\chi^{A_1}_{n = 4} + N_1\,\chi^{A_1}_{n = 0}
\eea
with $m_1 = N_1 - 245$.\\[1mm]


\subsubsection*{Admissible Solutions (iii)}
\bea
c=26, \quad m_1 \geq 0, \quad m_2 = 118105 + 8m_1, \quad m_3 = 18305456 + 17m_1
\eea
In this case we have:
\bea
p_1 &=& -\frac{(\text{$m_1$}-2210) (\text{$m_1$}-482)}{4 (\text{$m_1$}+247)} \label{26a1iii} \\ 
b_{4,1} &=& -\frac{(\text{$m_1$}-47138) (\text{$m_1$}-2210)}{52
   (\text{$m_1$}+247)} \label{26b41iii}
\eea
This is also the solution obtained by the quasi-character method:
\bea
\chi^{A_2}_{n = 6} + N_1\,\chi^{A_2}_{n = 0}
\eea
with $m_1 = N_1 - 247 $.\\[0mm]


\subsubsection*{Admissible Solutions (iv)}

\bea
c= \frac{134}{5},\quad m_1 \geq 0,\quad m_2 = 106731 + 14\,m_1,\quad m_3 = 19112822 + 42\,m_1
\eea
In this case we have:
\bea
p_1 &=& -\frac{2 (\text{$m_1$}-1876) (\text{$m_1$}-436)}{7
   \text{$m_1$}+1742} \label{26a1iv}\\
  b_{4,1} &=& -\frac{2 (\text{$m_1$}-1876) (7 \text{$m_1$}-206092)}{67 (7
   \text{$m_1$}+1742)}   \label{26b41iv}
\eea

This is also the solution obtained by the quasi-character method:
\bea
\frac17\big(\chi^{LY}_{n = 11} + N_1\,\chi^{LY}_{n = 1}\big)
\label{LYsum}
\eea
This is a curious case, already remarked upon in Section 5.2 of \cite{Chandra:2018pjq}. What happens here is that $\chi_{n=11}^{LY}$ has an integral $q$-expansion only if the first term of the identity character is normalised to 7, rather than 1. This is the normalisation chosen above. The first excited state ``degeneracy'' of this quasi-character is $-1742$ while all others are positive. Since the identity character of the sum will be  considered admissible only when its leading term is 1, we must divide the sum by 7 as shown above. As a result the degeneracy of the first excited state is $m_1 = \frac{N_1 - 1742}{7}$. This can be any integer, as long as we choose $N_1$ to be 1742 plus a multiple of 7. With this choice, the sum in \eref{LYsum} has all integral coefficients even after dividing by 7, which is a miracle of sorts since it means all the infinitely many coefficients become multiples of 7 even though neither of the terms in the sum has this property.


\subsubsection*{Admissible Solutions (v)}
\bea
c=28, \quad m_1 \geq 0, \quad m_2 = 97930 + 28m_1, \quad m_3 = 21891520 + 134m_1
\eea
In this case we have:
\bea
p_1 &=& -\frac{(\text{$m_1$}-1540) (\text{$m_1$}-388)}{3
   (\text{$m_1$}+252)} \label{26a1v}\\
b_{4,1} &=& -\frac{(\text{$m_1$}-17668) (\text{$m_1$}-1540)}{21
   (\text{$m_1$}+252)}    \label{26b41v}
\eea
This is also the solution obtained by the quasi-character method
\bea
\chi^{D_4}_{n = 2} + N_1\,\chi^{D_4}_{n = 0}
\eea
with $m_1 = N_1 - 252 $.


\subsubsection*{Admissible Solutions (vi)}

\bea
c= \frac{146}{5},\quad m_1 \geq 0,\quad m_2 = 96433 + 52\,m_1,\quad m_3 = 27102272 + 377\,m_1
\eea

\bea
p_1 &=& -\frac{3 (\text{$m_1$}-1314) (\text{$m_1$}-354)}{4 (
   2\text{$m_1$}+511)} \label{26a1vi}\\
b_{4,1} &=& -\frac{3 (\text{$m_1$}-1314) (13 \text{$m_1$}-157242)}{292 (2
   \text{$m_1$}+511)}   \label{26b41vi}
\eea
This is also the solution obtained by the quasi-character method
\bea
\half\Big(\chi^{LY}_{n = 12} + N_1\,\chi^{LY}_{n = 2}\Big)
\eea
with $m_1 = \half(N_1 - 511)$. Here the quasi-character, when normalised to be integral, starts with 2. For $m_1$ to be integral, we must choose $N_1$ to be an odd integer.


\subsubsection*{Admissible Solutions (vii)}

\bea
c=30, \quad m_1 \geq 0, \quad m_2 = 99675 + 78m_1, \quad m_3 = 32782900 + 729m_1
\eea

\bea
p_1 &=& -\frac{2 (\text{$m_1$}-1200) (\text{$m_1$}-336)}{5
   (\text{$m_1$}+258)} \label{26a1vii}\\
b_{4,1} &=& -\frac{2 (\text{$m_1$}-9840) (\text{$m_1$}-1200)}{25
   (\text{$m_1$}+258)}   \label{26b41vii}
\eea
This is also the solution obtained by the quasi-character method
\bea
\chi^{A_2}_{n = 7} + N_1\,\chi^{A_2}_{n = 1}
\eea
with $m_1 = N_1 - 258 $.


\subsubsection*{Admissible Solutions (viii)}

\bea 
c=31, \quad m_1 \geq 0, \quad m_2 = 110980 + 133m_1, \quad m_3 = 44696513 + 1673m_1
\eea

\bea
p_1 &=& -\frac{3 (\text{$m_1$}-1085) (\text{$m_1$}-317)}{7
   \text{$m_1$}+1829} \label{26a1viii}\\
b_{4,1} &=& -\frac{3 (\text{$m_1$}-1085) (7 \text{$m_1$}-55211)}{31 (7
   \text{$m_1$}+1829)}   \label{26b41viii}
\eea
This is also the solution obtained by the quasi-character method
\bea
\chi^{A_1}_{n = 5} + N_1\,\chi^{A_1}_{n = 1}
\eea
with $m_1 = \frac17(N_1 - 1829)$ and $N_1$ must be taken to be 1829 plus a multiple of 7.


\subsubsection*{Admissible Solutions (ix)}

\bea
c= \frac{158}{5},\quad m_1 \geq 0,\quad m_2 = 124741 + 190\, m_1,\quad m_3 = 56937196 + 2831\, m_1
\eea

\bea
p_1 &=& -\frac{4 (\text{$m_1$}-1027) (\text{$m_1$}-307)}{9
   \text{$m_1$}+2370} \label{26a1ix}\\
b_{4,1} &=& -\frac{4 (\text{$m_1$}-1027) (19 \text{$m_1$}-133273)}{237 (3
   \text{$m_1$}+790)}    \label{26b41ix}
\eea
This is also the solution obtained by the quasi-character method
\bea
\chi^{LY}_{n = 13} + N_1\,\chi^{LY}_{n = 3}
\eea
with $m_1 = \frac13(N_1 - 790)$ and $N_1$ has to be chosen to be 790 plus a multiple of 3.

\subsection{Analysis of the accessory equation}

Let us now analyse the accessory equation \eref{b41condition} in the case of $\ell=6$ in some more detail. This equation can be re-written as a quadratic:
\be
p_1^2+\alpha_0\alpha_1(p_1-b_{4,1})^2-\frac56 p_1(p_1-b_{4,1})-1728p_1+576(p_1-b_{4,1})=0
\ee
As noted below \eref{b41condition}, this is a hyperbola for all values of $\alpha_0\alpha_1$ except when $\alpha_0\alpha_1=\frac{25}{144}$, corresponding to $c=10$, when it degenerates to a parabola. Remaining away from $c=10$, we now analyse the hyperbola in some detail. We will see, among other things, that all $(2,6)$ solutions with $N_1>0$ lie on one branch of the hyperbola, with the other branch corresponding to negative values of $N_1$.

To illustrate this, we pick an example. Consider the $(2,6)$ solution with $c=25$. In this case, we have: $N_1 = m_1+245$ (see previous section). For this, the hyperbola is given below.
\begin{figure}[H]
	\begin{center}
		\includegraphics[width=0.7\textwidth]{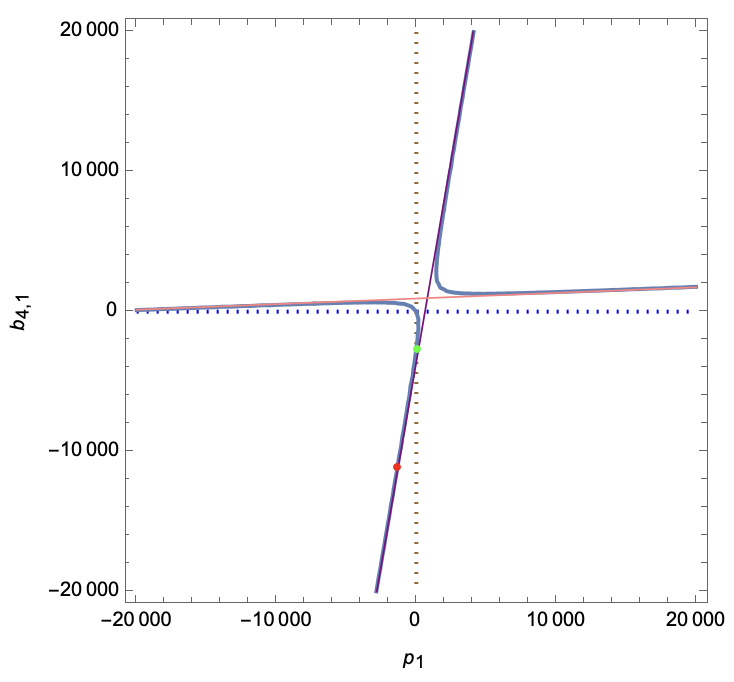}
	\end{center}
	\caption{Hyperbola in the $b_{4,1}$ vs $p_1$ plane corresponding to the $(2,6)$ solution with $c=25$.}\label{c25_l6}
\end{figure}
In this case, from \eref{26a1ii} and \eref{26b41ii}, we get, 
\begin{equation}
\begin{split}
    &b_{4,1}=-\frac{(N_1-3120)(N_1-118320)}{125 N_1}, \\
    &p_1=-\frac{(N_1-816)(N_1-3120)}{5 N_1}.
\end{split}
\label{b41_p1_c25_l6} 
\end{equation}
This gives, $\frac{b_{4,1}}{p_1}=\frac{N_1-118320}{25(N_1-816)}$. The asymptotes to the above hyperbola are: $b_{4,1} = -\frac{89856}{25} + \frac{29}{5}p_1$ (drawn in purple) and $b_{4,1} = \frac{117504}{125} + \frac{1}{25} p_1$ (drawn in pink). Note that, the origin lies on the lower branch. This can be seen from the fact that when $N_1\to 0$, we have $\frac{b_{4,1}}{p_1}\to \frac{29}{5}$ (whose slope is equal to the purple asymptote), which intersects the lower branch at the origin. This means that the point $N_1\to 0$ lies on the bottom end of the lower branch. Also, note that when $N_1\to\infty$, we have $\frac{b_{4,1}}{p_1}\to\frac{1}{25}$, whose slope is equal to the pink asymptote. This means that the point $N_1\to \infty$ lies on the left end of the lower branch.

The lower branch of the hyperbola corresponds to characters with $N_1>0$ and the upper branch corresponds to characters with $N_1<0$. To see this note the following argument. Using \eref{b41_p1_c25_l6}, we can see that both $b_{4,1}$ and $p_1$ can only be positive (which happens on the upper branch) if $N_1<0$. On the contrary, both $b_{4,1}$ and $p_1$ can never be positive (which happens on the lower branch) if $N_1>0$.

$N_1$ increases as we trace the lower branch from below. The red dot on the lower branch is where $m_1=0$. Starting from this red dot and tracing towards the left end of the lower branch we obtain all the admissible solutions. The green dot on the lower branch corresponds to the point where $m_1=571$ implying $N_1=816$. This in turn implies $p_1=0$ and $b_{4,1}\neq 0$. This is a factorised solution of the form $j^{\frac{1}{3}}\chi^{(2,2)}$ where $\chi^{(2,2)}$ is a $(2,2)$ CFT with $c=17$ (see \cite{Gaberdiel:2016zke}).  

In the next plot, we have all the hyperbolas corresponding to each of the $(2,6)$ solutions with central charges in the admissibility range: $24 < c^{(\ell=6)} < 32$.
\begin{figure}[H]
	\begin{center}
		\includegraphics[width=0.7\textwidth]{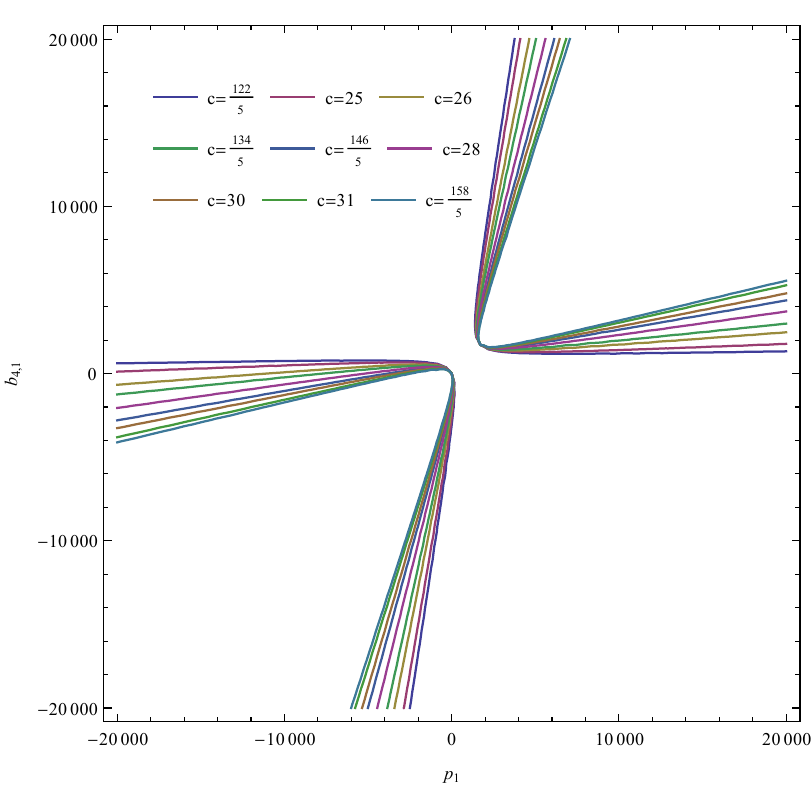}
	\end{center}
	\caption{All hyperbolas in the $b_{4,1}$ vs $p_1$ plane corresponding to $(2,6)$ solutions with central charges: $24 < c^{(\ell=6)} < 32$.}\label{l6_all}
\end{figure}
As explained in the $c=25$ example, for each hyperbola, we have the characters with $N_1>0$ on the lower branches and $N_1<0$ on the upper branches.

Now let us consider $(2,8)$ solutions. The accessory equation now becomes (see \eref{gen_quad_2_6rp2}),
\be
p_1^2+\alpha_0\alpha_1(p_1-b_{4,1})^2-\frac76 p_1(p_1-b_{4,1})-1728p_1+1152(p_1-b_{4,1})=0
\ee

Let us consider the solution with $c=23$. In this case, we have: $N_1 = \frac{m_1-69}{5}$ (see section \ref{wierd28} in appendix: case viii) with $c=23$). For this, the hyperbola is given below.

\begin{figure}[H]
	\begin{center}
		\includegraphics[width=0.7\textwidth]{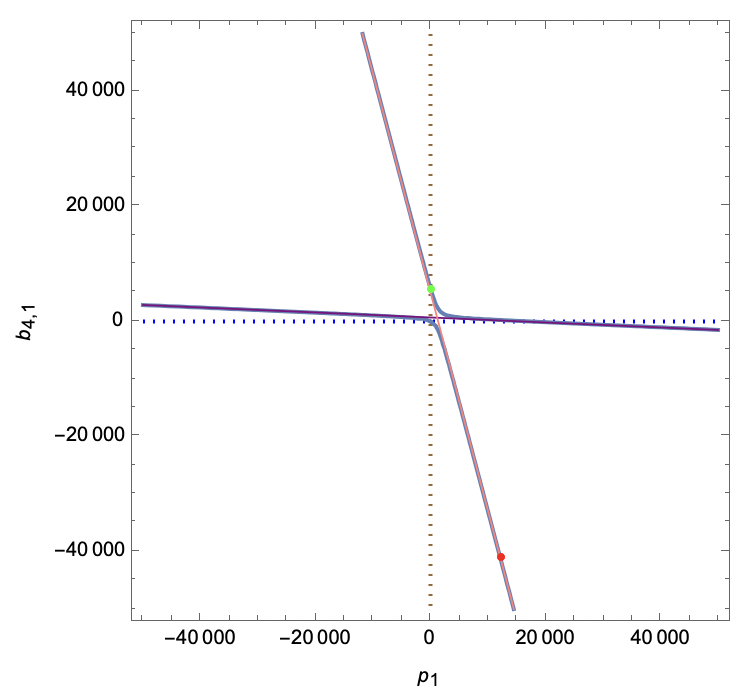}
	\end{center}
	\caption{Hyperbola in the $b_{4,1}$ vs $p_1$ plane corresponding to the $(2,8)$ solution with $c=23$.}\label{c23_l8}
\end{figure}

In this case, from \eref{28a1iii} and \eref{28b41iii}, we get, 
\begin{equation}
\begin{split}
    &b_{4,1}=-\frac{(5N_1-48944)(5N_1+4048)}{345 N_1}, \\
    &p_1=\frac{(5N_1-560)(5N_1+4048)}{15 N_1}.
\end{split}
\label{b41_p1_c23_l8} 
\end{equation}
This gives, $\frac{b_{4,1}}{p_1}=-\frac{5N_1-48944}{23(5N_1-560)}$. The asymptotes to the above hyperbola are: $b_{4,1} = -\frac{16128}{23} - \frac{1}{23}p_1$ (drawn in purple) and $b_{4,1} = \frac{25344}{5} - \frac{19}{5} p_1$ (drawn in pink). 

The red dot on the lower branch is where $m_1=0$, impying $N_1=-\frac{69}{5}$. Starting from this red dot and tracing towards the bottom end of the lower and then continuing from the top end of the upper branch we obtain all the admissible solutions. The green dot on the upper branch corresponds to the point where $m_1=629$ implying $N_1=112$. This in turn implies $p_1=0$ and $b_{4,1}\neq 0$. This is a factorised solution of the form $j^{\frac{2}{3}}\chi^{(2,0)}$ where $\chi^{(2,0)}$ is a $(2,0)$ MMS CFT with $c=7$. 

So, we see that in the $(2,8)$ case, admissible solutions lie on both the branches while in the $(2,6)$ case, admissible solutions appear only in the lower branch.

In the next plot, we have all the hyperbolas corresponding to each of the $(2,8)$ solutions with central charges in the admissibility range: $16 < c^{(\ell=8)} < 24$.

\begin{figure}[H]
	\begin{center}
		\includegraphics[width=0.7\textwidth]{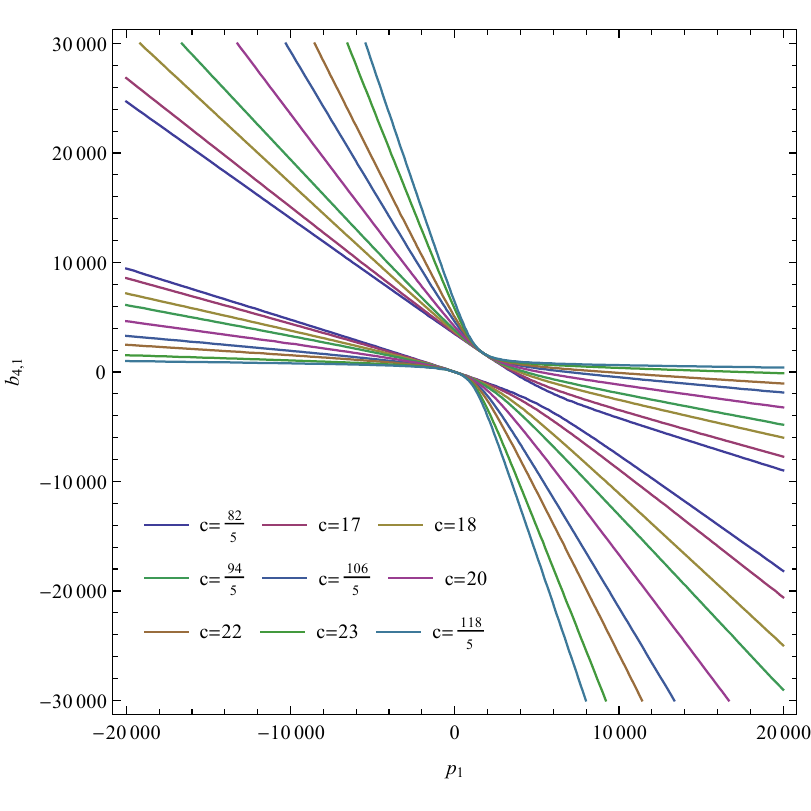}
	\end{center}
	\caption{All hyperbolas in the $b_{4,1}$ vs $p_1$ plane corresponding to $(2,8)$ solutions with central charges: $16 < c^{(\ell=8)} \leq 24$.}\label{l8_all}
\end{figure}

\section{Discussion of the case of two movable poles}

\label{sec212}

We now turn to the case of $\ell=12$. This is important because there are two independent poles $p_1,p_2$ and correspondingly two accessory parameters. A number of novel features will emerge in 
this setting that were not visible for $\ell<12$.

\subsection{The $(2,12)$ MLDE and constraints on accessory parameters}

The $(2, 12)$ MLDE in the $\tau$ plane is given by
\be
\left(D^2 + E_4^2 E_6\left(\frac{1}{E_4^3 - p_1\Delta} + \frac{1}{E_4^3 - p_2\Delta}\right)D + \frac{\alpha_0\alpha_1\,E_4\,(E_4^3 - b_{4, 1}\, \Delta)(E_4^3 - b_{4, 2}\, \Delta)}{(E_4^3 - p_1\, \Delta)(E_4^3 - p_2\, \Delta)}\right) \chi(\tau) = 0 \label{212MLDEt}
\ee
In the $j$-coordinate, the same MLDE is given by:
\be 
\begin{aligned}
\left(\partial_j^2 + \left(\frac{2}{3j} + \frac{1}{2(j-1728)} - \frac{1}{(j-p_1)}-\frac{1}{(j-p_2)}\right)\partial_j + \frac{\alpha_0\alpha_1(j-b_{4, 1})(j-b_{4, 2})}{j(j-1728)(j-p_1)(j-p_2)}\right) \chi(j) = 0 
\label{MLDE212j} 
\end{aligned}
\ee  
The accessory equations for $\ell=12$ can be read off from \eref{gen_quad_26r_0} and after some rationalisation of denominators they reduce to:
\be
\begin{split}
 576 -\frac{5\, p_1}{6} + \frac{\alpha _0\, \alpha _1 \left(p_1-b_{4,1}\right)
   \left(p_1-b_{4,2}\right)}{p_1-p_2}-p_1\left(p_1-1728\right)  
   \left(\frac{2}{p_1-p_2} - \frac{1}{p_1 - b_{4,1} }-\frac{1}{p_1 - b_{4,2} } \right) &= 0\\ 
576 -\frac{5\, p_2}{6}+ \frac{\alpha _0\, \alpha _1 \left(p_2 - b_{4,1}\right)
   \left(p_2-b_{4,2}\right)}{p_2-p_1}-p_2 \left(p_2-1728\right)  
   \left(\frac{2}{p_2-p_1} - \frac{1}{p_2 - b_{4,1} } - \frac{1}{p_2 - b_{4,2} } \right) &= 0\label{eq6} \\
\end{split}
\ee

When one examines the coefficient functions of the MLDEs, both in the $\tau$-space and the $j$-space, one finds that the poles and the accessory parameters appear in symmetric combinations.  Hence, we work with the symmetric parameters:
\be
\begin{split}
P_1 \equiv p_1 + p_2, &\quad P_2 = p_1 p_2\\
B_1 \equiv b_{4,1} + b_{4,2}, &\quad B_2 = b_{4,1} b_{4,2}
\end{split}
\label{symm_212}
\ee
More generally, for $\ell = 6r$, we would have $P_k, B_k$ with $k = 1, 2, \cdots r$, where $P_k$ denotes the $k$th symmetric polynomial in the movable poles and $B_k$ denotes the $k$th symmetric polynomial in the accessory parameters. We will see below that these symmetric parameters always turn out to be rational for admissible character solutions, while the individual poles and accessory parameters need not be. 

Now, the sum and difference of the two equations in \eref{eq6} can be written in terms of the symmetric parameters :
\begin{align}
&\frac{(P_1 - B_1)(2P_2 + 2B_2 - P_1B_1)}{P_2^2 + P_2(B_1^2 - P_1B_1 - 2B_2)+ B_2(P_1^2 - P_1B_1 + B_2)} \nn\\
&\qquad\qquad+ \alpha_0 \alpha_1 \left(\frac{1728B_2 + B_1P_2 - 1728 P_2 - B_2P_1}{P_2(P_2 - 1728P_1 + 1728^2)}\right) \nonumber\\
&\qquad\qquad + \left(\frac{576 (P_1^2 - 2P_2) - \frac{5}{6}P_1P_2 - 1728\cdot576P_1 + 2880P_2}{P_2(P_2 - 1728P_1 + 1728^2)}\right) = 0  \nonumber \\[5mm]
&\alpha_0 \alpha_1 \left(\frac{2P_2^2 + 2\cdot1728P_2B_1 + B_2P_1^2 - 1728P_1(P_2 + B_2) - P_2(P_1B_1 + 2B_2)}{P_2(P_2 - 1728P_1 + 1728^2)}\right) \nonumber\\
&\qquad\qquad + (P_1^2 - 4P_2)\left(\frac{2P_2 - P_1B_1 + B_1^2 - 2B_2}{P_2B_1(P_1 - B_1) + 2P_2B_2 - P_2^2 - B_2(P_1^2 - P_1B_1 + B_2)}\right) \nonumber\\
&\qquad \qquad + (P_1^2 - 4P_2)\left(\frac{576\cdot1728 - 576P_1 + \frac{5}{6}P_2}{P_2(P_2 - 1728P_1 - 1728^2)}\right) = 4 \label{eq6a}
\end{align}
The accessory equations for the more general case for $\ell = 6r$ (\eref{gen_quad_2_6r}) can also similarly be recast in terms of the $P_k$’s and the $B_k$’s. Although these look more complicated than \eref{eq6}, we will soon see that the $P_k$ and $B_k$ are real while the same does not hold for the $p_I,b_{4,I}$.

Now, we will discuss in general terms how one solves the $(2,12)$ MLDE. We start out as we did for the $(2,6)$ MLDE. The first step is to obtain the first few orders of the Frobenius solution for the identity character. Here we have four parameters and hence we need  four orders beyond the indicial equation. Thus there are five equations, the analogues of \eref{ord0a}-\eref{ord2a}: the first one is  simply $\alpha_0\alpha_1 = \frac{c(44-c)}{576}$ and four others for the Fourier coefficients of the identity character $m_1, m_2, m_3, m_4$ (to be consistent with our earlier notation these should have a superscript $(12)$ to denote the $\ell=12$ case, but we drop it to simplify the notation). These are four linear equations for the parameters $P_1, P_2, B_1, B_2$ and we can solve them and obtain the analogues of \eref{nr1a}-\eref{nr2a}. Each of the symmetric parameters is a rational function of $m_1, m_2, m_3, m_4$ with coefficients being rational functions of $c$. Hence we have shown, as promised, that the symmetric parameters are rational. This is the correct generalisation of the observation in \cite{Naculich:1988xv} that a single movable pole is rational.

Thus there are three possibilities for the movable poles: (i) they are both rational, (ii) they are both real and irrational and lie in a quadratic field extension of the rationals, (iii) they are complex conjugates of each other. The accessory parameter follow the same pattern. Notice that the possibility of complex or real irrational poles/accessory parameters occurs for the first time at $\ell= 12$.

For the general case of $\ell = 6r$ the analogs of \eref{ord0a}-\eref{ord2a} would be $2r + 1$ equations $\alpha_0\alpha_1 = \frac{c(4(6r -1)-c)}{576}$ and $2r$ more for the Fourier coefficients $m_1, \ldots m_{2r}$.  Similar to the $\ell = 6$ and $\ell = 12$ cases, we would solve these equations for the $2r$ variables $P_k, B_k$ and solve for each of them as a rational function of $m_1, \ldots m_{2r}$ with coefficients being rational functions of $c$. Again, we can conclude that the $P_k$ and the $B_k$ are rational numbers. Thus, any given movable pole is either complex (and occurs with its conjugate) or is real irrational (and occurs with its Galois conjugates) or is rational, and the same for any accessory parameter.
 
The next step is to bring in the accessory equations \eref{eq6a}. There are two of them and we substitute into them the symmetric parameters in terms of their rational expressions (of $m_1, m_2, m_3, m_4$), leading to two equations that contain only $c$ and the $m_1, m_2, m_3, m_4$. We then expect to solve for $m_3, m_4$ in terms of $m_1, m_2$ and $c$. From the previous example, our expectation is that the dependence will be linear in $m_1$ and $m_2$, leading to an analogue of $\eref{m2mldea}$. We would then solve for higher Fourier coefficients $m_k, k \geq 5$ and expect to obtain a linear dependence in $m_1$ and $m_2$ thus making contact with the quasi-character theory. 
Unfortunately this procedure becomes extremely tedious, so we employ an alternate route below.
 
\subsection{$(2,12)$ admissible characters}

We will use quasi-character theory \cite{Chandra:2018pjq} to obtain $(2,12)$ solutions and make contact with the above analysis. In general terms, quasi-character theory informs us  that $(2,6r)$ admissible character solutions can be found by taking $r+1$ summands, each of which is a $(2,0)$ quasi-character. The summation will contain $r$ quasi-character parameters, which are non-negative and subject to further restrictions.  The precise details and systematics of this procedure has never been worked out for $r \geq 2$, and will be addressed in \cite{DS:upcoming}. Here we will content ourselves with working out one example in full detail.

\subsubsection*{$(2,12)$ solution with $c=25, h=\frac{1}{4}$}

According to quasi-character theory, we can pick three $(2,0)$ quasi-characters in the $A_1$ class and sum them to obtain a $(2,12)$ solution: $c=25, h=\frac{1}{4}$ in the following way :
\bea
\chi = \chi_{n=4}^{A_1} + N_1\, \chi_{n=0}^{A_1} + N_2\, \chi_{n=-4}^{A_1} \label{Quasi212}
\eea
where the $n=4$ and $n=-4$ terms are quasi-characters (integral but not positive) while the $n=0$ term is an admissible character (for the $A_{1,1}$ WZW model).
The leading behaviour in the $q$-series expansion of the identity character corresponds to $c = 25$ and that of the non-identity character gives $h = \frac14$. These numbers ensure that $\ell = 12$. But it is not yet clear that these are admissible characters. For that we examine the $q$-series. For the identity character, we have 
\bea
&\chi_{n=4; 0}^{A_1} + N_1\, \chi_{n=0; 0}^{A_1} + N_2\, \chi_{n=-4; 0}^{A_1} =q^{-\frac{25}{24}}\left(1 + (-245 + N_1) q + (142640 + 3\, N_1 + 26752\, N_2) q^2 \right.\nonumber\\
&\left.+ (18615395 + 4\, N_1 + 1734016\, N_2) q^3 + (837384535 + 7\, N_1 + 46091264\, N_2) q^4 + \ldots\right)  \label{Id}
\eea
and for the non-identity character, we have
\bea
&\chi_{n=4; 1}^{A_1} + N_1\, \chi_{n=0; 1}^{A_1} + N_2\, \chi_{n=-4; 1}^{A_1} = q^{-\frac{19}{24}} \left(N_2 + (2\, N_1 - 247\, N_2) q + (565760 + 2\, N_1 - 86241\, N_2) q^2 \right. \nonumber\\
&\left.+ (51745280 + 6\, N_1 - 4182736\, N_2) q^3 + (1965207040 + 8\, N_1 - 96220123\, N_2) q^4 + \ldots\right) \label{NonId}
\eea
Requiring admissibility of the above $q$-series up to order $q^5$, we find the following restrictions on the quasi-character parameters $N_1$ and $N_2$.
\bea
N_1 = 245 + m_1,\, N_2 \leq \frac{490 + 2m_1}{247},\quad m_1\in \mathbb{Z}^{\geq 0} \label{admcon} 
\eea 
(we have denoted the integer by $m_1$ anticipating that it will be the degeneracy of the first excited state in the identity character). 

Let us consider the relations given in \eref{admcon}. The first expression relates the quasi-character parameter $N_1$ to $m_1$ which is the dimension of the Kac-Moody algebra (if any) of the final theory. On the other hand, the second relation serves as a restriction on $N_2$ for any fixed $N_1$. This restriction will be modified at every order in $q$. So, to ascertain admissibility of $\chi$ in \eref{Quasi212}, we need to look at the asymptotic growth of the coefficients in the $q$-series of this character. This is done by considering the Rademacher expansion (see \cite{Cheng:2012qc} and appendix A of \cite{Chandra:2018pjq}). This assures us that the asymptotic growth of the negative-type quasi-character, $\chi_{n=-4;1}^{A_1}$ is sub-leading as compared to that of the positive-type quasi-characters $\chi_{n=4;1}^{A_1}$ and $\chi_{n=0;1}^{A_1}$. Thus, 
\eref{Quasi212} will be an admissible character for all $N_1$ satisfying \eref{admcon}, i.e. $N_1\ge 245$, and finitely many $N_2$ satisfying some upper bound (not necessarily the one in \eref{admcon}.

Quasi-character theory claims that the two non-negative integral $q$-series above with $c=25, h=\frac{1}{4}$ are in fact $(2,12)$ admissible characters. We will check this claim by showing that they solve the $(2,12)$ MLDE. In particular, first we will compute the point in parameter space where the solution \eref{Quasi212} lives, in other words, we will determine the poles and accessory parameters as functions of the quasi-character parameters $N_1$ and $N_2$. Then we will show that this point satisfies the accessory equations \eref{eq6} -- \eref{eq6a}.

For this, we substitute \eref{Quasi212} into the MLDE \eref{212MLDEt}. The two-derivative terms simplify on using the fact that the summands in \eref{Quasi212} are solutions to the $(2, 0)$ MLDE:
\bea \label{20MLDE3}
 D^2 \chi_{n=4}^{A_1} &=& \frac{725}{576}\, E_4\,\chi_{n=4}^{A_1}, \nonumber \\      D^2 \chi_{n=0}^{A_1} &=& \frac{5}{576}\, E_4\,  \chi_{n=0}^{A_1}, \quad  \nonumber \\  D^2 \chi_{n=-4}^{A_1} &=& \frac{437}{576}\, E_4\, \chi_{n=-4}^{A_1}.
\eea
The $q$-series of the MLDE, at the leading order for both the identity and non-identity character, determines the rigid-parameter in \eref{Quasi212} to be the expected $\alpha_0\alpha_1 = \frac{475}{576}$. To obtain the poles and accessory parameters  we consider the first and second order terms in the $q$-series for the identity and non-identity characters. This gives us four equations for the four variables (the two poles and the two accessory parameters). For the identity character this results in:
\bea
898560 + 432\, N_1 - 125\, P_1 - 475\, B_1  = 0\label{Id1}
\eea
and 
\bea
&& -155450880 + 51984\, N_1 + 53932032\, N_2 + (572905 +451 N_1) \, P_1 + (469775 - 475 N_1)\, B_1   \nonumber\\
&& + 725\, P_2     - 125\, P_1^2 = 0\nonumber \\
\label{Id2}
\eea
For the non-identity character we find instead:
\bea
 1440 N_1 + 612864 N_2 + 19 N_2\, P_1 - 475 N_2\,B_1 = 0\label{NonId1}
\eea
and
\bea
&&  1466449920 - 468576 N_1 - 499484160 N_2 + (1190 N_1 + 287603 N_2)\,P_1 +(- 950 N_1 + 470725 N_2)\,B_1
 \nonumber\\
 && + 437 N_2\, P_2 + 475 N_2\, B_2 + 19 N_2\,P_1^2 - 475 N_2\,P_1\,B_1  = 0 \label{NonId2}
\eea
Solving  \eref{Id1}, \eref{Id2}, \eref{NonId1}, \eref{NonId2}, we obtain solutions for the symmetric parameters in terms of the quasi-character parameters:
\bea \label{Sol4S}
P_1  &=&  \frac{1984 N_2 + 3 N_1\,N_2 - 10 N_1}{N_2}, \quad  \quad P_2 = \frac{8  (816-N_1+88 N_2) (3120- N_1 - 1064 N_2)}{N_2} \nonumber \\
B_1 &=& \frac{650560\, N_2 + 57 N_1\, N_2 + 1250 N_1 }{475\, N_2} \quad  B_2 = \frac{8 (N_1 + 1064 N_2 - 3120) (-5 N_1 + 38456 N_2 + 591600)}{475 N_2} \nonumber \\
\eea
We see right away that the symmetric parameters are all rational. 

The equations \eref{Sol4} give the point in the MLDE parameter space where the quasi-character sum \eref{Quasi212} lives. Now a necessary condition for this to be an admissible character is that this point solves the accessory equations  \eref{eq6}. We have checked that this is indeed the case. Now, we know the exact $(2,12)$ MLDE that the quasi-character sum \eref{Quasi212} is expected to solve, namely \eref{212MLDEt} with the parameters given by \eref{Sol4} and $\alpha_0\alpha_1 = \frac{475}{576} $. We then just substitute the $q$-series expansions \eref{Id} and \eref{NonId} and verify. We have done so for high-enough order to convince us that the quasi-character sum \eref{Quasi212} is indeed a $(2,12)$ admissible character for all values of $N_1,N_2$ satisfying \eref{admcon}.

Now we can solve for the poles and accessory parameters in terms of the symmetric parameters, to find:
\bea
p_1 &=& \frac{1984\, N_2 + N_1\, (\, 3\, N_2 - 10\,) \pm \sqrt{A}}{2\, N_2}, \qquad p_2 = \frac{1984\, N_2 + N_1\, (\, 3\, N_2 - 10\,) \mp \sqrt{A}}{2\, N_2} \nonumber \\
b_{4, 1} &=& \frac{650560\, N_2 + N_1\, (  57\, N_2 + 1250) \pm \sqrt{B}}{950\, N_2},    b_{4, 2} = \frac{650560\, N_2 + N_1\, (  57\, N_2 + 1250) \mp \sqrt{B}}{950\, N_2} \label{Sol4} \nonumber \\ 
\eea
where 
\bea
A &=& (N_2 - 2)\left(4096\, N_1\, N_2 + N_1^2 (\,9\, N_2 - 50\,) + 512\, N_2\, (\,1463\, N_2 + 19890)\right) \nonumber \\
B &=& (19\, N_2 + 250)\left(-2723840\, N_1\, N_2 + N_1^2\, (\,171\, N_2 + 6250) - 243200\, N_2\, (\, 33649\, N_2 -115362)\right). \nonumber \\
\eea
This exemplifies our claim about the nature of movable poles and accessory parameters. When $A  > 0$ and is a perfect square, both the movable poles are rational. When $A > 0$ and not a perfect square the poles are both real and irrational and lie in the field extension $\mathbf{Q}[\sqrt{A}]$. When $A < 0$, the movable poles are complex conjugates. Also when $A = 0$ (which happens for $N_2=2)$ both poles coincide and we will discuss this in the next section. 

\subsubsection*{$(2,12)$ solution with $c=31, h=\frac{3}{4}$}

Following the discussion of the previous subsection, we can add three $(2,0)$ quasi-characters in the $A_1$ class to obtain a $(2,12)$ solution: $c=31, h=\frac{3}{4}$ in the following way :
\bea
\chi = \frac17\left(\chi_{n=5}^{A_1} + N_1\, \chi_{n=1}^{A_1} + N_2\, \chi_{n=-3}^{A_1}\right) \label{Quasi212_2}
\eea
where the $n=5$ and $n=-3$ terms are quasi-characters while the $n=1$ term is an admissible character (for the $E_{7,1}$ WZW model). Now let us consider the leading behaviour in the $q$-series expansions. The identity character corresponds to $c = 31$ and that of the non-identity character gives $h = \frac34$. Using the Riemann-Roch relation \eref{RRformula}, we see that these numbers correspond to $\ell = 12$. However, at this stage, it is not yet clear that the $q$-series expansions are those of admissible characters. For that we examine the $q$-series. For the identity character, we have 
\bea
&\frac17\left(\chi_{n=5; 0}^{A_1} + N_1\, \chi_{n=1; 0}^{A_1} + N_2\, \chi_{n=-3; 0}^{A_1}\right) = \frac{1}{7}q^{-\frac{31}{24}}\left(7 + (N_1-1829) q + (533603 + 133\, N_1 + 39\, N_2) q^2 \right.\nonumber\\
&\left.+ (309815674 + 1673\, N_1 + 1547\, N_2) q^3 + \ldots\right)  \label{Id299}
\eea
and for the non-identity character, we have
\bea
&\frac{1}{7}\left(\chi_{n=5; 1}^{A_1} + N_1\, \chi_{n=1; 1}^{A_1} + N_2\, \chi_{n=-3; 1}^{A_1}\right) = \frac{1}{7}q^{-\frac{13}{24}} \left(N_2 + (56\, N_1 - 377\, N_2) q + (40641 + 968\, N_1 - 22126\, N_2) q^2 \right. \nonumber\\
&\left.+ (4836279 + 7504\, N_1 - 422123\, N_2) q^3 + \ldots\right) \label{NonId299}
\eea
Requiring admissibility of the above $q$-series up to order $q^5$,
we find the following relations:
\bea
N_1 = 7\,m_1 + 1829,\quad N_2 \leq \frac{7678672813 + 1363376 m_1}{41490618},\quad m_1\in \mathbb{Z}^{\geq 0} \label{admcon2} 
\eea 
(we have denoted the integer by $m_1$ anticipating that it will be the degeneracy of the first excited state in the identity character).

Proceeding as before, let us note that, the first of the relations in \eref{admcon2} relates the quasi-character parameter $N_1$ to the dimension $m_1$ of the Kac-Moody algebra (if any) of the final theory. However the second relation should be viewed as a restriction on $N_2$ for any fixed $N_1$. This restriction will be modified at higher orders in $q$. As in the previous sub-section, the Rademacher expansion (see \cite{Cheng:2012qc} and appendix A of \cite{Chandra:2018pjq}) assures us that the asymptotic growth of the negative-type quasi-character, $\chi_{n=-3;1}^{A_1}$ is sub-leading as compared to that of the positive-type quasi-characters $\chi_{n=5;1}^{A_1}$ and $\chi_{n=1;1}^{A_1}$. Thus, \eref{Quasi212_2} will be an admissible character for all $N_1$, that is, $N_1\geq 1829$ satisfying \eref{admcon2} and finitely many $N_2$ satisfying some upper bound.

Now proceeding as before, we can express the symmetric parameters of the MLDE, namely, $P_1, P_2, B_1, B_2$ in terms of the quasi-character parameters only. However, in this case these expressions are quite lengthy and hence we do not report them here.

\section{Beyond the genericity assumption: merging of poles}

\label{mergepoles}

In this section we consider what happens when poles in the original MLDE merge. This first  happens when $p_1\to 0$ or 1728 in the $(2,6)$ case. Once we reach the value $\ell=12$, we can also have two movable poles $p_1,p_2$ merging. We will analyse some special cases below and then draw general conclusions at the end.

\subsection{$(2,6)$ solutions as $p_1\to 0$}

\label{merge26}

We want to investigate what happens to the $(2,6)$ solutions if we set $p_1=0$, corresponding to the point $\tau=\rho$. This is our first example of a case violating the genericity assumption. Let us take this limit directly in \eref{twosixMLDE}. It becomes: 
\be
\del_j^2\chi(j)+\bigg(\frac{1}{2(j-1728)}-\frac{1}{3j}\bigg)\del_j\chi(j)+\frac{\alpha_0\alpha_1(j-b_{4,1})}{j^2(j-1728)}\chi(j)=0. \label{26_p10_mlde}
\ee
Now we insert the expansion:
\be
\chi_i(j)=j^{\alpha_i^{(\rho)}}\sum_{k=0}^\infty a_{i,k}^{(\rho)}\,j^k
\ee
We find the indicial equation:
\be
\alpha_i^{(\rho)}(\alpha_i^{(\rho)}-1)-\frac13\alpharho_i
+\frac{\alpha_0\alpha_1 b_{4,1}}{1728}=0
\label{indicial26p0}
\ee
Since the last term in general contributes to the indicial equation, we do not immediately get the values of the exponents. Instead the equation tells us that \be
\alpha_0^{(\rho)}+\alpha_1^{(\rho)}=\frac43
\ee
This in fact already follows from \eref{indi_r}, since after taking $p_1\to 0$ we have $\ell_\rho=6$. Since the $\alpharho$ must be distinct non-negative multiples of $\frac13$, the possible solutions to the above equation are $(0,\frac43)$ or $(\frac13,1)$. 

The first choice of exponents lead to either $\alpha_0\alpha_1=0$ or $b_{4,1}=0$. First let us look at the case when $\alpha_0\alpha_1=0$. From \eref{26_p10_mlde} we notice that in this case the non-derivative term (the last term) vanishes. So $\del_j\chi(j)$ solves a first order MLDE. We can now integrate this to get a first-order MLDE for $\chi(j)$ itself. This means the solution space becomes $1$-dimensional and thereby we rule this case out \footnote{Note that this argument is independent of the exact form of the MLDE or even its order, hence it can be readily generalised to an $n^{th}$ order MLDE where it implies that the solution space is $(n-1)$-dimensional when $\alpha_0\alpha_1=0$.}.

The other choice, $b_{4,1}=0$, is a possibility and in this case indeed we have the exponents $(0,\frac43)$. 
As noted at the end of sub-section \ref{MLDEjspace}, when the lower of the two exponents is 0, it means we cannot extract some positive power of $j$ from the solution and still get a sensible expansion in powers of $j$. Thus such solutions, if they exist, would be non-factorisable.

In fact we have already shown that they do exist. In Section \ref{quasiMLDEcomp}, we have listed the values of $p_1$ and $b_{4,1}$ for all admissible solutions of the $(2,6)$ MLDE as functions of the degeneracy $m_1$  of the first excited state in the identity character. In Eqs. (\ref{26a1}, \ref{26b41}) we see that there is a common value $m_1=3538$ such that $p_1$ and $b_{4,1}$ both vanish keeping their ratio fixed. This is the unique solution with indices $(\alpharho_0,\alpharho_1)= \left(0,\frac43\right)$ within this family. A similar situation holds for each one of the subsequent pairs, Eqs (\ref{26a1ii}, \ref{26b41ii}), (\ref{26a1iii}, \ref{26b41iii}), (\ref{26a1iv}, \ref{26b41iv}), (\ref{26a1v}, \ref{26b41v}), (\ref{26a1vi}, \ref{26b41vi}), (\ref{26a1vii}, \ref{26b41vii}), (\ref{26a1viii}, \ref{26b41viii}), (\ref{26a1ix}, \ref{26b41ix}) -- for each case, there is a unique value of $m_1$ that makes both $p_1$ and $b_{4,1}$ vanish together. This, then, is the full list of admissible solutions with indices 
$\left(0,\frac43\right)$. 

The other alternative is that the exponents are $(\frac13,1)$. These arise in the same solutions listed in the previous paragraph, by choosing $m_1$ to be the value that makes $p_1$ vanish but $b_{4,1}\ne 0$. For example in Eqs (\ref{26a1}, \ref{26b41}) this value is 658. Each of the other cases is similar.

Inserting this in \eref{indicial26p0} leads to the constraint:
\be
b_{4,1}=\frac{576}{\alpha_0\alpha_1}
\ee
The reader may verify that this equation agrees with the values obtained as in the previous paragraph, by making $p_1$ vanish with $b_{4,1}\ne 0$ in each of our explicit solutions.

Now we come to a key point. Since the lower of the two exponents has shifted from 0 (when $p_1$ was a generic point) to $\frac13$ (after $p_1$ goes to 0) we can make the change of variable:
\be
\chi_i(j)=j^\frac13 \zeta_i(j)
\label{j13change}
\ee
where the function $\zeta(j)$ has a sensible expansion in power of $j$. Indeed, we find that $\zeta$ satisfies the $(2,2)$ MLDE, the middle equation of \eref{twocharj} with $r=0$. 
Recall that the parameters in that equation are $\alpha_0,\alpha_1$, the exponents around $\tau\to\infty$ (not to be confused with the $\alpharho_i$ above!). We find the relation:
\be
(\alpha_0\alpha_1)^{\ell=6}=(\alpha_0\alpha_1)^{\ell=2}+\frac16
\label{alphaprod13}
\ee
In fact from \eref{j13change} we already know the exponents of the solution $\zeta(j)$ must be:
\be
\alpha_i^{\ell=2}=\alpha_i^{\ell=6}+\frac13
\ee
and using the valence formula on both sides, it is easy to check that \eref{alphaprod13} agrees with this.

In this case we can come to the same conclusion by solving the accessory equation \eref{b41condition} as $p_1\to 0$. One solution is $b_{4,1}=0$ and the other is:
\be
b_{4,1}=\frac{576}{\alpha_0\alpha_1}
\ee
Inserting this into the MLDE in terms of $\tau$:
\bea
\left(D^2  + \frac{1}{3}\frac{E_6}{E_4}D  + \frac{(\alpha_0\alpha_1 - \frac{1}{6} )E_4^3 + (576 - \alpha_0\alpha_1\, b_{4, 1})\Delta}{E_4^2}\right)\zeta(\tau) = 0 \label{2622MLDEt}
\eea
and performing the change of dependent variable:
\be
\chi(\tau) = j^{\frac{1}{3}}\, \zeta(\tau)
\ee
we get the $(2, 2)$ MLDE:
\bea
\left(D^2  + \frac{1}{3}\frac{E_6}{E_4}D  + \left(\alpha_0\alpha_1-\frac16\right) E_4    \right)\zeta(\tau) = 0  
\eea 
In this equation we see the relation $(\alpha_0\alpha_1)^{\ell=2} = (\alpha_0\alpha_1)^{\ell=6} - \frac{1}{6}$. 

Thus we learn that, in the cases where the lower of the critical exponents is nonzero, sending the movable pole to the point $p_1=0$ causes the solution to factorise into a product of solutions of an MLDE with lower value of $\ell$ (in this case a pair of characters with $\ell=2$) times a single meromorphic character $j^\frac13$ which also has $\ell=2$. A priori this may not seem like a ``merger'' of poles since there was no pole at $p_1=0$ to begin with. However it does count as a merger because a single pole at $\tau=\rho$ is three times the minimum allowed pole at that point.

As we will see later, the reason we could simply take the limit in the accessory equation like \eref{b41condition} is that this limit does not create any new constraint, which in turn is because the new exponents do not differ from each other by an integer. Below we will see examples where merging of poles leads to new exponents that differ by an integer and consequently a novel constraint equation arises. In such cases, merging the poles in the original constraint equation can give incorrect results. Instead one has to start afresh from the MLDE where the poles have merged.

Let us consider the relation \eref{alphaprod13}, valid for factorised characters of the form $\chi^{(\ell=6)}=j^{\frac{1}{3}}\chi^{(\ell=2)}$. Next we  use the $\ell=6$ and $\ell=2$ valence formula \eref{RRformula} on the left and right of this equation respectively. Replacing everything in terms of the central charges, we get the following possibilities:
\begin{align}
    c^{(\ell=6)} = 12 - c^{(\ell=2)}, \qquad  \qquad c^{(\ell=6)} = c^{(\ell=2)} + 8 \label{c2c6_relns}
\end{align}
We know from \cite{Hampapura:2015cea} that admissible solutions for the $(2,2)$ case lie in the range $16 < c^{(\ell=2)} < 24$. Thus, admissibility of $(2,6)$ solutions rules out the first case in \eref{c2c6_relns}. Then from the second equation above, we have $24 < c^{(\ell=6)} < 32$. 
We already know the allowed central charges for admissible $(2,6)$ solutions from Sec. \ref{cent_relns}. Thus we see that tensor-product $(2,6)$ CFTs follow the exact same range and not a subset of it. We will encounter examples later where the product theories occupy a smaller range of central charges.

We will now look at some more examples that present different features, and then turn to the general case.

\subsection{$(2,6)$ solutions as $p_1\to 1728$}

Let us now investigate what happens to the $(2,6)$ solutions if we set $p_1=1728$. This corresponds to the point $\tau=i$ in the upper half plane. Taking this limit in \eref{twosixMLDE} we get: 
\be
\del_j^2\chi(j)+\bigg(\frac{2}{3j}-\frac{1}{2(j-1728)}\bigg)\del_j\chi(j)+\frac{\alpha_0\alpha_1(j-b_{4,1})}{j(j-1728)^2}\chi(j)=0. \label{p1_1728_26}
\ee
Let us now insert the following expansion in the MLDE \eref{p1_1728_26}:
\be
\chi_l(j)=(j-1728)^{\alpha_l^{(i)}}\sum_{k=0}^\infty a_{l,k}^{(i)}\,j^k
\ee
We find the indicial equation to be,
\be
\alpha_l^{(i)}(\alpha_l^{(i)}-1)-\frac12\alpha_l^{(i)}
+\alpha_0\alpha_1\left(1 - \frac{b_{4,1}}{1728} \right)=0
\label{indicial26p1728}
\ee
The indicial equation tells us that, 
\be
\alpha_0^{(i)}+\alpha_1^{(i)}=\frac32
\ee
One could have already deduced this fact from \eref{ind_i}, since after taking $p_1\to 1728$ we have $\ell_i=6$. Now the $\alpha^{(i)}$s must be distinct non-negative multiples of $\frac12$. Hence, the possible solutions to the above equation are $(0,\frac32)$ or $(\frac12,1)$. 

The exponents $(0,\frac32)$ correspond to either $\alpha_0\alpha_1=0$ or $b_{4,1}=1728$. The former can be ruled out since in this case the solution space is $1$-dimensional, as argued in the previous sub-section. However $b_{4,1}=0$ is a possibility and in this case indeed we have the exponents $(0,\frac32)$. Solutions, with these exponents, if they exist, would be non-factorisable. Once can see this by following similar arguments of regularity of solution around $\tau=i$ as described in the previous sub-section. 

The other alternative is that the exponents are $(\frac12,1)$. Inserting this in \eref{indicial26p1728} leads to the constraint:
\be
b_{4,1}=\frac{864\left(2\alpha_0\alpha_1-1\right)}{\alpha_0\alpha_1}
\ee
As noted in the previous sub-section, since the lower of the two exponents has shifted from 0 (when $p_1$ was a generic point) to $\frac12$ (after $p_1$ goes to 1728) we can make the following change of variable:
\be
\chi(j) = (j-1728)^\frac12 \zeta(j)
\label{j12change}
\ee
where the function $\zeta(j)$ is regular around $\tau=i$. Furthermore, we find that $\zeta$ satisfies the $(2,0)$ MLDE and this yields the following relation,
\be
(\alpha_0\alpha_1)^{\ell=6}=(\alpha_0\alpha_1)^{\ell=0}+\frac16
\label{alphaprod32}
\ee
In fact from \eref{j12change} we already know the exponents of the solution $\zeta(j)$ must be:
\be
\alpha_l^{\ell=0}=\alpha_l^{\ell=6}+\frac12
\ee
and using the valence formula on both sides, it is easy to check that \eref{alphaprod32} agrees with this.

Thus, as seen before in the previous sub-section, in the cases where the lower of the critical exponents is nonzero, sending the movable pole to the point $p_1=1728$ causes the solution to factorise into a product of solutions of an MLDE with lower value of $\ell$ (in this case a pair of characters with $\ell=0$) times $(j-1728)^\frac12$ which is a solution to the first order MLDE with $\ell=3$. 

In this case also, we can come to the same conclusions as above by solving the accessory equation \eref{b41condition} as $p_1\to 1728$. This is because, as before, taking $p_1\to 1728$ in \eref{b41condition} doesn't create any new constraint since the new exponents about $\tau=i$ do not differ by an integer.

Now we make a comment about the factorised case: $\chi^{(\ell=6)}=(j-1728)^{\frac{1}{2}}\chi^{(\ell=0)}$, analogous to the discussion following \eref{c2c6_relns}. Using the valence formula in \eref{alphaprod32} and replacing every exponent in terms of central charges we get:
\begin{align}
    c^{(\ell=6)} = 8 - c^{(\ell=0)}, \qquad
    c^{(\ell=6)} = 12 + c^{(\ell=0)} \label{c0c6_relns_in}
\end{align}
Each of these conditions, together with the known range for admissible $c^{(\ell=0)}$ solutions, implies that $c^{(\ell=6)} < 20$ for such a factorised solution. However admissible $(2,6)$ solutions have $c^{(\ell=6)}>24$, therefore there are none with this factorised form.

\subsection{$(2,12)$ solutions as $p_1\to p_2$}

Next we consider the case where two movable poles coalesce but remain away the points $0$ and $1728$. This possibility arises for the first time in the $(2,12)$ MLDE \eref{MLDE212j}. So we put $p_2=p_1$ in this equation, to get:
\begin{align}
    \partial_j^2\chi + \left(\frac{1}{2(j-1728)}+\frac{2}{3j}-\frac{2}{j-p_1}\right)\partial_j\chi + \frac{\alpha_0\alpha_1}{j(j-1728)}\frac{(j-b_{4,1})(j-b_{4,2})}{(j-p_1)^2} = 0. \label{212_p1eqp2}
\end{align}
Now we expand the characters about $j=p_1$ as in \eref{Iexp}:
\begin{align}
    \chi_i(j) = (j-p_1)^{\alpha^{(1)}_i}\sum\limits_{k=0}^{\infty}a_{i,k}^{(1)}(j-p_1)^k, \label{coeff_exp212_p1ep2}
\end{align}
Due to the double pole, the last term in \eref{212_p1eqp2} contributes to the indicial equation,
which becomes:
\be
\alpha_i^{(1)}(\alpha_i^{(1)}-3)+\frac{\alpha_0\alpha_1(p_1-b_{4,1})(p_1-b_{4,2})}{p_1(p_1-1728)}=0
\label{indi03}
\ee
Again we cannot read off the exponents directly, but from the above equation we have:
\be
 \alpha_0^{(1)}+\alpha_1^{(1)}=3, \label{sum_sigma}
\ee
Since $p_1$ is a regular point in moduli space, $\alpha_0^{(1)},\alpha_1^{(1)}$ must be distinct non-negative integers, so the only possibilities are $(0,3)$ and $(1,2)$. The former leads to either $\alpha_0\alpha_1=0$ or $p_1=b_{4,1}$ or $p_1=b_{4,2}$. 

This in fact already follows from \eref{ind_genr}, since after taking $p_2\to p_1$ we have $\ell_\tau=12$. Since the $\alpha^{(I)}$ must be distinct non-negative integers, the possible solutions to the above equation are $(0,3)$ or $(1,2)$. 

With $\alpha_0\alpha_1=0$, as noted before, we get the solution space to be $1$-dimensional. Thus, we can only have the exponents $(0,3)$ if $p_1=p_2=b_{4,1}$ or $p_1=p_2=b_{4,2}$. At such points, one factor cancels from the numerator and denominator of the $(2,12)$ MLDE. Thus we get the equation:
\be
\del_j^2\chi(j)+\bigg(\frac{1}{2(j-1728)}+\frac{2}{3j}-\frac{2}{j-p_1}\bigg)\del_j\chi(j)+\frac{\alpha_0\alpha_1(j-b_{4,2})}{j(j-1728)(j-p_1)}\chi(j)=0
\ee
In this case there is a constraint equation at third order, which is left as an exercise for the reader.

Now we return to the other possible set of exponents, namely $(\alpha_0^{(1)},\alpha_1^{(1)})=(1,2)$. From \eref{indi03} we then immediately find the condition:
\be
\frac{\alpha_0\alpha_1(p_1-b_{4,1})(p_1-b_{4,2})}{p_1(p_1-1728)}=2
\label{indi_corr_212}
\ee

Because the indices now differ by 1 (rather than 2 in the generic case), there is a potential logarithmic singularity in the character $\chi_0(j)$ manifested by a constraint arising at first order beyond the indicial equation (as against second order in the generic case). The mechanism has been discussed before -- at this order the coefficient $a_{0,1}$ will not appear and instead we will get a constraint. 

From the MLDE \eref{212_p1eqp2}, this constraint is found to be:
\begin{align}
    \frac{p_1}{2} + \frac{2(p_1-1728)}{3} - 2(2p_1-1728) + \alpha_0\alpha_1(2p_1-B_1) = 0,\label{constr1_212_p1ep2}
\end{align}
which is written in terms of the symmetric polynomial basis (see \eref{symm_212}).

In the symmetric polynomial basis, we can write \eref{indi_corr_212} as:
\begin{align}
    \frac{\alpha_0\alpha_1}{p_1(p_1-1728)}(p_1^2-B_1p_1+B_2) = 2, \label{constr2_212_p1ep2}
\end{align}
Now we can solve for $B_1$ and $B_2$ using \eref{constr1_212_p1ep2} and \eref{constr2_212_p1ep2} to get:
\begin{align}
    &B_1 = \frac{13824 - 17p_1 + 12\alpha_0\alpha_1 \, p_1}{6\alpha_0\alpha_1}, \label{R_212} \\
    &B_2 = \frac{p_1(6\alpha_0\alpha_1 \, p_1 - 5p_1 - 6912)}{6\alpha_0\alpha_1}, \label{S_212}
\end{align}

We now show that the solution is factorised, with one factor being a meromorphic character and the other being a solution (not necessarily admissible) of the $(2,0)$ MLDE. Since the lower exponent is 1, we substitute:
\be
\chi(j)=(j-p_1)\zeta(j)
\label{fact212}
\ee
in \eref{212_p1eqp2} to get:
\be
\begin{split}
    (j-p_1)&\Bigg[\,\partial_j^2\zeta + \bigg(\frac{1}{2(j-1728)}+\frac{2}{3j}\bigg)\partial_j\zeta + \bigg(\frac{1}{2(j-1728)(j-p_1)}\\
    & +\frac{2}{3j(j-p_1)}-\frac{2}{j(j-p_1)^2} +\frac{\alpha_0\alpha_1}{j(j-1728)}\frac{j^2 - B_1j + B_2}{(j-p_1)^2}\bigg)\zeta\,\Bigg] = 0. \label{212_p1eqp2_1}
\end{split}
\ee
On inserting the values of $R$ and $S$ from \eref{R_212} and \eref{S_212}, \eref{212_p1eqp2_1} simplifies to:
\be
(j-p_1)\Bigg[\partial_j^2\zeta + \bigg(\frac{1}{2(j-1728)}+\frac{2}{3j}\bigg)\partial_j\zeta
+ \frac{1}{j(j-1728)}\bigg(\alpha_0\alpha_1-\frac{5}{6}\bigg)\zeta\Bigg] = 0 \label{20_212_p1ep2}
\ee
which means $\zeta(j)$ solves the $(2,0)$ MLDE if we identify:
\begin{align}
     (\alpha_0\alpha_1)^{\ell=12} = (\alpha_0\alpha_1)^{\ell=0} + \frac{5}{6}. \label{c0_c12_rel}
\end{align}
The above equation holds for factorised $(2,12)$ solutions of the form: $\chi^{(\ell=12)}=(j-p_1)\chi^{(\ell=0)}$. Now let us comment on tensor-product $(2,12)$ CFTs of the above factorised form. Using the valence formula in \eref{c0_c12_rel} and writing everything in terms of central charges, we get: $c^{(12)}=20-c^{(0)}$ or $c^{(12)}=c^{(0)}+24$. Since the admissible central charge range of $(2,0)$ solutions is: $0 < c^{(\ell=0)} < 8$ and unitarity implies $c^{(12)}>22$, the first possibility gets ruled out. So, we must have: $c^{(12)}=c^{(0)}+24$ implying $24 < c^{(\ell=12)} < 32$, for tensor-product $(2,12)$ CFTs. So, again we find an example where the central charge range for tensor-product solutions lie in a smaller range compared to the full admissible range.

In this sub-section, we have shown that when two movable poles $p_1,p_2$ coincide with each other (but not with an accessory parameter), the solutions of the MLDE factorise into a product of a meromorphic character and a pair of characters $\zeta_i(j)$ satisfying an MLDE with $\ell=0$. As we already discussed in sub-section \ref{MLDEjspace}, this factorisation can means one of two things for an admissible character $\chi_i(j)$: either $\zeta_i(j)$ is itself an admissible character, or $\zeta_i(j)$ is not admissible but becomes admissible upon multiplying by $(j-p_1)$.

To exemplify the above considerations, consider the example of $(2,12)$ characters studied in Section \ref{sec212}, \eref{Quasi212}. The conditions on the parameters $N_1,N_2$ are in \eref{admcon}. Now when $N_2 = 2$, we see that the parameter $A$ in \eref{Sol4} vanishes and the poles $p_1,p_2$ merge. Now the equations \eref{Id} and \eref{NonId} become:
\bea
&&  \chi_{n=4; 0}^{A_1} + N_1\, \chi_{n=0; 0}^{A_1} + 2\, \chi_{n=-4; 0}^{A_1} \nonumber\\
&=& q^{-\frac{25}{24}}\left(1 + (-245 + N_1) q + (196144 + 3\, N_1) q^2 + (22083427 + 4\, N_1) q^3 + (929567063 + 7\, N_1) q^4 + \ldots\right) \label{I31}\nonumber  \\ \\ 
&&\chi_{n=4; 1}^{A_1} + N_1\, \chi_{n=0; 1}^{A_1} + 2\, \chi_{n=-4; 1}^{A_1} \nonumber\\
&=& q^{-\frac{19}{24}} \left(2 + 2 (-247 + N_1) q + 2 (196639 + N_1) q^2 + 
 6 (7229968 + N_1) q^3 + (1772766794 + 8\, N_1) q^4+ \ldots\right)\label{I32}\nonumber \\ 
\eea
It is easily verified that the above two equations \eref{I31}, \eref{I32} are in the form:
\bea
\chi_{n=4 }^{A_1} + N_1\, \chi_{n=0 }^{A_1} + 2\, \chi_{n=-4 }^{A_1} = (j + N_1 - 992)\, \chi_{n=0 }^{A_1}, \label{crucial}
\eea  
thus they are factorised as in \eref{fact212}.

\subsection{Analogous considerations for $(2,8)$ and $(2,14)$ cases}
In this sub-section, we shall first take $p_1\to 0$ and $p_1\to 1728$ in the $(2,8)$ case, and then take $p_2\to p_1$ in the $(2,14)$ case. We shall see that, as a result of this procedure, the equations and expressions that come out will be very similar to the $(2,6)$ and $(2,12)$ cases discussed in the previous sub-sections. So we shall focus on the main results and shall be very brief about the intermediate steps.  

The $(2,8)$ MLDE is the second line of \eref{MLDEjspace} with $r=1$. The indicial equation obtained after taking $p_1\to 0$ in this equation results in the same equation as \eref{indicial26p0} with the second term being $\frac23$ instead of $\frac13$. Thus, now the sum of roots become, $\alpharho_0+\alpharho_1=\frac53$. So the choice of exponents are:
\be
(\alpharho_0,\alpharho_1)= \Big(0,\frac53\Big),~\Big(\frac13,\frac43\Big),~\Big(\frac23,1\Big)
\ee
The first of these cases corresponds to, as before, $b_{4,1}=0$ and non-factorised characters. Regarding the second case, we notice that the difference between the two critical exponents is an integer. This solution is ruled out, as already observed in \cite{Naculich:1988xv}, because the monodromy about $\tau=\rho$ would become reducible. The third case allows us to write $\chi(j)=j^\frac23\zeta(j)$ with $\zeta(j)$ having $\ell=0$, and in this case, $\alpha_0\alpha_1 b_{4,1}=1152$. One can show that $\zeta(j)$ satisfies a $(2,0)$ MLDE and this in turn gives the following relation,
\begin{align}
    (\alpha_0\alpha_1)^{\ell=8} = (\alpha_0\alpha_1)^{(\ell=0)} + \frac{1}{3}. \label{c0c8_rel}
\end{align}
\eref{c0c8_rel} is true for factorised $(2,8)$ solutions of the form: $\chi^{(\ell=8)}=j^{\frac{2}{3}}\chi^{\ell=0}$. Using the valence formula in \eref{c0c8_rel} and expressing everything in terms of central charges, we obtain the following two possibilities,
\begin{align}
    c^{(\ell=8)} = 12 - c^{(\ell=0)}, \qquad c^{(\ell=8)} = 16 + c^{(\ell=0)}
\end{align}
Now unitarity implies, $h^{(\ell=8)}=\frac{c^{(\ell=8)}-14}{12}>0$. Thus, $c^{(\ell=8)}>14$ and this rules out the first possibility. Hence, for such factorised solutions, we have: $c^{(\ell=8)} = 16 + c^{(\ell=0)}$. Since the admissibility range of $(2,0)$ solutions fall in, $0 < c^{(\ell=0)} < 8$, we conclude that tensor-product $(2,8)$ CFTs of the form $\chi^{(\ell=8)}=j^{\frac{2}{3}}\chi^{\ell=0}$ lie in the range: $16 < c^{(\ell=8)} < 24$. Recall that, from \ref{cent_relns}, we had the admissibility range for $(2,8)$ solutions as: $16 < c^{(\ell=8)} < 24$ and $40 < c^{(\ell=8)} < 48$. So, this is an example where the central charge range for tensor-product solutions lie in a smaller range compared to the full admissible range. This is in contrast to the tensor-product $(2,6)$ CFT case.  

Next we consider $p_1\to 1728$ in the $(2,8)$ MLDE. The analysis parallels that of the $(2,6)$ case with very slight modifications. In this case, the indicial equation remains the same and hence the choice of exponents also remain the same: $(0,\frac32)$ or $(\frac12,1)$. The first choice lead to, $b_{4,1}=1728$ and to non-factorised solutions while the second choice leads to solutions of the form: $\chi = (j-1728)^{\frac{1}{2}}\zeta(j)$. $\zeta(j)$ solves the $(2,2)$ MLDE which in turn leads to the following relation,
\begin{align}
    (\alpha_0\alpha_1)^{\ell=8} = (\alpha_0\alpha_1)^{\ell=2} + \frac13, \label{alphapord132822}
\end{align}
Let us make a comment about the factorised case: $\chi^{(\ell=8)}=(j-1728)^{\frac{1}{2}}\chi^{(\ell=2)}$. Using the valence formula in \eref{alphapord132822} and replacing every exponent in terms of central charges we get:
\begin{align}
    c^{(\ell=8)} = 16 - c^{(\ell=2)}, \qquad
    c^{(\ell=8)} = 12 + c^{(\ell=2)}. \label{c0c8_relns_in}
\end{align}
If we consider admissible $(2,8)$ solutions, for which $c^{(\ell=8)}>14$, then the first case is ruled out. The second case implies, $28 < c^{(\ell=8)} < 36$. However, from the central charge range of $(2,8)$ admissible solutions \eref{c_relns}, we know there are no admissible solutions in the above range. hence we conclude that there are no admissible solutions of the above factorised form.

Now we consider coalescing of poles $p_2\to p_1$ in the $(2,14)$ MLDE. This analysis parallels the $p_2\to p_1$ case in the $(2,12)$ MLDE with slight modifications. The $(2,14)$ MLDE obtained after taking $p_2\to p_1$ is similar to \eref{212_p1eqp2} with the only difference being in the second term inside the coefficient of the $\del_j\chi$ term. It is now $\frac{1}{3j}$ instead of $\frac{2}{3j}$. 

The indicial equation obtained from this is the same as in the $(2,12)$ case and hence the choice of exponents remains the same: $(0,3)$ or $(1,2)$. The first choice leads to $p_1=p_2=b_{4,1}$ or $p_1=p_2=b_{4,2}$ and to non-factorised solutions. Considering the second choice of exponents we recover \eref{indi_corr_212}. As explained before, since the indices now differ by 1 we get a constraint equation at first order beyond the indicial equation. Using, the symmetric parameters $B_1$ and $B_2$ (see \eref{symm_212}), this constraint equation becomes,
\begin{align}
    \frac{p_1}{2} + \frac{p_1-1728}{3} - 2(2p_1-1728) + \alpha_0\alpha_1(2p_1-B_1) = 0,\label{constr1_214_p1ep2}
\end{align}
Using the indicial equation \eref{indi_corr_212} and \eref{constr1_214_p1ep2}, we can solve for $B_1$ and $B_2$, as before, in terms of $\alpha_0\alpha_1$ and the movable pole $p_1$. 

Note that, since the lower exponent, at $j=p_1$, is $1$ we can have the following substitution:
\be
\chi(j)=(j-p_1)\zeta(j)
\ee
which we can plug in the $(2,14)$ MLDE (with $p_2\to p_1$) to get,
\be
\begin{split}
    (j-p_1)&\Bigg[\,\partial_j^2\zeta + \bigg(\frac{1}{2(j-1728)}+\frac{1}{3j}\bigg)\partial_j\zeta + \bigg(\frac{1}{2(j-1728)(j-p_1)}\\
    & +\frac{1}{3j(j-p_1)}-\frac{2}{j(j-p_1)^2} +\frac{\alpha_0\alpha_1}{j(j-1728)}\frac{j^2 - B_1j + B_2}{(j-p_1)^2}\bigg)\zeta\,\Bigg] = 0. \label{214_p1eqp2_1}
\end{split}
\ee
On inserting the values of $B_1$ and $B_2$ obtained above, \eref{214_p1eqp2_1} simplifies to:
\be
(j-p_1)\Bigg[\partial_j^2\zeta + \bigg(\frac{1}{2(j-1728)}+\frac{1}{3j}\bigg)\partial_j\zeta
+ \frac{1}{j(j-1728)}\bigg(\alpha_0\alpha_1-\frac{7}{6}\bigg)\zeta\Bigg] = 0 \label{22_214_p1ep2}
\ee
which means $\zeta(j)$ solves the $(2,2)$ MLDE if we identify:
\begin{align}
     (\alpha_0\alpha_1)^{\ell=14} = (\alpha_0\alpha_1)^{\ell=2} + \frac{7}{6}. \label{c2_c14_rel}
\end{align}
The above equation holds for factorised $(2,14)$ solutions of the form: $\chi^{(\ell=14)}=(j-p_1)\chi^{(\ell=2)}$. Now let us make a comment on tensor-product $(2,14)$ CFTs of the above factorised form. Using the valence formula in \eref{c2_c14_rel} and writing everything in terms of central charges, we get: $c^{(14)}=28-c^{(2)}$ or $c^{(14)}=c^{(2)}+24$. Since the admissible central charge range of $(2,2)$ solutions is: $16 < c^{(\ell=2)} < 24$ and unitarity implies $c^{(14)}>26$, the first possibility gets ruled out. So, we must have: $c^{(14)}=c^{(2)}+24$ implying $40 < c^{(\ell=14)} < 48$, for tensor-product $(2,14)$ CFTs. Here again we find an example where the central charge range for tensor-product solutions lie in a smaller range compared to the full admissible range.

\subsection{Merging of movable poles in the general case}

Here we examine the general case, namely $\ell=6r,6r+2,6r+4$ for arbitrarily large values of $r$, the number of movable poles. Now from the poles $p_1,p_2,\cdots p_r$, we send $p_r\to 0$, or $p_r\to 1728$, or $p_r\to p_{r-1}$. In each of these situations, the poles $p_1,p_2,\cdots,p_{r-2}$ are held fixed. 

The exponents around the special points $p_r=0,1728$ and around $p_r=p_{r-1}$ can now be derived by inspection of the general MLDEs \eref{MLDEjspace}. But in fact there is a simpler way to find the same results. The behaviour at special points for $\ell=0,2$ and at a generic isolated pole has long been understood (starting with \cite{Naculich:1988xv}) and we have listed the possible exponents in Eqs. (\ref{exp_2-char}, \ref{alpha_i_sum}, \ref{indi_sig}). Now when a pole merges with any of the above, we need to add 1 to either of the exponents, keeping the lower one distinct from the upper and also not allowing an integral difference between the exponents in the case of the points $(0,1728)$. Morever at these two points the added exponent of 1 can be broken into factors of $\frac13$ or $\half$ respectively. The result is as follows:
\begin{table}[H]
\begin{center}
\begin{tabular}{|c|c|c|}
\hline
$\ell$ & Limit & Exponents \\
\hline
$6r$ & $p_r\to 0$ & $(0,\frac43), (\frac13,1)$\\
& $p_r\to 1728$ &  $(0,\frac32), (\half,1)$\\
& $p_r\to p_{r-1}$ & $(0,3),(1,2)$\\
\hline
$6r+2$ & $p_r\to 0$ & $(0,\frac53), (\frac23,1)$\\
& $p_r\to 1728$ &  $(0,\frac32), (\half,1)$\\
& $p_r\to p_{r-1}$ & $(0,3),(1,2)$\\
\hline
$6r+4$ & $p_r\to 0$ & $(\frac13,\frac53), (\frac23,\frac43)$\\
& $p_r\to 1728$ &  $(0,\frac32), (\half,1)$\\
& $p_r\to p_{r-1}$ & $(0,3),(1,2)$\\
\hline
\end{tabular}
\end{center}
\end{table}

From the exponents we learn whether, and what, we can factorise from the solution. If the lower exponent is 0 we have a non-factorisable solution, while if it is $\frac13, \frac23,1$ we can factorise $j^\frac13, j^\frac23, (j-p_{r-1})$ respectively. In particular this means that if $p_r\to p_{r-1}$ then the character $\zeta$ after extracting $(j-p_{r-1})$ loses its dependence on the merged pole entirely.

\section{Illustrative examples of genuine CFTs}

\label{illustrative}

The main focus of this paper has been on finding admissible solutions to MLDE with $\ell\ge 6$. In general one expects that some, though not all, of the admissible solutions will be actual CFTs. Completely classifying these is a major project, perhaps an unachievable one, but one may check whether at least some illustrative CFTs can be found for each of the classes we have considered. We do this here.

There are two cases for which CFTs are already known: the $(2,6)$ and $(2,8)$ MLDEs with the movable pole away from the boundary of moduli space. For the $(2,6)$ case, this has been done in \cite{Chandra:2018ezv} making use of a relation derived in Eq.(3.6) of \cite{Gaberdiel:2016zke} that relates three Wronskian indices: ${\cal L}$ for a meromorphic theory which is the numerator in a coset relation, 
$\ell$ for the denominator theory and $\tilde\ell$ for the coset theory \footnote{The equation as written in \cite{Gaberdiel:2016zke} involves $N\equiv \frac{c}{24}$ where $c$ is the central charge of the meromorphic theory, but it is easily verified that $6N=\frac{c}{4}={\cal L}$.}. We write the equation for the case of two characters:
\be
\tilde\ell=2({\cal L}+1)-6n-\ell
\label{ellcoset}
\ee
Here $n\ge 1$ is an integer labelling the sum of dimensions of the non-trivial primary for the denominator and coset theories. One finds $n=2$ whenever the coset is of the type where a simple factor of a Kac-Moody algebra is deleted by a corresponding denominator, and $n=1$ for non-trivial embeddings of Kac-Moody algebras in the numerator. Now, \cite{Chandra:2018ezv} considers ${\cal L}=8$ and $\ell=0$, corresponding to cosets of a $c=32$ meromorphic theory, where the embedding is of the ``deletion'' type with $n=2$. This results in $\tilde\ell=6$. Nearly 150 CFTs of this form are listed in Appendix A of \cite{Chandra:2018ezv}. These belong to a class with complete Kac-Moody algebras, which means the stress tensor is pure Sugawara with no additional contribution. 

For $(2,8)$, a number of CFTs can be found in  \cite{Mukhi:2022bte}. Although Wronskian indices are not the main focus of this paper, some of the cosets considered are of meromorphic theories with $c=24$ (and hence ${\cal L}=6$) with a non-trivial embedding of the Kac-Moody algebra of a denominator with $\ell=0$. Thus $n=1$ and \eref{ellcoset} gives $\tilde\ell=8$. Several theories of this type are included in Table 1 of that paper. It may be mentioned that the method used in that work only reproduces theories in the range $c<25$. However for the $(2,8)$ case, \eref{c_relns} also allows the range $c\in (40,48)$. This can potentially be realised with ${\cal L}=12$ and $n=3$ in \eref{ellcoset}, but it is not clear if $n=3$ is allowed for two-character meromorphic cosets, so we leave this question for the future. 

Next we move on to the case of $(2,12)$ with generic poles. \eref{ellcoset} tells us that this can be realised by a non-trivial embedding of Kac-Moody algebras (with $n=1$) in a $c=32$ meromorphic CFT (for which ${\cal L}=8$). Such embeddings have not been completely classified, even for the complete KM algebra case, but one can start with the meromorphic CFTs corresponding to the 132 even, unimodular Kervaire lattices \cite{Kervaire:1994} and take a coset by non-trivially embedding $A_{1,1}$ into any of the simple factors of the numerator. This will result in the desired $(2,12)$ CFT and one expects that most of them will satisfy MLDEs with generic (non-coincident) poles.  As an example, take the Kervaire lattice with root system $A_{9,1}^2E_{7,1}^2$ and quotient by $A_{1,1}$, embedding it in $E_{7,1}$. The quotient theory has central charge $31$ and algebra $A_{9,1}^2D_{6,1}E_{7,1}$ with $m_1=397$.

Similarly, the coset of a $c=48$ meromorphic theory by $A_{1,1}$ with a trivial embedding of the KM algebra ($n=2$) will give $\ell=14$. We are not aware of a classification of even, unimodular lattices of dimension 48 even under the restriction of having complete root systems, but one possibility is $A_{1,1}^{48}$ and the trivial embedding deletes one of the 48 factors leaving an $\ell=14$ CFT with Kac-Moody algebra $A_{1,1}^{47}$. There will surely be many more examples, most of which should have generic poles.

Now we turn to a CFT example for $(2,6)$ with $p_1\to 0$, a limit studied above that effectively corresponds to coincident poles. As we saw in Sub-section \ref{merge26}, in this limit there are two possible sets of exponents at 0, namely $(0,\frac43)$ and $(\frac13,1)$. As explained there, these correspond respectively to characters of non-factorisable and factorisable type.

Within the factorisable or $(\frac13,1)$ type, we can easily find a set of tensor-product $(2,6)$ CFTs with characters of the form in \eref{j13change}, namely $\chi_i(j)=j^\frac13 \zeta_i(j)$ where $\zeta(j)$ are the characters of a $(2,2)$ CFT. The latter have central charges $c\in (16,24)$ and are enumerated in \cite{Gaberdiel:2016zke}. Since multiplication by $j^\frac13$ increases $c$ by 8, the result has central charge $c\in (24,32)$ consistent with the expected range from \eref{c_relns}. The question now is whether we can get a $(2,6)$ CFT with factorised characters where $\zeta_i(j)$ does {\em not} represent a CFT, though $\chi_i(j)$ does. This is easily answered. For $\chi_i(j)$ to be admissible, $\zeta_i(j)$ must at least be a quasi-character, as then it is possible for multiplication by $j^\frac13$ to turn the negative coefficients positive. However, the central charge associated to $\zeta$ must still lie in the range $c\in (16,24)$. But in this range there are no quasi-characters as we can see from \eref{quasi22}. We conclude that there are no factorised $(2,6)$ characters other than those of admissible CFT (this in contrast to the case of $(2,4)$ discussed in sub-section \ref{6rplus4}).

Turning now to the non-factorisable case, where the indices are $(0,\frac43)$ and the CFT is irreducible. As shown in sub-section \ref{merge26}, this case arises when $p_1$ and $b_{4,1}$ vanish together. As an example, in Eqs (\ref{26a1ii}, \ref{26b41ii}) this happens when $m_1=2875$. Similarly there is a unique $m_1$ that achieves this for all the other cases in sub-section \ref{quasiMLDEcomp}. Now  all non-factorisable examples that arise as cosets of 32d lattices having complete Kac-Moody algebras were listed in Appendix A of \cite{Chandra:2018ezv}.  The relation between ${\cal N}$ of that Appendix and the above $m_1$, which we temporarily denote $m_1^{\rm coset}$, is easily seen to be $m_1^{\rm coset}={\cal N}-m_1^{\rm denom}$ where $m_1^{\rm denom}$ is the dimension of the KM algebra of the denominator in the coset. With this one finds that in none of the cases can the character with indices $(0,\frac43)$ be associated with a coset CFT. We do not know of a deep reason why this should be the case.

We move on to $(2,8)$ with $p_1\to 0$. Here the possible indices are $(0,\frac53)$ and $(\frac23,1)$. We again see that factorised solutions corresponding to the latter case are trivially possible and they lie in the sub-range $c\in (16,24)$. To populate the other sub-range in \eref{c_relns}, namely $c\in (40,48)$, we now have a possibility: consider quasi-characters for $\ell=0$ in the range $24<c<32$, for example the $c=25$ quasi-character in the $A_1$ series. While this is not admissible, multiplying it by $j^\frac23$ makes it admissible and it has $c=41$. In this case, the result is a tensor product of $E_{8,1}$ times the exotic $c=33$ theory of \cite{Grady_2020}. But more general non-tensor-product theories could well exist.

Finally we look at the case of $(2,12)$. In the coincident limit $p_2\to p_1$, one can look for factorisable as well as non-factorisable characters. In the factorisable case one simply has $(j-p_1)\zeta_i^{(\ell=0)}$ where $\zeta_i$ is an MMS character hence lies in the range $c\in (0,8)$. The result is in the range $c\in (24,32)$ and will have indices $(1,2)$. Searching for the non-factorisable case with indices $(0,3)$ is more trivial and we leave it for the future.

\section{Discussion and conclusions}

Since this has been a lengthy discussion, let us review the chain of arguments that led to our understanding of MLDEs with movable poles. The first step is to suitably parametrise the MLDE. This was done in Section \ref{MLDEtau} in the $\tau$ parameter and Section \ref{MLDEj} in the $j$ parameter. The next several steps use the latter form of the equation. Assuming a generic location of poles, we examined the behaviour of solutions around each movable pole and thereby derived a constraint equation (the ``accessory equation'') relating the accessory parameters to the poles (sub-sections \ref{access26}, \ref{accessgen}). The accessory equations describe an $r$-dimensional sub-manifold of the $2r$-dimensional space of poles and accessory parameters (it is an algebraic variety if we rationalise all denominators to make the equations polynomial, though the form of the equations in the general case is simpler without doing so). We then took one of the poles to infinity and were thereby led to the boundary of this sub-manifold. The original accessory equations now reduced to two types of equations: accessory equations for the case with one less movable pole, but with a modified accessory parameter, and an equation that determines the ratio of the accessory parameter and the pole on the boundary. Together, these determined the critical exponents (hence $c,h$) of solutions with $r$ movable poles in terms of solutions with $r-1$ movable poles. Applying this recursively gave us the allowed central charges for any number of movable poles, displayed in Section \ref{cent_relns}.

Next, in Section \ref{sec26} we returned to the  single-pole MLDE as a function of $\tau$ and computed the Frobenius solution. Inserting the known allowed values of $c$ then determined both the characters completely up to an arbitrary integer. One cannot reduce further since it is known \cite{Harvey:2018rdc, Chandra:2018pjq} that there is exactly one free integer parameter for each movable pole. We were also able to relate our results precisely with the quasi-character construction of \cite{Chandra:2018pjq} which constructs admissible characters for generic Wronskian index without any use of the corresponding MLDE, and precise agreement was found. An analogous discussion in Section \ref{sec212} considered the case of two movable poles. In Section \ref{mergepoles} we considered what happens when one violates the genericity assumption by merging poles. The equation and solutions remain well-defined when one pair of poles is merged, though they become singular if we simultaneous merge more than two poles. Finally in Section \ref{illustrative} we gave just a few examples of CFTs for the various cases we considered, showing that they are populated by genuine CFTs, and leaving a more detailed analysis for the future.

Our analysis makes it clear that whenever there are movable poles, there is an equal number of free integer parameters in the admissible solutions. This fact has previously been noted in \cite{Chandra:2018pjq}, but here we have re-obtained it directly from MLDE. This means there is an infinite set of admissible characters for every $\ell\ge 6$. However it can be argued that the number of CFTs for a given $\ell\ge 6$ is finite. For example, a result of \cite{Mukhi:2022bte} implies that every $(2,6)$ CFT with $c<32$ is a coset of a $c=32$ meromorphic CFT. At the same time,  \eref{c_relns} makes it clear that there are no $(2,6)$ admissible characters (hence no CFT) for $c\ge 32$. So in fact, all $(2,6)$ CFTs are cosets of $c=32$ meromorphic theories. It is expected that the latter are finite in number (though the number is enormous) which implies that the former number is also finite.

We conclude with a discussion of some open questions. For any number $n$ of characters, movable poles are present for $\ell\ge 6$ but the corresponding MLDEs do not appear to have been studied at all. Even for low values like $n=3,4,5$, the classifications in the literature 
\cite{Mathur:1988gt, Gaberdiel:2016zke, Franc:2016, franc2020classification, Kaidi:2021ent, Das:2021uvd, Bae:2021mej, Das:2022uoe} are all at $\ell=0$. Things are only slightly better using the alternate approaches of Hecke operators and quasi-characters. Ref. \cite{Harvey:2018rdc} constructed some admissible characters for the $(3,6)$ case as Hecke images of the Ising model characters. Several sets of quasi-characters solving the $(3,0)$ MLDE were found in \cite{Mukhi:2020gnj} and their linear combinations were shown to provide admissible characters with $\ell=6$. A concrete example of a $(3,6)$ CFT was also provided in this reference. Beyond this, the space of admissible characters and CFT for $n\ge 3$ characters arising from MLDE with movable poles, is essentially unexplored. It should certainly be possible to gain some insights into at least the $(3,6)$ case from the MLDE following the methods used here.

Another space of MLDEs that is largely unexplored is the $(n,0)$ class -- with $n$ characters but no poles. Like the $(2,\ell)$ case studied in the present paper, $(n,0)$ also involves a proliferation of parameters for $n\ge 6$, however clearly these do not correspond to poles or accessory parameters and one has to find a useful interpretation. Moreover the number of exponents is $n$, if this is large the analysis may be quite difficult. Nevertheless, as the existing literature shows, a lot can be learned bout RCFT by exploring modular differential equations, and we hope to report on more of the open problems in the future.

\section*{Acknowledgements}

AD would like to thank Jishu Das and Naveen Balaji Umasankar for useful discussions on modular forms and MLDEs. He would also like to thank Sigma Samhita for helpful discussions regarding SageMath. CNG thanks Iosif Bena and gratefully acknowledges the hospitality of CEA Saclay where some part of this work was done. CNG also thanks Bobby Acharya, Paolo Creminelli, Atish Dabholkar and gratefully acknowledges the hospitality of the High-Energy Section of the ICTP where some part of this work was done. SM would like to thank the Department of Theoretical Physics at CERN, Geneva for its warm hospitality. JS would like to thank Suresh Govindarajan for valuable discussions. He gratefully acknowledges the hospitality of the School of Physical Sciences at NISER, Bhubaneswar. He would also like to acknowledge support from the Institute Postdoctoral fund of IIT Madras. 

\begin{appendix}

\section{Critical indices at the poles}\label{indices}

In this Appendix we review the leading behaviour of the character $\chi(j)$, about various points in the upper half plane, following \cite{Naculich:1988xv}. 

About $\tau=\rho$ we have $j\to 0$ and the leading behaviour of the characters are parametrised as:
\begin{align*}
    &\chi_0 \sim j^{\alpha_0^{(\rho)}} \\
    &\chi_1 \sim j^{\alpha_1^{(\rho)}}
\end{align*}
$\alpha_0^{(\rho)}$ and $\alpha_1^{(\rho)}$ must be non-negative multiples of $\frac{1}{3}$ to ensure regularity of the characters around $\tau=\rho$. Now let us compute the leading behaviour of the Wronskian $W(j)$ about $\tau=\rho$:
\begin{align}
    W(j) &\sim j^{\alpha_0^{(\rho)}}\left(-j\frac{E_6}{E_4}\right)\partial_j (j^{\alpha_1^{(\rho)}}) \nn\\
    &\sim j^{\alpha_0^{(\rho)}+\alpha_1^{(\rho)}}\left(\frac{E_6}{E_4}\right) \nonumber\\
& \sim 
j^{\alpha_0^{(\rho)}+\alpha_1^{(\rho)}-\frac{1}{3}} = j^{\frac{\ell_\rho}{6}}
\end{align}
where we used the fact that $E_4 = j^{\frac{1}{3}}\Delta^{\frac{1}{3}}$ and both $\Delta$ and $E_6$ are non-vanishing at $\tau=\rho$.
Thus, we get: 
\begin{align}
    \alpha_0^{(\rho)}+\alpha_1^{(\rho)}-\frac{1}{3}=\frac{\ell_\rho}{6}. \label{indi_r}
\end{align}

About $\tau=i$ we have $j\to 1728$ and the leading behaviour of characters is parametrised as:
\begin{align*}
    &\chi_0 \sim (j-1728)^{\alpha_0^{(i)}} \\
    &\chi_1 \sim (j-1728)^{\alpha_1^{(i)}}
\end{align*}
with $\alpha_0^{(i)}$ and $\alpha_1^{(i)}$ being non-negative multiples of $\frac{1}{2}$. This is to ensure regularity of characters around $\tau=i$. Now let us compute the leading behaviour of the Wronskian $W(j)$ about $\tau=i$:
\begin{align}
    W(j) &\sim (j-1728)^{\alpha_0^{(i)}}\left(-j\frac{E_6}{E_4}\right)\partial_j (j-1728)^{\alpha_1^{(i)}}  \nonumber\\
    &\sim (j-1728)^{\alpha_0^{(i)}}(j-1728+1728)(j-1728)^{\alpha_1^{(i)}-1}\left(\frac{E_6}{E_4}\right) \nonumber\\
    &\sim (j-1728)^{\alpha_0^{(i)}}\left[(j-1728)(j-1728)^{\alpha_1^{(i)}-1}+1728(j-1728)^{\alpha_1^{(i)}-1}\right]\left(\frac{E_6}{E_4}\right) \nonumber\\
    &\sim (j-1728)^{\alpha_0^{(i)}+\alpha_1^{(i)}}\left[1+1728(j-1728)^{-1}\right]\left(\frac{E_6}{E_4}\right) \nonumber\\
    \text{note,} \, & E_6 = (j-1728)^{\frac{1}{2}}\Delta^{\frac{1}{2}} \nonumber
\end{align}
From this we get:
\begin{align}
    &W(j) \sim (j-1728)^{\alpha_0^{(i)}+\alpha_1^{(i)}-\frac{1}{2}} (E_4^{-1}\Delta^{1/2}) \sim (j-1728)^{\alpha_0^{(i)}+\alpha_1^{(i)}-\frac{1}{2}} \sim (j-1728)^{\frac{\ell_i}{6}}
\end{align}
where we used the fact that $E_4$ and $\Delta$ are finite at $\tau=i$. Then we have:
\begin{align}
    \alpha_0^{(i)}+\alpha_1^{(i)}-\frac{1}{2} = \frac{\ell_i}{6}. \label{ind_i}
\end{align}

Next let us study the leading behaviour of the Wronskian about a movable pole say, $j = p_1$. We parametrise this by:
\begin{align*}
    &\chi_0 \sim (j-p_1)^{\alpha_0^{(1)}} \\
    &\chi_1 \sim (j-p_1)^{\alpha_1^{(1)}}
\end{align*}
with $\alpha_0^{(1)}$ and $\alpha_1^{(1)}$ being non-negative integers to ensure regularity of characters around $j=p_1$. Now the leading behaviour of the Wronskian $W(j)$ about $j=p_1$ is:
\begin{align}
    &W(j) \sim (j-p_1)^{\alpha_0^{(1)}+\alpha_1^{(1)}-1} \sim (j-p_1)^{\frac{\ell_\tau}{6}}
\end{align}
and we find:
\begin{align}
    \alpha_0^{(1)}+\alpha_1^{(1)}-1 = \frac{\ell_\tau}{6}. \label{ind_genr}
\end{align}

\section{$\ell$ is even for $2$-character solutions}
In this appendix we will show that for $2$-character solutions, $\ell$ is always even. This result was first obtained in \cite{Naculich:1988xv} using monodromy arguments for solutions around $\tau=i$. It was shown that if $\ell$ is odd then the monodromy is reducible, implying the solution space becomes one-dimensional and hence is not allowed. Here we will approach the problem in a slightly different way but will arrive at the same conclusion.

Using equations \eref{alldenoms} and \eref{phitwobasis} we write the $(2,\ell)$ MLDE in the $j$-plane for $\ell=6r+1, 6r+3$ and $6r+5$.
\begin{align*}   
&\ell=6r+1\!: \, \, \partial^2_j\chi(j)+\left[-\sum\limits_{I=1}^{r-1}\frac{1}{j-p_I}\right]\partial_j\chi(j)+
\frac{\alpha_0\alpha_1}{j^2(j-1728)}\frac{\prod\limits_{I=1}^r(j-b_{4,I})}{\prod\limits_{I=1}^{r-1}(j-p_{I})}\chi(j) = 0.
\end{align*}
\begin{align*}
&\ell=6r+3\!: \, \, \partial^2_j\chi(j)+\left[\frac{2}{3j}-\sum\limits_{I=1}^{r}\frac{1}{j-p_I}\right]\partial_j\chi(j)+
\frac{\alpha_0\alpha_1}{j(j-1728)}
\frac{\prod\limits_{I=1}^r(j-b_{4,I})}{\prod\limits_{I=1}^{r}(j-p_I)}\chi(j) = 0.
\end{align*}
\begin{align}
&\ell=6r+5\!: \, \, \partial^2_j\chi(j)+\left[\frac{1}{3j}-\sum\limits_{I=1}^{r}\frac{1}{j-p_I}\right]\partial_j\chi(j)+\frac{\alpha_0\alpha_1}{j(j-1728)}\frac{\prod\limits_{I=1}^{r}(j-b_{4,I})}{\prod\limits_{I=1}^{r}(j-p_I)}\chi(j) = 0
\label{twocharj_odd}
\end{align}
Comparing the above MLDEs to the ones given in \eref{twocharj}, we notice a striking difference, namely the term $\frac{1}{2(j-1728)}$ is missing in the first-derivative term. We shall see that the absence of this term is crucial to ruling out odd $\ell$ values.

Suppose we expand the characters around $\tau=i$ as given in \eref{ixp}. The indicial equation in all the above cases gives $(\alpha_0^{(i)}, \alpha_0^{(i)}) = (0,1)$. In fact we could have already found these values from \eref{ind_i}. 
Now the next order is interesting. Due to the absence of $\frac{1}{2(j-1728)}$ in the linear derivative term, we get no contribution from this term at this order. So for the character with exponent $\alpha_0^{(i)}=0$, we find at this order:
\begin{align}
    \alpha_0\alpha_1\prod\limits_{I=1}^{r}(1728-b_{4,I}) = 0, \label{constr_odd_l}
\end{align}
The above implies either $\alpha_0\alpha_1=0$ or $b_{4,I}=1728$ for some $1\leq I\leq r$. The second choice is ruled out since it leads to removal of the pole at $j=1728$ in the last term. Since $\frac{1}{2(j-1728)}$ is also absent in the middle term, the MLDE has now no poles about $j=1728$ and thus the expansion \eref{ixp} does not make sense. So we must rule out this possibility.

Now let us move on to the other possibility $\alpha_0\alpha_1=0$. From the footnote in section \ref{merge26} we know that whenever this happens, the solution space becomes $1$-dimensional. So this too is ruled out. 

Thus, the above considerations rule out all odd $\ell$ values for two-character solutions.


\section{Frobenius solutions of MLDEs}

\subsection{$(2,0)$ and $(2,2)$ MLDEs}\label{app2022}

Here we review the well-established method for recursively solving the MLDE in the cases $\ell=0,2$. The $(2,0)$ MLDE in the $\tau$-plane is given in \eref{MMSeqn}. It is a one-parameter MLDE, the only parameter is the rigid parameter: $\alpha_0\alpha_1 = -\frac{c(c+4)}{576}$. Its solutions are :
\bea \label{20solndefn}
\chi_0(q) = q^{-\frac{c}{24}}\,\sum_{k = 0}^{\infty} m^{(0)}_k(c)\,q^k, \qquad \qquad \chi_1(q). = q^{\frac{c+4}{24}}\,\mathsf{D}\,\sum_{k = 0}^{\infty} m^{(0)}_k(-c-4)\,q^k.
\eea
Here $\mathsf{D}$ is the apparent degeneracy of the non-identity character. The $m^{(0)}_k(c)$'s are rational functions of the central charge $c$; the superscript indicates the fact that these belong to the $l = 0$ solution. We give here the first few: we have  $m^{(0)}_0(c) = 1$ and then for $k \geq 1$:
\be \label{20mcs}
m^{(0)}_k(c) \equiv (-1)^k \frac{N^{(0)}_k(c)}{D^{(0)}_k(c)}.
\ee
The $D^{(0)}_k(c)$ are the denominator polynomials:
\be \label{20den}
D^{(0)}_k(c) = k!~\Pi_{l=0}^{k-1}(c - 10 - 12l)
\ee
and the $N^{(0)}_k(c)$'s are the numerator polynomials, of which the first few are:
\bea \label{20num}
N^{(0)}_1(c) &=& 5c^2 + 22 c \nonumber \\
N^{(0)}_2(c) &=& 25 c^4 + 175 c^3 + 508 c^2 + 804 c \nonumber \\
N^{(0)}_3(c) &=& 125 c^6 + 975 c^5 + 10330 c^4 + 68308 c^3 + 148872 c^2 + 33344 c \nonumber \\
N^{(0)}_4(c) &=& 625 c^8+4250 c^7 + 136475 c^6 + 1359450 c^5 + 6793624 c^4 + 22169872 c^3 \nonumber \\
&&+ 38327216 c^2 + 18775968 c \nonumber \\
N^{(0)}_5(c) &=& 3125 c^{10} + 12500 c^9 + 1464375 c^8 + 16026500 c^7 + 1629216204 c^6 + 1246732800 c^5 \nonumber \\  &&+ 5241174800 c^4 + 12353480000 c^3 + 14698399680 c^2 + 2755008000 c \nonumber \\ 
N_6^{(0)}(c) &=& 15625 c^{12} - 9375 c^{11} + 13815625 c^{10} + 132866875  c^9 + 2911676350  c^8 + 32677746940  c^7 \nonumber \\ && + 238017546040  c^6 + 1317574464400  c^5 + 4550303524000  c^4 +  8002202756160  c^3 \nonumber \\&& + 6057775308160  c^2  + 2846891980800 c
\eea
We note that $D^{(0)}_k(c)$ is a polynomial of degree $k$, $N^{(0)}_k(c)$ is a polynomial of degree $2k$ with the leading coefficient being $5^k$ and vanishing constant term. 
Examples that will be relevant to the main text are:
\be
\begin{split}
m_1^{(0)} & = -\frac{5c^2 + 22 c}{c-10} \\
m_2^{(0)} & = \frac{25 c^4 + 175 c^3 + 508 c^2 + 804 c}{2 (c-10) (c-22)} \\
m_3^{(0)} & =  - \frac{125 c^6 + 975 c^5 + 10330 c^4 + 68308 c^3 + 148872 c^2 + 33344 c}{6 (c-10) (c-22) (c-34)}
\end{split}
\label{app20m1m2}
\ee

The $(2,2)$ MLDE in the $\tau$-plane is given in \eref{leq2eqn}. It is a one-parameter MLDE, the only parameter is the rigid parameter : $\alpha_0\alpha_1 = -\frac{c(c-4)}{576}$. It's solutions are :
\bea \label{20solndefn}
\chi_0(q) = q^{-\frac{c}{24}}\,\sum_{k = 0}^{\infty} m^{(2)}_k(c)\,q^k, \qquad \qquad \chi_1(q). = q^{\frac{c - 4}{24}}\,\mathsf{D}\,\sum_{k = 0}^{\infty} m^{(2)}_k(-c + 4)\,q^k.
\eea
Here, as before, $\mathsf{D}$ is the apparent degeneracy of the non-identity character. The $m^{(2)}_k(c)$ are rational functions of the central charge $c$ and the superscript indicates the fact that these belong to the $l = 2$ solution. We give here the first few: again we start with $m^{(2)}_0(c) = 1$ and then for $k \geq 1$ we get:
\be \label{22mcs}
m^{(2)}_k(c) \equiv (-1)^k \frac{N^{(2)}_k(c)}{D^{(2)}_k(c)}.
\ee
$D^{(2)}_k(c)$'s are the denominator polynomials
\be \label{20den}
D^{(2)}_k(c) = k!~\Pi_{l=0}^{k-1}(c - 14 - 12l)
\ee
and the $N^{(2)}_k(c)$'s are the numerator polynomials, the first few are:
\bea \label{20num}
N^{(2)}_1(c) &=& 5c^2  - 142 c \nonumber \\
N^{(2)}_2(c) &=& 25 c^4 - 1465 c^3 + 8980 c^2 - 45420 c \nonumber \\
N^{(2)}_3(c) &=& 125 c^6 - 11325 c^5 + 159550 c^4 - 1931740 c^3 + 7672440 c^2 - 19603520 c \nonumber \\
N^{(2)}_4(c) &=& 625 c^8 - 77750 c^7 + 1850075 c^6 - 37721430 c^5 + 336005080 c^4 - 2291330800 c^3 \nonumber \\
&&+ 7121862320 c^2 - 12830855520 c  \nonumber \\
N^{(2)}_5(c) &=& 3125 c^{10}-500000 c^9+17589375 c^8-523745000 c^7+7785543020 c^6-93188748960 c^5 \nonumber \\
&&+632315675600 c^4-3135002595200 c^3+8096425231680 c^2-11243426250240 c \nonumber \\
N^{(2)}_6(c) &=& 15625 c^{12}-3084375 c^{11}+148590625 c^{10}-5971718125 c^9+130232057350
   c^8\nn\\
   &&-2331666656740 c^7 
    +26060719253080 c^6-228682548002800 c^5\nn\\
    &&+1273810283284000
   c^4-5062375605466560 c^3  +11295987759233920 c^2\nn\\
   && -13145822789068800 c
\eea
Examples that will be relevant to the main text are:
\be
\begin{split}
m_1^{(2)} & = -\frac{5c^2  - 142 c}{c-14} \\
m_2^{(2)} & = \frac{25 c^4 - 1465 c^3 + 8980 c^2 - 45420 c}{2 (c-14) (c-26)} \\
m_3^{(2)} & =  - \frac{125 c^6 - 11325 c^5 + 159550 c^4 - 1931740 c^3 + 7672440 c^2 - 19603520 c}{6 (c-14) (c-26) (c-38)}
\end{split}
\label{app20m1m2C}
\ee

We will now use the above results to understand the case of MLDEs with movable poles, specifically $(2,6)$ and $(2,8)$.

\subsection{$(2,6)$ and $(2,8)$ MLDEs.}\label{app2628}

Now we exhibit the Frobenius solution of MLDEs with one movable pole. This sub-section contains formulae that will be referred to in the main text of the paper.
In section \ref{twosixsoln}, we solved the $(2,6)$ MLDE.  In the first step, one computes the first three orders of the Frobenius solution for the identity character and obtains \eref{ord1a} and \eref{ord2a} where the functions $f_1(c, p_1, b_{4,1})$ and $f_2(c, p_1, b_{4,1})$ are given by :
\bea
\label{ord1b}
f_1(c, p_1, b_{4,1}) &=&   -\frac{1}{48 (c-22)}(240 c(c-94)+c (c+4)\, p_1 - c (c-20)\,b_{4,1})   \\
\label{ord2b}
f_2(c, p_1, b_{4,1}) &=& \frac{1 }{96 (c-34)} \left(-720 c (243 c-17294) -240 \left(c^2+50 c-1392\right) m_1^{(6)} + \right. \nonumber \\
&& \left. + \left( 24\,c(c+8) - (c-20)(c-24)m_1^{(6)} \right) p_1 + c (c-20) (216 + m_1^{(6)}) b_{4,1} \right). \nonumber \\
\eea
In the next step, we solved for  three parameters in terms of objects associated to the identity character viz. the central charge $c$, the Fourier coefficients $m_1^{(6)}$ and $m_2^{(6)}$. For the non-rigid parameters, we obtained the equations \eref{nr1a} and \eref{nr2a} where $f_3(c, m_1^{(6)}, m_2^{(6)})$ and $f_4(c, m_1^{(6)}, m_2^{(6)})$ are given by:
\bea \label{nr1b}
f_3(c, m_1^{(6)}, m_2^{(6)}) &=& \frac{1}{c(5c + 22) + (c-10) m_1^{(6)} } \Big( 285 c (9 c-554) + 24 (21 c - 92) m_1^{(6)}\nonumber \\  &&  - (c-22) (m_1^{(6)})^2   + 2\,(c-34)m_2^{(6)}\Big)  \\
\label{nr2b}
f_4(c, m_1^{(6)}, m_2^{(6)}) &=& \frac{1 }{ c (c-20)( c(5c + 22) + (c-10) m_1^{(6)})} \Big( 15 c^2 \left(251 c^2-17010 c-75192\right) \nonumber \\
&&  + 24 c (41\, c^2 - 1224\, c + 8064) m_1^{(6)} +2\,c (c+4)(c-34)m_2^{(6)}\nn\\
&& -(c-20)(c-22)(c-24)(m_1^{(6)})^2 \Big)   \nonumber \\
\eea
We then used the accessory equation and obtained a relation between $m_2^{(6)}$, $m_1^{(6)}$ and $c$ in \eref{m2mldea} where the $A_2(c)$ and $B_2(c)$ are:
\bea
\label{m2mldeb}
A_2(c)  &&= -\frac{25 c^4 - 2135 c^3 + 41140 c^2 + 224940 c}{2(c-22)(c-34)},\quad B_2(c) = -\frac{(c-24)(5c - 98)}{c-34}
\eea

Next we rewrote the pole and accessory parameters only in terms of $c$ and $m_1^{(6)}$; after substituting \eref{m2linear} in \eref{ord1a} and \eref{ord1b}. We obtained \eref{nr1.2a} and \eref{nr2.2a} where the $f_5(c, m_1^{(6)})$ and $f_6(c, m_1^{(6)})$ are given in:
\bea \label{nr1.2b}
f_5(c, m_1^{(6)}) &&=  -\frac{[ c(5c - 94) + (c-22) m_1^{(6)}] [ 5 c^2 - 470 c + 6912 + (c-22) m_1^{(6)}]}{(c-22)[ c(5c + 22) + (c-10) m_1^{(6)}]},  \\
\label{nr2.2b}
f_6(c, m_1^{(6)}) &&=  -\frac{[ c(5c - 94) + (c-22) m_1^{(6)}][c \left(5 c^2-590 c-2544\right)+ (c-22) (c-24) m_1^{(6)}]}{c(c-22)[ c(5c + 22) + (c-10) m_1^{(6)}]}, \nonumber \\
\eea
In the next step, we obtained the third Fourier coefficient of the identity character in \eref{m3mlde} where $A_3(c)$ and $B_3(c)$ are given by: 
\bea \label{m3mldeb}
A_3(c) &&= \frac{250 c^6-32700 c^5 +1373240 c^4 -18801040 c^3 + 90660480 c^2 + 892610560 c}{6 (c-46) (c-34) (c-22)}, \nonumber \\ 
B_3(c) &&= \frac{(c-24)\left(25 c^3-1625 c^2+35308 c-256188\right)}{2 (c-46) (c-34) }. 
\eea
We also obtained the fourth Fourier coefficient of the identity character in \eref{mklinear} (with $k=4$), where $A_4(c)$ and $B_4(c)$ are given by: 
\bea \label{m4mldeb}
A_4(c) &=& \frac{-1875 c^8+333750 c^7-21989925 c^6+680543850 c^5-12260107560 c^4+97916677200
   c^3}{24 (c-58) (c-46) (c-34) (c-22)}, \nonumber \\
   &&+ \frac{-87462415440 c^2-3618704872800 c}{24 (c-58) (c-46) (c-34) (c-22)} \nonumber \\
B_4(c) &=& -\frac{ (c-24)  \left(125 c^5-14025 c^4+636730 c^3-14585852 c^2+168166728
   c-778842496\right)}{6 (c-58) (c-46) (c-34) }. 
\eea

We now give  formulae that will be referred to in the main text of the paper, for the $(2,8)$ MLDE.  In the first step, one computes the first three orders of the Frobenius solution for the identity character and obtains \eref{ord1A} and \eref{ord2A} where the functions $\widetilde{f_1}(c, p_1, b_{4,1})$ and $\widetilde{f_2}(c, p_1, b_{4,1})$ are given by :
\bea
\label{ord1B}
\widetilde{f_1}(c, p_1, b_{4,1}) &=&   -\frac{1}{48 (c-26)}(48 c(5 c - 634)+c (c-4)\, p_1 - c (c-28)\,b_{4,1})   \\
\label{ord2B}
\widetilde{f_2}(c, p_1, b_{4,1}) &= &\frac{1 }{96 (c-38)} \left(-144 c (1615 c-167794)-48 \left(5 c^2+326 c-13104\right) m_1^{(8)}  \right. \nonumber \\
 && \left. - \left( 24 c (9 c+208) + (c-24)(c-28)m_1^{(8)} \right) p_1 + c (c - 28) (456 + m_1^{(8)})  b_{4,1} \right). \nonumber \\
\eea
In the next step, we solved for  three parameters in terms of objects associated to the identity character viz. the central charge $c$, the Fourier coefficients $m_1^{(8)}$ and $m_2^{(8)}$. For the non-rigid parameters, we obtained the equations \eref{nr1A} and \eref{nr2A} where $\widetilde{f_3}(c, m_1^{(8)}, m_2^{(8)})$ and $\widetilde{f_4}(c, m_1^{(8)}, m_2^{(8)})$ are given by:
\bea \label{nr1B}
&& \widetilde{f_3}(c, m_1^{(8)}, m_2^{(8)}) = \frac{1}{c(5c - 142) + (c-14) m_1^{(8)} } \left( 3 c (855 c - 71426) + 24 (21 c-52) m_1^{(8)} \right. \nonumber \\ 
&& \qquad \qquad \qquad \qquad \left.- (c-26) (m_1^{(8)})^2   + 2\,(c-38)m_2^{(8)}\right)  \\
\label{nr2B}
&& \widetilde{f_4}(c, m_1^{(8)}, m_2^{(8)}) =   \frac{1}{c (c - 28)(c(5c - 142) + (c-14) m_1^{(8)}) } \left( 3 c^2 \left(1255 c^2-136926 c+1726152\right) \right. \nonumber \\
&& \left. + 24 c \left(41 c^2-2088 c+25344\right)m_1^{(8)} + 2 c (c-38) (c-4)m_2^{(8)} -((c-28) (c-26) (c-24))(m_1^{(8)})^2  \right) \nonumber \\
\eea

Next we rewrote the pole and accessory parameters only in terms of $c$ and $m_1^{(6)}$; after substituting \eref{m2linear8} in \eref{nr1A} and \eref{nr2A}. We obtained \eref{nr1.2A} and \eref{nr2.2A} where the $\widetilde{f_5}(c, m_1^{(6)})$ and $\widetilde{f_6}(c, m_1^{(6)})$ are given in:
\bea \label{nr1.2B}
\widetilde{f_5}^{(8)}(c, m_1^{(8)}) &&=  -\frac{[ c(5c - 634) + (c-26) m_1^{(8)}] [ 5 c^2 - 634 c + 13824 + (c-26) m_1^{(8)}]}{(c-26)[ c(5c - 142) + (c-14) m_1^{(8)}]},  \\
\label{nr2.2B}
\widetilde{f_6}^{(8)}(c, m_1^{(8)}) &&=  -\frac{[ c(5c - 634) + (c-26) m_1^{(8)}][c \left(5 c^2 - 754 c + 8304\right)+ (c-26) (c-24) m_1^{(8)}]}{c(c-26)[ c(5c - 142) + (c-14) m_1^{(8)}]}, \nonumber \\
\eea

\section{Some (2, 8) MLDE solutions and quasi-characters}\label{wierd28}

Here we analyse the $(n,\ell)=(2, 8)$ MLDE in the same was as was done for $(2,6)$ in Section \ref{sec26}. The MLDE in the $j$-coordinate is given by:
\begin{eqnarray}
\left(\partial_j^2  + \left(\frac{1}{3j} + \frac{1}{2(j-1728)} - \frac{1}{(j-p_1)}\right)\partial_j + \frac{\alpha_0 \alpha_1(j-b_{4, 1})}{j(j-1728)(j-p_1)}\right) \chi(j) = 0 \label{MLDE28j}
\end{eqnarray} 
Using the series expansion $\chi_i = \sum\limits_{k=0}^{\infty} a_{i, k}^{(1)}\, (j-p_1)^{k+\alpha_i^{(1) }},\, a_{i, 0}^{(1)} \neq 0$, the indicial equation around $j=p_1$ is: 
\begin{eqnarray}
 \alpha_i^{(1) }\, (\alpha_i^{(1)}-2) = 0
\end{eqnarray}
At first subleading order, for the solution $\alpha_0^{(1)} = 0$ we get:
\begin{eqnarray}
a_{0, 1}^{(1)} = \frac{\alpha_0 \alpha_1(p_1 - b_{4, 1})}{p_1(p_1-1728)}
\end{eqnarray}
At second order beyond this, we find (as expected) that the $a_{0, 2}^{(1))}$ terms cancel resulting in a constraint equation:
\begin{eqnarray}
\alpha_0 \alpha_1  (p_1 - b_{4, 1})^2 + \left(1152 - \frac{7p_1}{6}\right)(p_1 - b_{4, 1}) + p_1(p_1-1728) = 0 \label{acceqn_28} 
\end{eqnarray} 


Now one analyses the Frobenius solution by going back to the $(2, 8)$ MLDE in the $\tau$-plane:
\bea
\left(D^2 + \Bigg(\frac{E_6}{3E_4}+\frac{E_4^2E_6}{E_4^3-p_I\Delta}\Bigg)D + \frac{\alpha_0 \alpha_1\, E_4\left(E_4^3 - b_{4, 1}\, \Delta\right)}{\left(E_4^3 - p_1\, \Delta\right)}\right) \chi(\tau) = 0 \label{MLDE28}
\eea
and using the methods explained in Section \ref{sec26}, we obtain:
\bea
p_1 &=& \frac{-738720 \alpha _0^2-12 \alpha _0
   \left(\left(m_1-504\right) m_1-2
   m_2+214278\right)-13 m_1^2+624 m_1+38
   m_2}{\left(12 \alpha _0+7\right) m_1-24 \alpha _0
   \left(60 \alpha _0+71\right)} \nonumber\\ 
b_{4, 1} &=& \frac{1}{\alpha _0 \left(6 \alpha _0+7\right) \left(1440
   \alpha _0^2+1704 \alpha _0-12 \alpha _0 m_1-7
   m_1\right)}\bigg(6505920 \alpha _0^4+29576016 \alpha _0^3\nn\\[2mm]
   &&+15535368
   \alpha _0^2 +72 \alpha _0^3 m_1^2-70848 \alpha
   _0^3 m_1-144 \alpha _0^3 m_2+234 \alpha _0^2
   m_1^2-150336 \alpha _0^2 m_1\nn\\
   &&-252 \alpha _0^2
   m_2+253 \alpha _0 m_1^2 - 76032 \alpha _0 m_1-38
   \alpha _0 m_2+91 m_1^2\bigg)   \label{28parameters}
\eea

Next we exhibit the quasi-characters for $\ell=6r+2$ cases, for which the initial quasi-characters solve the $\ell=2$ MLDE. These solutions exist for the following values of $c$:
\be
\begin{split}
\hbox{dual Lee-Yang family:}\quad c &=\frac{2(6n-1)}{5},~n\ne 1 \hbox{ mod }5\\
\hbox{dual } A_1 \hbox{ family:}\quad c &=6n-1\\
\hbox{dual } A_2 \hbox{ family:}\quad c &=4n-2,~n\ne 1\hbox{ mod }3\\
\hbox{dual } D_4 \hbox{ family:}\quad c &=12n-4
\end{split}
\label{quasi22}
\ee
Of these, the central charges: 
\be
c=\frac{82}{5},17,16,\frac{94}{5},20,\frac{106}{5},22,23,\frac{118}{5}
\label{ghmlist}
\ee
correspond to admissible characters \footnote{Again these all correspond to CFTs, except for the first and last cases that are Intermediate Vertex Operator Algebras \cite{Kawasetsu:2014}. A new feature here is that a single set of admissible characters corresponds to more than one CFT.}  with $\ell=2$ \cite{Gaberdiel:2016zke}. As before, linear combinations of these quasi-characters make up admissible characters with increasing values of $\ell$, this time in the family $\ell=6r+2$, and all such characters are generated.

We can now list the admissible $(2,8)$ solutions and express them in terms of quasi-characters.

\subsubsection*{Admissible Solutions (i)}

\bea
c=\frac{82}{5}, \quad m_1 = 410 +   87\, m,\quad m_2 = 64739 + 5510\, m,\quad m_3 = 2089934 + 95323\, m,\quad 0 < m \leq 2 \nonumber
\eea
For this case,
\bea
p_1 &=& \frac{4 (\text{$m_1$}-497) (\text{$m_1$}+943)}{\text{$m_1$}-410} \label{28a1viii} \\
b_{4, 1} &=& -\frac{4 (\text{$m_1$}+943) (19 \text{$m_1$}-11603)}{41
   (\text{$m_1$}-410)} \label{28b41viii}
\eea
The equations \eref{28a1viii} and \eref{28b41viii} satisfy \eref{acceqn_28}. 
This solution is equal to the following sum of quasi-characters:
\begin{align*}
    \chi^{\ell=8} = \chi^{\widetilde{LY}}_{n = 7} + N_1\,\chi^{\widetilde{LY}}_{n = -3}
\end{align*}
with the identification $m=N_1$.

\subsubsection*{Admissible Solutions (ii)} 

\bea
c=17, \quad m_1 = 323 +   11\, m,\quad m_2 = 60860 + 649\, m,\quad m_3 = 2158575 + 10480\, m,\quad 0 < m \leq 40 \nonumber
\eea
For this case,
\bea
p_1 &=& \frac{3 (\text{$m_1$}-499)
   (\text{$m_1$}+1037)}{\text{$m_1$}-323} \label{28a1iv} \\
b_{4,1} &=& -\frac{3 (\text{$m_1$}+1037) (7 \text{$m_1$}-5797)}{17
   (\text{$m_1$}-323)} \label{28b41iv}  
\eea
The equations \eref{28a1iv} and \eref{28b41iv} satisfy \eref{acceqn_28}.
This solution is equal to the following sum of quasi-characters:
\begin{align*}
    \chi^{\ell=8} = \chi^{\tilde{A}_1}_{n = 3} + N_1\,\chi^{\tilde{A}_1}_{n = -1}
\end{align*}
with the identification, $m=N_1$. This solution appears in \cite{Mukhi:2022bte}.

\subsubsection*{Admissible Solutions (iii)}

\bea
c=18, \quad m_1 = 234 +   5\, m,\quad m_2 = 59805 + 258\, m,\quad m_3 = 2482242 + 3690\, m,\quad 0 < m \leq 171 \nonumber 
\eea
For this case,
\bea
p_1 &=& \frac{2 (\text{$m_1$}-504)
   (\text{$m_1$}+1224)}{\text{$m_1$}-234} \label{28a1vii} \\
b_{4, 1} &=& -\frac{2 (\text{$m_1$}-1368) (\text{$m_1$}+1224)}{3
   (\text{$m_1$}-234)} \label{28b41vii}
\eea
The equations \eref{28a1vii} and \eref{28b41vii} satisfy \eref{acceqn_28}. This solution is equal to the following sum of quasi-characters:
\begin{align*}
    \chi^{(8)} = \chi^{\tilde{A}_2}_{n=5} + N_1\chi^{\tilde{A}_2}_{n=-1}
\end{align*}
with the identification $m=N_1$.

\subsubsection*{Admissible Solutions (iv)} 

\bea
c=\frac{94}{5}, \quad m_1 = 188 +   46\, m,\quad m_2 = 62087 + 2093\, m,\quad m_3 = 2923494 + 27002\, m,\quad 0 < m \leq 26 \nonumber
\eea
For this case,
\bea
p_1 &=& \frac{3 (\text{$m_1$}-510) (\text{$m_1$}+1410)}{2
   (\text{$m_1$}-188)} \label{28a1v}\\
b_{4,1} &=&  -\frac{3 (\text{$m_1$}+1410) (13 \text{$m_1$}-26790)}{94
   (\text{$m_1$}-188)} \label{28b41v}
\eea
The equations \eref{28a1v} and \eref{28b41v} satisfy \eref{acceqn_28}. This solution is equal to the following sum of quasi-characters,
\begin{align*}
    \chi^{\ell=8} = \chi^{\tilde{LY}}_{n = 8} + N_1\,\chi^{\tilde{LY}}_{n = -2}
\end{align*}
with the identification $m=N_1$. This solution appears in \cite{Mukhi:2022bte}.

\subsubsection*{Admissible Solutions (v)} 

\bea
c=20, \quad m_1 = 140 +   m,\quad m_2 = 69950 + 36\, m,\quad m_3 = 3983800 + 394\, m,\quad 0 < m \leq 1807 \nonumber 
\eea
For this case,
\bea
p_1 &=& \frac{(\text{$m_1$}-524) (\text{$m_1$}+1780)}{(\text{$m_1$}-140)} \label{28a1i}\\
b_{4,1} &=& -\frac{(\text{$m_1$}-3980) (\text{$m_1$}+1780)}{5
   (\text{$m_1$}-140)}\label{28b41i}
\eea

The equations \eref{28a1i} and \eref{28b41i} satisfy \eref{acceqn_28}. The quasi-character sum for this solution is:
\begin{align}
    \chi^{(8)} = \chi^{\tilde{D}_4}_{n=2} + N_1\chi^{\tilde{D}_4}_{n=0},
\end{align}
with the identification $m=N_1$. This solution appears in \cite{Naculich:1988xv} with $m = 960$.

\subsubsection*{Admissible Solutions (vi)} 

\bea
c=\frac{106}{5}, \quad m_1 = 106 +   17\, m,\quad m_2 = 84429 + 442\, m,\quad m_3 = 5825442 + 4063\, m,\quad 0 < m \leq 155 \nonumber
\eea
For this case,
\bea
p_1 &=& \frac{2 (\text{$m_1$}-548) (\text{$m_1$}+2332)}{3
   (\text{$m_1$}-106)} \label{28a1vi} \\
b_{4,1} &=&   -\frac{2 (\text{$m_1$}+2332) (7 \text{$m_1$}-59996)}{159
   (\text{$m_1$}-106)} \label{28b41vi}
\eea

The equations \eref{28a1vi} and \eref{28b41vi} satisfy \eref{acceqn_28}. This is equal to the following sum of quasi-characters:
\begin{align*}
    \chi^{\ell=8} = \chi^{\widetilde{LY}}_{n = 9} + N_1\,\chi^{\widetilde{LY}}_{n = -1}
\end{align*}
with the identification $m=N_1$. This solution appears in \cite{Mukhi:2022bte}.

\subsubsection*{Admissible Solutions (vii)} 

\bea
c= 22, \quad  m_1 = 88 +   m,\quad m_2 = 99935 + 19\, m,\quad m_3 = 7846300 + 155\, m,\quad 0 < m \leq 3436\quad  \nonumber
\eea
For this case,
\bea
p_1 &=& \frac{(\text{$m_1$}-574) (\text{$m_1$}+2882)}{2
   (\text{$m_1$}-88)} \label{28a1ii} \\ 
b_{4,1} &=& -\frac{(\text{$m_1$}-16126) (\text{$m_1$}+2882)}{22
   (\text{$m_1$}-88)} \label{28b41ii}
\eea

The equations \eref{28a1ii} and \eref{28b41ii} satisfy  \eref{acceqn_28}. This is equal to the following sum of quasi-characters:
\begin{align*}
    \chi^{(8)} = \chi^{\tilde{A}_2}_{n = 6} + N_1\,\chi^{\tilde{A}_2}_{n = 0}
\end{align*}
with the identification, $m=N_1$. This solution appears in \cite{Naculich:1988xv} with $m = 1782$.

\subsubsection*{Admissible Solutions (viii)} 

\bea \label{n10}
c = 23,\quad  m_1 = 69 + 5\, m,\quad m_2 = 131905 + 49\, m,\quad m_3 = 12195106 + 345\, m,\quad 0 < m \leq 996\quad  \nonumber
\eea
For this case,
\bea
p_1 &=& \frac{(\text{$m_1$}-629) (\text{$m_1$}+3979)}{3
   (\text{$m_1$}-69)} \label{28a1iii} \\
b_{4,1} &=& -\frac{(\text{$m_1$}-49013) (\text{$m_1$}+3979)}{69
   (\text{$m_1$}-69)} \label{28b41iii}
\eea

The equations \eref{28a1iii} and \eref{28b41iii} satisfy \eref{acceqn_28}. This solution is equal to the following sum of quasi-characters:
\begin{align*}
    \chi^{(8)} = \chi^{\tilde{A}_1}_{n = 4} + N_1\,\chi^{\tilde{A}_1}_{n = 0}
\end{align*}
with the identification, $m=N_1$. This solution appears in \cite{Mukhi:2022bte}.

\subsubsection*{Admissible Solutions (ix)}

\bea
c=\frac{118}{5}, \quad m_1 = 59 +   11\, m,\quad m_2 = 164315 + 44\, m,\quad m_3 = 16778125 + 285\, m,\quad 0\leq m \leq 591 \nonumber
\eea
For this case,
\bea
p_1 &=& \frac{(\text{$m_1$}-686) (\text{$m_1$}+5074)}{4
   (\text{$m_1$}-59)} \label{28a1ix}\\
b_{4, 1} &=&   - \frac{(\text{$m_1$}-164846) (\text{$m_1$}+5074)}{236
   (\text{$m_1$}-59)} \label{28b41ix}
\eea
The equations \eref{28a1ix} and \eref{28b41ix} satisfy \eref{acceqn_28}. This solution is equal to the following sum of quasi-characters:
\begin{align*}
    \chi^{\ell=8} = \chi^{\widetilde{LY}}_{n = 10} + N_1\,\chi^{\widetilde{LY}}_{n = 0}
\end{align*}
with the identification $m=N_1$.

\end{appendix}

\bibliographystyle{JHEP}

\bibliography{nonRigid}

\end{document}